\documentclass[sigconf, authorversion]{acmart}

\usepackage[utf8]{inputenc}
\usepackage[T1]{fontenc}

\usepackage{booktabs} 
\usepackage{amsmath}
\usepackage{amssymb}
\usepackage{amsfonts}
\usepackage{xspace}
\usepackage{url}
\usepackage{microtype}
\usepackage[binary-units]{siunitx}
\usepackage{listings}

\usepackage{algorithm}
\usepackage{algorithmicx}
\usepackage[noend]{algpseudocode}

\usepackage{tikz}
\usetikzlibrary{shapes.geometric,arrows,calc,decorations.pathreplacing}
\usepackage{pgf-umlsd}
\usepackage{placeins}
\usepackage{enumitem}

\usepackage{tagging}

\setcopyright{none}

\renewcommand\footnotetextcopyrightpermission[1]{} 
\newcommand{\productname}[1]{#1} 

\newcommand{\skampi}{\productname{SKaMPI}\xspace}
\newcommand{\reprompi}{\productname{ReproMPI}\xspace}
\newcommand{\reprompiurl}{\url{https://github.com/hunsa/reprompi}}

\newcommand{\osu}{\productname{OSU Micro-Benchmarks}\xspace}

\newcommand{\openmpi}{\productname{Open\,MPI}\xspace}

\newcommand{\intelmpi}{\productname{Intel\,MPI}\xspace}
\newcommand{\mvapich}{\productname{MVAPICH}\xspace}
\newcommand{\mpich}{\productname{MPICH}\xspace}
\newcommand{\infiniband}{IB\xspace}
\newcommand{\gcc}{gcc\xspace}
\newcommand{\icc}{icc\xspace}
\newcommand{\mpirun}{\texttt{mpirun}\xspace}

\newcommand{\List}{Listing\xspace}
\newcommand{\fig}{Figure\xspace}

\newcommand{\tab}{Table\xspace}

\newcommand{\alg}{Algorithm\xspace}

\newcommand{\append}{Appendix\xspace}
\newcommand{\Sec}{Section\xspace}

\newcommand{\Equs}{Equations\xspace}
\newcommand{\eg}{e.g.\xspace}
\newcommand{\ie}{i.e.\xspace}

\newcommand{\etcet}{etc.\xspace}
\newcommand{\cf}{cf.\xspace}

\newcommand{\runtime}{run-time\xspace}

\newcommand{\pgmpi}{PGMPI\xspace}
\newcommand{\pgmpiurl}{\url{https://github.com/hunsa/pgmpi}}
\newcommand{\pgtunelib}{PGMPITuneLib\xspace}
\newcommand{\mockup}{mock-up\xspace}
\newcommand{\mockups}{mock-ups\xspace}

\newcommand{\Mockups}{Mock-ups\xspace}

\newcommand{\pgtunecore}{PGMPITune\xspace}
\newcommand{\pglibcli}{PGMPITuneCLI\xspace}
\newcommand{\pglibtuned}{PGMPITuneD\xspace}
\newcommand{\pgprofile}{\textit{profile}\xspace}
\newcommand{\pgprofiles}{\textit{profiles}\xspace}
\newcommand{\pgdefault}{\textit{Default}\xspace}
\newcommand{\pgtuned}{\textit{Tuned}\xspace}

\newcommand{\Bytes}{Bytes\xspace}
\newcommand{\Byte}{Byte\xspace}

\newcommand{\guidelt}{\ensuremath{\preceq}\xspace}

\newcommand{\glmono}{monotony\xspace}
\newcommand{\glsplit}{split-robustness\xspace}
\newcommand{\glpattern}{pattern\xspace}

\newcommand{\gldesc}[2]{\hspace{1 mm}\\\hspace*{.5em}#1 \ #2}

\newcommand{\mpia}{\texttt{MPI\_A}\xspace}

\newcommand{\msize}{\textit{msize}\xspace}

\newcommand{\nrep}{\textit{nrep}\xspace}

\newcommand{\func}{\textit{func}\xspace}
\newcommand{\nmpiruns}{\textit{nmpiruns}\xspace}

\newcommand{\exectimes}{\ensuremath{l}}

\newcommand{\mycomment}[1]{\textit{//\ #1}\xspace}
\renewcommand*\Call[2]{\textproc{#1}(#2)}

\newcommand{\repRSE}{\ensuremath{RSE}\xspace}

\newcommand{\jupitermvapich}{\productname{MVAPICH2-2.2}\xspace}

\newcommand{\jupiteropenmpilatest}{\productname{Open\,MPI~2.1.0}\xspace}

\newcommand{\vscintelmpi}{\productname{Intel\,MPI~2017~(Update~2)}\xspace}

\newcommand{\juqueenmpi}{\productname{IBM~BG~MPI}\xspace}

\newcommand{\machjupiter}{\emph{Jupiter}\xspace}
\newcommand{\machvsc}{\mbox{\emph{VSC-3}}\xspace}
\newcommand{\machjuqueen}{\mbox{\emph{JUQUEEN}}\xspace}

\newcounter{tempcounter}
\newcounter{regcounter}

\newenvironment{myalign}
 {%
  \setcounter{tempcounter}{\value{equation}}%
  \setcounter{equation}{\value{regcounter}}%
  \align
 }
 {%
  \endalign
  \setcounter{regcounter}{\value{equation}}%
  \setcounter{equation}{\value{tempcounter}}%
 }

\makeatletter
\def\lst@MSkipToFirst{%
    \global\advance\lst@lineno\@ne
    \ifnum \lst@lineno=\lst@firstline
        \def\lst@next{\lst@LeaveMode \global\lst@newlines\z@
        \lst@OnceAtEOL \global\let\lst@OnceAtEOL\@empty
        \lst@InitLstNumber 
        \lsthk@InitVarsBOL
        \c@lstnumber=\numexpr-1+\lst@lineno 
        \lst@BOLGobble}%
        \expandafter\lst@next
    \fi}
\makeatother

\newcommand{\mpibarrier}{\texttt{MPI\_Barrier}\xspace}
\newcommand\mpibcast{\texttt{MPI\_Bcast}\xspace}
\newcommand\mpiscatter{\texttt{MPI\_Scatter}\xspace}
\newcommand\mpiscatterv{\texttt{MPI\_Scatterv}\xspace}
\newcommand\mpigather{\texttt{MPI\_Gather}\xspace}
\newcommand\mpigatherv{\texttt{MPI\_Gatherv}\xspace}
\newcommand\mpiallgather{\texttt{MPI\_Allgather}\xspace}
\newcommand\mpiallgatherv{\texttt{MPI\_Allgatherv}\xspace}
\newcommand\mpialltoall{\texttt{MPI\_Alltoall}\xspace}
\newcommand\mpialltoallv{\texttt{MPI\_Alltoallv}\xspace}

\newcommand\mpireduce{\texttt{MPI\_Reduce}\xspace}
\newcommand\mpiallreduce{\texttt{MPI\_Allreduce}\xspace}
\newcommand\mpireducescatter{\texttt{MPI\_Reduce\_scatter}\xspace}
\newcommand\mpireducescatterblock{\texttt{MPI\_Reduce\_scatter\_block}\xspace}
\newcommand\mpiscan{\texttt{MPI\_Scan}\xspace}
\newcommand\mpiexscan{\texttt{MPI\_Exscan}\xspace}
\newcommand\mpireducelocal{\texttt{MPI\_Reduce\_local}\xspace}

\newcommand{\mpiinit}{\texttt{MPI\_Init}\xspace}

\newcommand{\mpiint}{\texttt{MPI\_INT}\xspace}
\newcommand{\mpibor}{\texttt{MPI\_BOR}\xspace}

\newcommand{\appparagraph}[1]{\vspace*{0pt}\paragraph{#1}}

\begin{document}

\usetag{arxiv}

\tagged{hpcasia}{%
\title{Autotuning MPI Collectives using Performance Guidelines}
}

\tagged{arxiv}{
\title{Tuning MPI Collectives by Verifying Performance Guidelines}
}

\author{Sascha Hunold}
\email{hunold@par.tuwien.ac.at}
\affiliation{%
  \institution{TU Wien, Faculty of Informatics}
  \streetaddress{Favoritenstrasse 16}
  \city{Vienna} 
  \state{Austria} 
}

\author{Alexandra Carpen-Amarie}
\email{carpenamarie@par.tuwien.ac.at}
\affiliation{%
  \institution{TU Wien, Faculty of Informatics}
  \streetaddress{Favoritenstrasse 16}
  \city{Vienna} 
  \state{Austria} 
}

\begin{abstract}
  MPI collective operations provide a standardized interface for
  performing data movements within a group of processes. The
  efficiency of collective communication operations depends on the
  actual algorithm, its implementation, and the specific communication
  problem (type of communication, message size, number of
  processes). Many MPI libraries provide numerous algorithms for
  specific collective operations.  The strategy for selecting an
  efficient algorithm is often times predefined (hard-coded) in MPI
  libraries, but some of them, such as \openmpi, allow users to change
  the algorithm manually. Finding the best algorithm for each case is
  a hard problem, and several approaches to tune these algorithmic
  parameters have been proposed. We use an orthogonal approach to the
  parameter-tuning of MPI collectives, that is, instead of testing
  individual algorithmic choices provided by an MPI library, we
  compare the latency of a specific MPI collective operation to the
  latency of semantically equivalent functions, which we call the
  \mockup implementations. The structure of the \mockup
  implementations is defined by self-consistent performance
  guidelines. The advantage of this approach is that tuning using
  \mockup implementations is always possible, whether or not an MPI
  library allows users to select a specific algorithm at \runtime. We
  implement this concept in a library called \pgtunelib, which is
  layered between the user code and the actual MPI
  implementation. This library selects the best-performing algorithmic
  pattern of an MPI collective by intercepting MPI calls and
  redirecting them to our \mockup implementations. Experimental
  results show that \pgtunelib can significantly reduce the latency of
  MPI collectives, and also equally important, that it can help
  identifying the tuning potential of MPI libraries.
\end{abstract}


\maketitle

\section{Introduction}
\label{sec:introduction}

The Message Passing Interface (MPI) is still the most prominent and
probably the most frequently used programming model for
supercomputers, for example, MPI is driving most of the machines on
the TOP500 list. The scalability of parallel applications running on
these large platforms is therefore directly dependent on the
performance of the underlying MPI implementations. The performance of
MPI libraries is therefore of utmost importance for the overall
efficiency of the software stack.

In the present article, we address the problem of optimizing the
performance of MPI libraries, that is, we want to minimize the latency
of a given MPI function for a given payload and a specific number of
processes. The performance of MPI libraries can be improved in
different ways. One possibility is to devise better algorithms for
various communication patterns. 
Another possibility is to better exploit current hardware, \eg, by
aligning memory segments to cache lines or by respecting ccNUMA
domains when allocating memory chunks. Altogether, typical open-source
MPI implementations, such as \mpich, \mvapich, or \openmpi, provide
several algorithms for each MPI function, and each of these individual
implementations may be able to leverage some hardware-specific
optimizations.

Now, the problem is that potentially all provided algorithmic and
hardware parameters that an MPI library provides must be considered
when tuning on a given parallel machine. The goal of such a tuning
processes is to select the best possible algorithm for a given message
size and number of processes (and possibly other factors like the
process to core mapping, \etcet). Since libraries allow developers to
control and vary hundreds of parameters (\eg, \openmpi), the search
space can be extremely large and tuning will be extremely costly.
Moreover, parameter tuning may suffer from the fact that tuning is
done for individual MPI functions, often in isolation and without a
baseline implementation. Thus, having found the best set of parameters
for a specific function (\eg, \mpibcast) will not guarantee its
efficiency.

Self-consistent performance guidelines can help to provide such a
performance baseline. An MPI performance guideline states that the
currently inspected, specialized MPI functionality, say functionality
$A$, should not be slower than a less-specialized, but semantically
equivalent functionality, say $B$ ($A \guidelt B$).  For example, the
specialized \mpigather function, which only works with equal-sized
messages, should not take longer than more generic \mpigatherv
function on the same equal-sized problem.

\begin{taggedblock}{hpcasia}
\citet{pgmpi2016} have shown that many MPI libraries available on
production systems violate performance guidelines for several blocking
MPI collective operations. They have also demonstrated that guideline
violations can be avoided by changing the algorithm used in a specific
case.%
\end{taggedblock}
\begin{taggedblock}{arxiv}
In previous work~\cite{pgmpi2016}, we have shown that many MPI
libraries available on
production systems violate performance guidelines for several blocking
MPI collective operations. We have also demonstrated that guideline
violations can be avoided by changing the algorithm used in a specific
case.%
\end{taggedblock}
However, not all guideline violations could be fixed by changing
the algorithm. First, only some MPI libraries contain multiple
algorithmic strategies for each MPI function. Second, many proprietary
libraries do not expose algorithmic variants in form of adjustable
parameters to the programmer. In both cases, performance violations of
MPI libraries cannot be repaired at library level, and in these cases,
a programmer would have to adapt the application code (\eg, switching
from \mpiallgather to \mpiallgatherv).

\begin{taggedblock}{hpcasia}
To overcome these limitations, we employ self-consistent performance
guidelines for tuning MPI libraries, and we make the following
contributions:
\begin{enumerate}[topsep=0pt,itemsep=-1ex,partopsep=1ex,parsep=1ex,leftmargin=*]
\item We propose the \emph{library \pgtunelib}, which can be used to
  \emph{automatically tune} the performance of \emph{any} MPI
  library. \linebreak \pgtunelib can \emph{replace the default
    implementation of an MPI function} with one of the semantically
  equivalent \mockup versions, if the corresponding performance
  guideline is violated.
\item To the best of our knowledge, we are the \emph{first} to use
  \emph{self-consistent performance guidelines for automatically
    tuning MPI libraries}, in contrast to the traditional way of
  tuning library-provided algorithmic parameters. As the DRAM memory
  per core is a scarce resource on larger parallel machines, our
  implementation of the tuning library takes special care about
  additional memory requirements.
\item The experiments provide evidence that our tuning strategy can
  \emph{automatically repair all guideline violations} on three
  different test machines, including an IBM BlueGene/Q.
\item We also propose an \emph{algorithm for solving the NREP
    problem}, \ie, obtaining reproducible experimental results by
  keeping the number of measurements small.
\item Our proposed tuning process is also able to reveal algorithmic
  variants that have not been considered in MPI libraries. In the
  present paper, we \emph{show how a new and faster algorithm for
    \mpiallreduce in \openmpi can be derived}.
\end{enumerate}%
\end{taggedblock}%
\begin{taggedblock}{arxiv}
In order to address this problem, we make the following
contributions. We propose the library \pgtunelib, which can be used to
improve the performance of any MPI library. \pgtunelib replaces the
default implementation of an MPI function with its semantically
equivalent \mockup version, if the corresponding performance guideline
has been violated.  We propose a tuning strategy that allowed us to
automatically repair all guideline violations on three different test
machines including a BlueGene/Q.%
\end{taggedblock}%

\section{Background and Related Work}
\label{sec:related-work}

Due to the diversity of parallel hardware, it is not surprising that
MPI libraries only provide implementations of the MPI standard in a
best-effort manner, \ie, the decision which underlying algorithm to
use for a given case is predefined in a library. Nonetheless, as
systems are, among themselves, usually very heterogeneous, it is
necessary to adapt/tune MPI libraries to hardware. This tuning process
is very difficult for two main reasons: first, the number of
parameters that MPI libraries (\eg, \openmpi) expose for tuning can be
very large. In addition, theoretically one would need to examine all
possible variants of mapping processes to cores and all possible
message sizes, which would simply be infeasible. Second, the
optimization functions are often not convex (for minimizing the
\runtime), which makes it harder to find the optimal value as the
problems may become intractable. Previous work on library tuning faced
these problems, and we will summarize three approaches.

\citet{ChaarawiSGF08} developed the Open Tool for Parameter
Optimization (OTPO), whose task is to find a good set of parameter
values for a given number of processes and an MPI function. It
basically performs a brute-force search over all specified parameters
and their ranges in \openmpi. A related method was proposed by
\citet{PjesivacGrbovicMPISelection}, in which a quadtree scheme is
used to encode the best collective algorithm for a given pair of
(number of processes, message size).  The quadtree is the internal
data-structure for allowing a fast lookup of the best-suited
algorithm. Since the quadtree can be limited in its depth and
granularity, this tuning approach avoids a full enumeration of the
search space. A different method was proposed by \citet{Sikora2016},
where a user can specify parameters and their ranges that should be
tuned. Then, a plugin of the Periscope Tuning Framework tries to find
the best configuration of these parameters by applying a
meta-heuristic, in this case a genetic algorithm. In contrast to
previous approaches, the tool of \citet{Sikora2016} benchmarks and
optimizes the \runtime of entire MPI applications instead of
optimizing individual MPI functions.

The mentioned previous approaches try to optimize the \runtime of MPI
functions for different message sizes but using a fixed number of
processes. It is also possible to search for optimization potential by
looking at the scalability behavior of MPI functions, as it was done
by \citet{Shudler15}. In general, MPI functions have an expected and
an actual performance, and the expected performance depends on the
theoretical lower bound of an algorithm, which can be obtained
analytically for different network topologies~\cite{ChanHPG07}. The
approach of \citet{Shudler15} compares the expected scalability curve
of an MPI function to the actual, measured scalability curve. A
mismatch indicates that an MPI function has tuning potential.

Performance guidelines (previously called ``performance
requirements'') can be used to verify the consistency of an MPI
library.  In MPI, several communication patterns can be expressed in a
semantically equivalent way. For example, the specialized MPI function
\mpiallreduce can also be implemented by chaining calls to \mpireduce
and \mpibcast together. The user's expectation is that the composition
of the latter two functions should not be faster than executing the
specialized one. In a more formal definition~\cite{TraffGT10}, a
performance guideline is defined between two functionalities $A$ and
$B$, which semantically implement the same operation. If functionality
$A$ is the more specialized of the two, we can state that
$\mathtt{MPI}\_A(n) \guidelt \mathtt{MPI}\_B(n)$, which means that $A$
should complete faster than $B$ for a comparable communication
volume~$n$. The communication volume $n$ should be understood as the
amount of ``actual'' data items. It is possible that functionality $B$
needs to transfer messages of larger size, \eg, $pn$, to mimic
functionality $A$ with $n$ data items and $p$ processes. However, as
$B$ mimics $A$, only a communication volume of size $n$ is
relevant. The majority of MPI performance guidelines are defined for a
fixed number of processes and for the same communicator. Guidelines
for different communicators can also be devised, but they are not
considered in this work.

\begin{taggedblock}{hpcasia}
\citet{pgmpi2016} have implemented and tested
several performance guidelines for blocking, collective MPI
operations, such as \mpibcast. Their goal was to get an overview of how
many libraries violate such guidelines in practice. For that task, they 
have implemented the toolkit \pgmpi,
which distinguishes three classes of performance guidelines: \glmono,
\glsplit, and \glpattern. The \glmono guideline ensures that
increasing the message size(s) also increases the \runtime. The goal
of the \glsplit guideline is to ensure that splitting a communication
operation into smaller chunks does not improve the overall
performance. Last, \glpattern guidelines are defined between
semantically equivalent operations, \eg,
$\mpiallreduce \guidelt \mpireduce + \mpibcast$. \citet{pgmpi2016}
have shown that all tested MPI libraries (\mvapich, \openmpi,
\intelmpi) violated performance guidelines in various cases. In
addition, they have demonstrated how violations of performance
guidelines can be fixed by selecting a better underlying algorithm for
a specific communication operation.
\end{taggedblock}
\begin{taggedblock}{arxiv}
In previous work~\cite{pgmpi2016}, we have implemented and tested
several performance guidelines for blocking, collective MPI
operations, such as \mpibcast. Our goal was to get an overview of how
many libraries violate such guidelines in practice. For that task, we 
have implemented the toolkit \pgmpi\footnote{\pgmpiurl},
which distinguishes three classes of performance guidelines: \glmono,
\glsplit, and \glpattern. The \glmono guideline ensures that
increasing the message size(s) also increases the \runtime. The goal
of the \glsplit guideline is to ensure that splitting a communication
operation into smaller chunks does not improve the overall
performance. Last, \glpattern guidelines are defined between
semantically equivalent operations, \eg,
$\mpiallreduce \guidelt \mpireduce + \mpibcast$. We 
have shown that all tested MPI libraries (\mvapich, \openmpi,
\intelmpi) violate performance guidelines in various cases. In
addition, we have demonstrated how violations of performance
guidelines can be fixed by selecting a better underlying algorithm for
a specific communication operation.
\end{taggedblock}

In the present paper, we combine the detection of
performance-guideline violations with the tuning of MPI libraries.
\tagged{hpcasia}{The work of \citet{pgmpi2016} pointed out two problems: first,}%
\tagged{arxiv}{Our previous work~\cite{pgmpi2016} pointed out two problems: first,}
several, often vendor-provided MPI implementations lack
user-controlled parameters for algorithmic tuning. Second, some MPI
libraries only provide a small set of algorithmic choices for several
MPI functions. In such cases, even though a performance violation has
been detected, it cannot be repaired due to the limited number of
algorithms provided.  

Therefore, we propose to use performance-guideline variants as
possible replacement implementations. The idea is the following: a
guideline may state $\mpigather \guidelt \mpigatherv$, which is a
natural and almost trivial requirement. If an MPI library violates
this guideline and if no other (or faster) variant (algorithm)
implementing \mpigather is available, the scientific programmer either
needs to accept inferior performance, or she could refactor the code
and replace the call to \mpigather with a call to \mpigatherv. Yet, as
this optimization might only be useful on machine $Z$ with $K$
processes, it does not seem to be good strategy, in general. We solve
this problem by introducing \pgtunelib, which sits between the MPI
user code and the MPI library. By using the \texttt{PMPI}-interface,
it intercepts calls to a specific MPI function, say \mpigather, and
redirects them to an internally implemented \mpigather function, which
uses \mpigatherv as its base implementation. Our approach is in the spirit
of the approaches of \citet{PjesivacGrbovicMPISelection} and
\citet{FarajYL06}. The latter authors proposed the STAR-MPI library, which
selects an algorithm for a collective operation (online) after
benchmarking (timing) several algorithmic variants during the \runtime
of an application.  Our \pgtunelib library instead provides several
implementations of a specific collective $\mathtt{MPI}\_C$, and each
variant corresponds to one performance guideline. We then profile MPI
functions in isolation (\emph{offline}) and check for
performance-guideline violations. If violations occur, we record these
cases and later redirect MPI calls (\emph{online}) to faster
implementations during application runs. Instead of quadtrees,
\pgtunelib uses a combination of hash functions and binary searches,
ensuring an efficient lookup of algorithmic variants for a given
number of processes and message size, which in our case can be done in
time $O(\log m)$, where $m$ denotes the largest message size that may
occur.

\section{Autotuning MPI Libraries with \pgtunelib}
\label{sec:pgtunelib}

Now, we describe our approach for autotuning blocking MPI collective
operations using \pgtunelib. First, we show all performance guidelines
that our library comprises and give a short explanation for each of
them. Second, we discuss implementation details of the library and
describe the tuning process.

\subsection{Performance Guidelines and Semantics}

Currently, \pgtunelib contains implementations of the performance
guidelines listed in
\Equs~\eqref{gl:allgather_lt_gather_bcast}--\eqref{gl:scatter_lt_scatterv},
some of which were introduced
before~\cite{TraffGT10,pgmpi2016}.

\begin{footnotesize}
\begin{myalign}
	\mpiallgather(n) & \guidelt \mpigather(n) + \mpibcast(n) \label{gl:allgather_lt_gather_bcast} \\
        \mpiallgather(n)  & \guidelt \mpialltoall(n) \label{gl:allgather_lt_alltoall} \\
        \mpiallgather(n)  & \guidelt \mpiallreduce(n) \label{gl:allgather_lt_allreduce} \\
        \mpiallgather(n)  & \guidelt \mpiallgatherv(n) \label{gl:allgather_lt_allgatherv} 
\end{myalign}
\begin{myalign}
	\mpiallreduce(n) & \guidelt \mpireduce(n) + \mpibcast(n) \label{gl:allreduce_lt_reduce_bcast} \\
        \mpiallreduce(n) & \guidelt \mpireducescatterblock(n) + \mpiallgather(n) \label{gl:allreduce_lt_reducescatterblock_allgather} \\
        \mpiallreduce(n) & \guidelt \mpireducescatter(n) + \mpiallgatherv(n) \label{gl:allreduce_lt_reducescatter_allgatherv} 
\end{myalign}
\begin{myalign}
	\mpialltoall(n ) & \guidelt \mpialltoallv(n) \label{gl:alltoall_lt_alltoallv} 
\end{myalign}
\begin{myalign}
	\mpibcast(n) & \guidelt \mpiallgatherv(n) \label{gl:bcast_lt_allgatherv} \\
        \mpibcast(n) & \guidelt \mpiscatter(n) + \mpiallgather(n) \label{gl:bcast_lt_scatter_allgather}
\end{myalign}
\begin{myalign}
	\mpigather(n) & \guidelt \mpiallgather(n) \label{gl:gather_lt_allgather} \\
        \mpigather(n) & \guidelt \mpigatherv(n) \label{gl:gather_lt_gatherv} \\
        \mpigather(n) & \guidelt \mpireduce(n) \label{gl:gather_lt_reduce} 
\end{myalign}
\begin{myalign}
	\mpireduce(n) & \guidelt \mpiallreduce(n) \label{gl:reduce_lt_allreduce} \\
        \mpireduce(n) & \guidelt \mpireducescatterblock(n) + \mpigather(n) \label{gl:reduce_lt_reducescatterblock_gather} \\
        \mpireduce(n) & \guidelt \mpireducescatter(n) + \mpigatherv(n) \label{gl:reduce_lt_reducescatter_gatherv} 
\end{myalign}
\begin{myalign}
	\mpireducescatterblock(n) & \guidelt \mpireduce(n) + \mpiscatter(n) \label{gl:reducescatterblock_lt_reduce_scatter} \\
        \mpireducescatterblock(n) & \guidelt \mpireducescatter(n) \label{gl:reducescatterblock_lt_reducescatter} \\
        \mpireducescatterblock(n) & \guidelt \mpiallreduce(n) \label{gl:reducescatterblock_lt_allreduce} 
\end{myalign}
\begin{myalign}
	\mpiscan(n) & \guidelt \mpiexscan(n) + \mpireducelocal(n) \label{gl:scan_lt_exscan_reducelocal} 
\end{myalign}
\begin{myalign}
         \mpiscatter(n) & \guidelt \mpibcast(n) \label{gl:scatter_lt_bcast} \\ 
         \mpiscatter(n) & \guidelt \mpiscatterv(n) \label{gl:scatter_lt_scatterv}
\end{myalign}
\end{footnotesize}

\begin{table*}[t!]
  \caption{\label{tab:gl_overview}Performance guidelines implemented
    in \pgtunelib.  Variable $n$ denotes the number of elements of
    basetype in the send count of an operation, $p$ denotes the
    number of processes in the communicator, and $I$ denotes the size
    of \mpiint.}
\begin{footnotesize}  
\begin{tabular}{lllll}
    \toprule
    MPI collective  & max memory  & guidel.  & \mockup & add. mem. requirement \\ 
                    & requ. per proc. & & &  \\
    \midrule
    \mpiallgather   &  $n$ + $pn$       & \ref{gl:allgather_lt_gather_bcast} & \mpigather $+$ \mpibcast   & none \\
                    &                  & \ref{gl:allgather_lt_alltoall} & \mpialltoall   & $pn$ ($p$ times larger send buffer)  \\
                    &                  & \ref{gl:allgather_lt_allreduce} & \mpiallreduce  & $pn$ ($p$ times larger send buffer)  \\
                                       
                    &                   & \ref{gl:allgather_lt_allgatherv}  & \mpiallgatherv & $2pI$ (displs, recvcounts) \\ 
    \midrule 
    \mpiallreduce   & $2n$             & \ref{gl:allreduce_lt_reduce_bcast} & \mpireduce $+$ \mpibcast  & none \\
                    &                  & \ref{gl:allreduce_lt_reducescatterblock_allgather} & \mpireducescatterblock $+$ \mpiallgather & $(n+c)+(n+c)/p$ (small $c$ for padding) \\
                    &                  & \ref{gl:allreduce_lt_reducescatter_allgatherv} & \mpireducescatter $+$ \mpiallgatherv & $max\{ \lfloor n/p \rfloor + C, C\}$ (chunk size $C$) + $2pI$ (displs, recvcounts) \\
    \midrule 
    \mpialltoall    & $2pn$            & \ref{gl:alltoall_lt_alltoallv} & \mpialltoallv   & $2pI$ (displs, recvcounts) \\
    \midrule
    \mpibcast       & $n$            & \ref{gl:bcast_lt_allgatherv} & \mpiallgatherv  & $2pI$ (displs, recvcounts)  + $n$ (for recv buf)\\ 
                    &                & \ref{gl:bcast_lt_scatter_allgather} & \mpiscatter $+$ \mpiallgather & $(n+c)+(n+c)/p$ (small $c$ for padding) \\
                                     
    \midrule
    \mpigather      & $n$ + $pn$       & \ref{gl:gather_lt_allgather} & \mpiallgather   & none on root, $pn$ on other processes \\
                    &                  & \ref{gl:gather_lt_gatherv} & \mpigatherv     & $2pI$ (displs, recvcounts) \\
                    &                  & \ref{gl:gather_lt_reduce} & \mpireduce      & $pn$ (for new send buf) \\
    \midrule
    \mpireduce      & $n$ + $n$ (on root) & \ref{gl:reduce_lt_allreduce}   & \mpiallreduce  & extra $n$ (on processes other than root) \\
                    &                  &  \ref{gl:reduce_lt_reducescatterblock_gather}  & \mpireducescatterblock $+$ \mpigather & $(n+c) + (n+c)/p$ ($c$ for padding) \\
                    &                  &  \ref{gl:reduce_lt_reducescatter_gatherv}  & \mpireducescatter $+$ \mpigatherv & $max\{ \lfloor n/p \rfloor + C, C\}$ (chunk size $C$) + $2pI$ (displs, recvcounts)\\
    \midrule
    \mpireducescatterblock & $n$ + $n/p$ & \ref{gl:reducescatterblock_lt_reduce_scatter}  & \mpireduce $+$ \mpiscatter & $n$ (for first reduce) \\
                           &            & \ref{gl:reducescatterblock_lt_reducescatter} & \mpireducescatter & $pI$ (recvcounts) \\
                           &            & \ref{gl:reducescatterblock_lt_allreduce}  & \mpiallreduce     & $n$ (for new recv buffer)\\
    \midrule       
    \mpiscan        & $2n$              & \ref{gl:scan_lt_exscan_reducelocal} & \mpiexscan $+$ \mpireducelocal & none \\
    \midrule
    \mpiscatter      & $n$ + $n/p$      & \ref{gl:scatter_lt_bcast}  & \mpibcast    & extra $n$ (on processes other than root)  \\
                     &                 & \ref{gl:scatter_lt_scatterv} & \mpiscatterv & $2pI$  (displs, recvcounts) \\
    \bottomrule                   
  \end{tabular}
\end{footnotesize}  
\end{table*}

All inequalities have a regular blocking MPI collective on the
left-hand side. Regular means that all processes use the same send
buffer size (\eg, in \mpiallreduce, \mpigather), and in case of
\mpibcast, equal-sized receive buffers. We only included regular
collectives as their use cases seemed to be better defined. 
However, in the irregular case (\texttt{$\ast{}$v} functions), we have another
degree of freedom (how much data each process contributes), which makes
global tuning harder. 
Thus, it seems unrealistic that offline-tuned irregular collectives
will actually be (re-)used. For such irregular collectives, an online
tuning approach (such as done by STAR-MPI) seems to be more promising.

We want to stress the fact that \pgtunelib contains an actual
implementation of the right-hand side of each performance guideline
shown above. That means, in some cases (detailed below) it is
necessary to allocate additional buffer space and to perform data
movements between buffers for obtaining a semantically equivalent
implementation of the right-hand side. A brief description of the
semantics and some implementation details of the considered
performance guidelines is given in \append~\ref{sec:appendix}.
\tab~\ref{tab:gl_overview} summarizes the additional memory
requirements for each guideline implementation.  When referring to
guidelines of the form $\mathtt{MPI}\_A \guidelt \mathtt{MPI}\_A'$, we
say that an implementation of $A'$ is a \mockup version of
functionality $A$. Our notation of $n$ and $p$ in
\tab~\ref{tab:gl_overview} is as follows.  The number $p$ denotes the
number of processes involved in the MPI collective.  As in most cases
the memory requirements are different among the processes, the table
lists the maximum memory requirement of any process, which is often
the root process for rooted operations like Gather and Scatter.  The
number $n$ denotes the number of elements that are placed in the send
buffer by a process. For example, in \mpialltoall, $n$ elements are
sent from each process to every other process, and thus overall, each
process sends $pn$ elements. Since a process also receives $pn$
elements, the total memory requirement for \mpialltoall is~$2pn$.

\mpireducescatterblock is a special case, as it has no explicit send
count.  Therefore, the initial send buffer holds $n$ elements, on
which the reduction will be performed. After the scatter step, each
process receives $n/p$ elements, and overall, the memory requirement
for this routine is $n + n/p$.

When referring to $n$ as the memory space needed for one process, we
would have to say precisely $nE$, where $E$ is the extent of the base
datatype used. However, for the sake of a better readability we omit
$E$ and simply say $n$ for the memory requirement.
\tab~\ref{tab:gl_overview} also uses variable $I$, which denotes the
extent of \mpiint, which is commonly needed when specifying the
displacement and the receive (send) count vectors.

\begin{taggedblock}{arxiv}
In the following, we will give a brief summary of the idea and
implementation behind each performance guideline.

\appparagraph{\mpiallgather and its \Mockups}
\gldesc{\eqref{gl:allgather_lt_gather_bcast}}{ trivially composes
  \mpigather with \mpibcast to obtain a functionally equivalent
  version of \mpiallgather.}
\gldesc{\eqref{gl:allgather_lt_alltoall}}{uses a $p$ times larger send
  buffer, in which each process puts $p$ copies of its own buffer
  contents. Then, \mpialltoall is called to mimic \mpiallgather.}
\gldesc{\eqref{gl:allgather_lt_allreduce}}{uses, similar to the
  \mpialltoall \mockup, a $p$ times larger send buffer. This larger
  buffer is initialized with zeros, and the actual message of each
  process $i$ is copied into the large buffer starting at index
  $i\cdot{}n$. Then, an \mpiallreduce is applied to all buffers and a
  bit-wise or-operation ensures that the result is semantically
  equivalent to the result of \mpiallgather.}
\gldesc{\eqref{gl:allgather_lt_allgatherv}}{calls \mpiallgatherv
  instead, and therefore needs to allocate two additional buffers of
  size $p$ ($p$ elements of type \mpiint) for the receive counts and
  the displacements. \textbf{Note} that we will \emph{not further
    comment} on any other \mockup implementation using an
  \emph{irregular operation} (\eg, \mpigatherv, \mpialltoallv,
  \etcet), as they are all straightforward to implement.}

\appparagraph{\mpiallreduce and its \Mockups}

\gldesc{\eqref{gl:allreduce_lt_reduce_bcast}}{composes
  straightforwardly \mpireduce and \mpibcast.}
\gldesc{\eqref{gl:allreduce_lt_reducescatterblock_allgather}}{first
  calls \mpireducescatterblock, which needs equal-sized blocks in the
  Scatter phase. As the send buffer of the original \mpiallreduce
  function does not have to be a multiple of the number of processes,
  our \mockup version will add (a maximum of $p-1$) dummy elements of
  the send type as additional padding.  It is now possible to perform
  an \mpiallgather on the receive buffer of the previous stage. After
  this \mpiallgather has been completed, the \mockup version only
  copies the first $n$ elements (ignoring the padded elements) back to
  the original receive buffer.}
\gldesc{\eqref{gl:allreduce_lt_reducescatter_allgatherv}}{applies a
  similar strategy as the previous \mockup function.  Since
  \mpireducescatter and \mpiallgatherv work with send buffers of
  arbitrary size, the \mockup function only needs to allocate and
  properly handle buffers for the receive counts and for the
  displacements. We also introduce a variable $C, 1 \le C \le n$,
  which denotes the minimum size of chunks that are distributed to
  each process in the scatter phase. Thus, if $C=1$, each process
  receives roughly $n/p$ elements in the scatter phase. If $C=n$, only
  one process receives elements in the scatter phase.}

\appparagraph{\mpibcast and its \Mockups}

\gldesc{\eqref{gl:bcast_lt_allgatherv}}{denotes one specific process
  to be the root process of the broadcast operation. This process
  (\eg, rank 0) allocates a send buffer of the size of the original
  broadcast operation. All other processes contribute zero bytes to
  the result of the allgather operation. Then, a call to
  \mpiallgatherv copies the buffer contents of the (fake) root rank to
  all other processes in a broadcast-to-all fashion. As \mpiallgatherv
  works with different send and receive buffers, an additional receive
  buffer is needed, holding $n$ elements. Additionally, each process
  must allocate two buffers of size $p$ for the count and displacement
  information.}

\appparagraph{\mpigather and its \Mockups}

\gldesc{\eqref{gl:gather_lt_allgather}}{This \mockup simply uses
  \mpiallgather behind the \mpigather-interface. As now every process
  needs to receive $n$ elements of the base datatype from every other
  process, we need to allocate a buffer with space for $p\cdot n$
  elements of basetype.}%
\gldesc{\eqref{gl:gather_lt_reduce}}{Similarly to
  \eqref{gl:allgather_lt_allreduce}, we allocate a $p$ times larger
  send buffer on all processes. Each process copies its contents of
  the send buffer into the larger temporary buffer.  Then a call to
  \mpireduce with an $\mpibor$-operation results in the desired
  emulation of \mpigather.}

\appparagraph{\mpireduce and its \Mockups}

\gldesc{\eqref{gl:reduce_lt_allreduce}}{uses the same strategy as
  guideline~\eqref{gl:gather_lt_allgather}. Every process, except the
  root process of the operation, needs to allocate a receive buffer
  that can accommodate $n$ basetype elements. Calling \mpiallreduce
  will not only give the root the result but also the other processes,
  which simply ignore the result.}%
\gldesc{\eqref{gl:reduce_lt_reducescatterblock_gather}}{This \mockup
  is a rather heavyweight replacement of \mpireduce. It first performs
  an \mpireducescatterblock on the send buffer. For the scatter part,
  the vector size must be a multiple of the number of processes, as
  \mpireducescatterblock requires send buffers of the same size. To
  achieve that, an extra padding is added to the end of the send
  buffer. Two new buffers are allocated, one holding the new, padded
  send buffer and another one for the result of the
  \mpireducescatterblock operation, which is exactly $p$ times smaller
  than the new send buffer. Upon completion of this operation, we can
  call \mpigather on the result buffers of \mpireducescatterblock,
  which finally gives us an emulated version of \mpireduce.}%
\gldesc{\eqref{gl:reduce_lt_reducescatter_gatherv}}{The idea of this
  \mockup is similar to the one above. The only difference is that we
  do not need the additional padding, as \mpireducescatter works on
  vectors of arbitrary size. However, to accomplish an emulation, we
  need to allocate two buffers for the displacement and the count
  information that will be used for \mpireducescatter and the
  following \mpigatherv. The chunk size $C$ has the same meaning as in
  guideline \eqref{gl:allreduce_lt_reducescatter_allgatherv}, \ie,
  chunks of size $C$ are assigned to processes in round-robin
  fashion.}

\appparagraph{\mpireducescatterblock and its \Mockups}

\gldesc{\eqref{gl:reducescatterblock_lt_reduce_scatter}}{This \mockup
  function uses a straightforward composition of \mpireduce and
  \mpiscatter.  As the result of the first step (\mpireduce) requires
  a receive buffer of size $n$ (elements), we need to allocate this
  additional buffer between the two calls.}%
\gldesc{\eqref{gl:reducescatterblock_lt_reducescatter}}{is a trivial
  emulation using the irregular counterpart, for which an additional
  buffer holding the receive counts is required.}%
\gldesc{\eqref{gl:reducescatterblock_lt_allreduce}}{The
  \mpireducescatterblock functionality can also be emulated with
  \mpiallreduce. We need to allocate an additional receive buffer on
  each but the root process with space for $n$ elements. Allreduce
  will then distribute the reduction result to all processes. Now,
  each process picks its part of the reduction result, which it would
  have received from a scatter operation. This completes the
  emulation of \mpireducescatterblock.}

\appparagraph{\mpiscan and its \Mockups}

\gldesc{\eqref{gl:scan_lt_exscan_reducelocal}}{This \mockup version
  performs first an exclusive scan on the same data as the inclusive
  scan would have performed. In order to obtain the same result as the
  inclusive scan, we need to perform a local reduction operation on
  all processes but the root. Overall, no additional buffers are
  needed.}

\appparagraph{\mpiscatter and its \Mockups}

\gldesc{\eqref{gl:scatter_lt_bcast}}{This version allocates on all
  processes but the root an additional receive buffer for the $n$
  elements of the root process. The root process then broadcasts all
  its data to the others. Now, every process (also the root) copies
  its part of the data ($n/p$ elements) to the receive buffer of the
  scatter operation.}

\end{taggedblock}

\subsection{Library Design and Implementation}

\subsubsection{General Design}

As the library \pgtunelib makes use of the PMPI interface of MPI, it
is layered between the MPI user code and the MPI library. If the user
code calls an MPI function, in our case a blocking MPI collective,
\pgtunelib intercepts the call and may select one of the \mockup
implementations. \fig~\ref{fig:allreduce_example} shows an example:
the MPI user code calls \mpiallreduce, which is intercepted by
\pgtunelib. Internally, \pgtunelib uses performance \pgprofiles
containing identifiers of possible replacement algorithms for various
message sizes. Therefore, \pgtunelib searches for a replacement algorithm
for \mpiallreduce. If such a replacement algorithm can be found,
\pgtunelib emulates the original call by using its replacement, which
is in our example the combination of \mpireduce and \mpibcast.
\begin{figure}[t]
  \centering
  \begin{footnotesize}
  \begin{sequencediagram} 
    \newinst{mpiusercode}{:MPI\_User\_Code}
    \newinst[0.5]{pgtunelib}{:PGTuneLib}
    \newinst[0.5]{pgprofile}{:PGTuneProfile}
    \newinst[0.5]{mpilib}{:MPI\_Library}
    
    \begin{call}{mpiusercode}{\mpiallreduce}{pgtunelib}{}
      \begin{call}{pgtunelib}{\tiny getImpl(MPI\_ALLREDUCE)}{pgprofile}{\tiny \mpireduce$+$\mpibcast} 
      \end{call}
      \begin{call}{pgtunelib}{\texttt{P}\mpireduce}{mpilib}{} 
      \end{call}      
      \begin{call}{pgtunelib}{\texttt{P}\mpibcast}{mpilib}{} 
      \end{call}      
    \end{call}
  \end{sequencediagram}
  \end{footnotesize}
  \caption{\label{fig:allreduce_example}Example of intercepting
    \mpiallreduce and replacing it with calls to \mpireduce and
    \mpibcast.}
\end{figure}
If no replacement algorithm is found, \pgtunelib uses the default
implementation, \ie, it calls \texttt{P\mpiallreduce}.

\subsubsection{Tuning Workflow and Modes of Operation}

\pgtunelib provides two modes of operation, which are encapsulated in
different libraries and which can be linked with an arbitrary MPI
application (or benchmark). \fig~\ref{fig:pgtune_arch} shows the
general architecture, where \pgtunecore provides the basic API.
\begin{figure}[t]
  \centering

  \tikzset{
    col1/.style = {draw, rectangle, minimum width=3cm, minimum height=1cm, text centered, text width=2cm, font=\footnotesize, draw=black, fill=yellow!30},
    col2/.style = {draw, rectangle, minimum width=6.1cm, minimum height=1cm, text centered, text width=5.0cm, font=\footnotesize, draw=black, fill=green!10},
    col3/.style = {draw, rectangle, minimum width=6.1cm, minimum height=1cm, text centered, text width=5.0cm, font=\footnotesize, draw=black, fill=gray!30},
  }
  
  \begin{tikzpicture}[node distance=1.1cm,transform shape,scale=0.9]
    \node (a1) [col1] {\pglibcli};
    \node (a2) [col1, right of=a1, xshift=2cm] {\pglibtuned};
    \node (b) [col2, below of=a1, xshift=1.55cm] {\pgtunecore};
    \node (c) [col3, below of=b] {MPI Library (\openmpi, \mvapich, \mpich)};

    \draw (a1);
    \draw (a2);
    \draw (b);
    \draw (c);
  \end{tikzpicture}

  \caption{\pgtunelib architecture.}
  \label{fig:pgtune_arch}
\end{figure}
On top of that core API, two different libraries exist. One is the
library called \pglibcli (CLI stands for command line interface),
which is used for benchmarking the performance of \mockup
implementations. To that end, MPI developers link their applications
against the CLI version of \pgtunelib. It is now possible to select a
\mockup version for a specific MPI function as follows:
\begin{lstlisting}[basicstyle=\ttfamily\footnotesize,language=bash]
mpicc *.c -o mympicode -lpgmpitunecli -lmpi
mpirun -np 2 ./mympicode 
   --module=allgather:alg=allgather_as_gather_bcast
\end{lstlisting}
In this example, all calls to \mpiallgather will be replaced with the
\mockup implementation of
guideline~\eqref{gl:allgather_lt_gather_bcast}. By using this CLI
version of \pgtunelib, we can analyze the latency of all implemented
collective algorithms for different message sizes. Any MPI benchmark
suite can be used to measure the latency of MPI collective
operations. Benchmarking allows us to discover the message sizes for
which the performance guidelines are violated. When violations occur,
\pgtunelib stores for which message sizes a possible replacement
\mockup has been found. After scanning over all collectives and
selected message sizes, \pgtunelib writes a performance \pgprofile for
each MPI collective. Performance \pgprofiles contain the replacement
algorithms for specific message ranges.
\begin{lstlisting}[float,floatplacement=ht,label={lst:profile},caption={Profile of \mpiscatter on \machjuqueen.},basicstyle=\ttfamily\footnotesize,frame=single,numbers=left,xleftmargin=3em,escapechar=|]
# pgtune profile
MPI_Scatter
1024 # nb. of. processes |\label{line:nbprocs}|
2    # nb. of mock-up impl. 
2 scatter_as_bcast
3 scatter_as_scatterv
8    # nb. of ranges
1 1 2 # byte_range_start byte_range_end alg_id
8 8 2
32 32 2
64 64 2
100 100 2
512 512 2
1024 1024 2
10000 10000 3  
\end{lstlisting}
\List~\ref{lst:profile} shows a sample \pgprofile for \mpiscatter that
was recorded with \num{64x16} processes on \machjuqueen.
Each \pgprofile only contains message ranges for which violations have
occurred and for which a replacement algorithm should be used. As we
have measured in this example for discrete message sizes, the sample
profile uses the same message size for the start and the end of a
message range. For example, algorithm~2 (\texttt{scatter\_as\_bcast})
should be applied for the message ranges \SIrange{1}{1}{\Byte},
\SIrange{8}{8}{\Bytes}, and so on.

After the performance \pgprofiles have been written, any MPI
application can use these profiles. A developer simply needs to link
their application against the \pglibtuned library. Similar to
\pglibcli, the \pglibtuned library intercepts MPI calls and redirects
them to the \mockup versions implemented in the core library.  
\pglibtuned reads in all performance \pgprofiles from
disk, which happens transparently when intercepting \mpiinit. Then,
\pglibtuned has all the information required to select a (possibly)
better \mockup version for a collective MPI operation at \runtime.

\subsubsection{Implementation Details} 

It seems obvious that a tuned MPI library should be implemented as
efficiently as possible. We have therefore tried to keep the overhead
incurred by \pgtunelib very low. As mentioned before, some \mockup
implementations require the allocation of additional memory, \eg, for
padded data or for displacement or send/receive count vectors.  First,
\pgtunelib avoids additional system calls (\eg, \texttt{malloc}) and
allocates two memory chunks at the start of the MPI program, one for
additional message buffers and one for displacement or send/receive
count vectors.  The size of both buffers can be controlled by the user
with the variables \texttt{size\_msg\_buffer\_bytes} and
\texttt{size\_int\_buffer\_bytes}, which can be set in the
configuration file of \pgtunelib.
Another advantage of this additional memory management is that users
can accurately control how much extra memory they want to dedicate for
possibly faster MPI functions. Now, cases may occur, where a
replacement algorithm was found to be faster than the default
implementation provided by the MPI library, but such a \mockup would
need too much extra memory and will therefore not be selected.

For providing an efficiently tuned library, it is also important to
perform fast look-ups to check whether a replacement algorithm is
available. Currently, performance profiles are read for a specific
number of processes only.  Thus, \pgtunelib can look up the right
performance \pgprofile for a certain collective and can check whether
the profile is compatible with the current number of processes in time
$O(1)$. Then, \pgtunelib only needs to verify whether the profile
contains a replacement algorithm for the current message size. As we
sort the $M$ different message ranges at program start, such a lookup
of the replacement algorithm is performed in time $O(\log M)$ using
binary search.

\section{Experimental Evaluation}
\label{sec:exper-eval}

\subsection{Hardware Setup}

We have evaluated \pgtunelib on three different machines, whose
characteristics are summarized in \tab~\ref{tab:machines}.
\begin{table}[t]
  \centering
  \caption{Parallel machines used in our experiments.}
  \label{tab:machines}
  \begin{scriptsize}
  \begin{tabular}{l@{\hskip .1in}l@{\hskip .1in}ll}
    \toprule
    Name & Hardware & MPI Libraries & Compiler \\
    \midrule 
    \machjupiter & 36 $\times$ Dual Opteron 6134 @ \SI{2.3}{\giga\hertz}  & \jupitermvapich & \gcc 4.4.7 \\
             &  \infiniband QDR MT26428  & \jupiteropenmpilatest  &  \\
    \midrule 
    \machvsc &  \num{2000} $\times$ Dual Xeon E5-2650V2 @ \SI{2.6}{\giga\hertz} & \vscintelmpi &  \icc 16.0.4  \\
             & \infiniband QDR-80  \\
    \midrule 
    \machjuqueen    &  \num{28672} $\times$ IBM PowerA2 @ \SI{1.6}{\giga\hertz} & \juqueenmpi &  IBM XL \\
             & IBM-BlueGene/Q, 5D Torus interconnect  \\ 
    \bottomrule
  \end{tabular}
  \end{scriptsize}
\end{table}
The systems \machjupiter and \machvsc are rather similar when
comparing their hardware setup. The advantage of having similar
architectures is that the reproduction of phenomena on other systems
increases the confidence in the significance of our findings.  A
BlueGene/Q called \machjuqueen allows us to study a vendor-provided,
tailor-made MPI library on an actual supercomputer. 
\begin{taggedblock}{hpcasia}
Notice that we do
not present experimental data from \machvsc due to space
restrictions. However, we emphasize that our tuning approach has also
been successfully been applied on this machine~\cite{arxiv_pgtunelib}.
\end{taggedblock}

\subsection{Tuning Workflow}

The tuning process of \pgtunelib first checks whether the performance
guidelines defined for blocking collective MPI operations are
fulfilled. If violations occur, these cases are recorded and a
\pgprofile is written. In a subsequent execution, \pgtunelib can then
change to a different \mockup implementation at \runtime.

In order to automatically tune an MPI library, we need to benchmark
the latency of the default implementations of blocking collectives and
their \mockup versions that are part of \pgtunelib. For measuring the
latency, one could employ any type of MPI benchmark suite, \eg,
\osu~\cite{osu_benchmarks} or \skampi~\cite{ReussnerST02}. It is only
required that the benchmark suite is linked against \pgtunelib.

\begin{taggedblock}{arxiv}
For the analysis shown in the present paper, we have used our own
benchmark suite called \reprompi\footnote{\reprompiurl}, 
\end{taggedblock}
\begin{taggedblock}{hpcasia}
In the present paper, we have used the benchmark suite ``\reprompi'',
\end{taggedblock}
which allows to record raw data (the latency of every single
measurement) from each experiment~\cite{tpds16}. In contrast to other
benchmark suites, it refrains from performing any kind of data
aggregation (\eg, computation of means) or data removal (\eg,
discarding the first X measurements for ``warming up'' the system).
By using \reprompi, we can record every single measurement and perform
the data analysis in R, Python, or Julia later.

The auto-tuning process with \pgtunelib is divided into \emph{three}
steps. The \emph{first} and a critical step is to estimate the
number of repetitions (\nrep) of measurements that have to be conducted for 
a specific MPI function with a given message and communicator size
(number of processes). \emph{Second}, we benchmark the MPI collectives
and their \mockup counterparts using the CLI version of \pgtunelib
(\cf \fig~\ref{fig:pgtune_arch}).  From the performance data gathered,
we can then detect violations of the performance guidelines
\eqref{gl:allgather_lt_gather_bcast}--\eqref{gl:scatter_lt_scatterv}.
Among all \mockup functions for which guideline violations have
occurred, the \mockup version that performs best for a given message
range is selected and written into a performance \pgprofile (\cf
\List~\ref{lst:profile}).  \emph{In a last step}, we re-link the
\reprompi benchmark suite against the \emph{tuned} version of
\pgtunelib.  Now, \pgtunelib can replace individual MPI collectives
with their faster \mockup counterparts.

\begin{algorithm}[t]
  \begin{scriptsize}
  \caption{\label{alg:mpi_timing}MPI timing procedure (\cite{tpds16}).}
  \begin{algorithmic}[1]
    \Procedure{Time\_MPI\_function}{\func, \msize, \nrep} 
    \Statex \hfill \mycomment{\func\ - MPI function; \msize\ - message size; \nrep\ - nb. of observations}
    \State initialize time array \exectimes\ with $\nrep$ elements
    \For{\textit{obs} in $1$ to $\nrep$}
    \State \Call{\mpibarrier}{} \hfill \mycomment{use external dissemination barrier implementation}
    \State t = \Call{Get\_Time}{}
    \State execute \func(\msize)
    \State \exectimes[\textit{obs}] = \Call{Get\_Time}{} - t
    \EndFor  
    \EndProcedure
\end{algorithmic}
\end{scriptsize}
\end{algorithm}
\reprompi supports different synchronization strategies and different
ways (clocks) to measure the \runtime (latency) of collective calls.
For the presented experiments, we have used the timing procedure shown
in \alg~\ref{alg:mpi_timing}: Before every individual measurement,
processes are synchronized with a barrier. Here, we use a
dissemination barrier, which---due to its structure---ensures that
processes leave this barrier relatively synchronized (which would not
be the case for tree-based barriers for example).

The ``NREP problem'' consists of finding a suitable (and possibly
minimal) number of repetitions, such that the derived statistical
measures (mean, median) are reproducible~\cite{pgmpi2016}.

\begin{figure}[t]

  \begin{tikzpicture}[transform shape,scale=0.9,x=0.4cm,y=0.4cm]

    \xdef\lastx{0}
    \xdef\y{0}
    \foreach \x in {2,4,6,8,10,12,14,16,18,20}
    {
      \node at (\x, \y) {\textbullet};
      \draw (\lastx,\y) -- (\x,\y);
      \xdef\lastx{\x}
    }

    \node at (0,2) [anchor=west] {$MPI\_A$ with \SI{1}{\Byte} message};

    \node at (0.5,0.5) [anchor=west] {$1$};
    \node at (2.5,0.5) [anchor=west] {$2$};
    \node at (5.5,0.5) [anchor=west] {$\cdots$};
    \node at (17.0,0.3) [anchor=west,rotate=45] {$\nrep_1-1$};
    \node at (19.0,0.3) [anchor=west,rotate=45] {$\nrep_1$};

    \draw[<->,yshift=0pt] (0, \y-0.5) -- (20, \y-0.5) node [midway,yshift=-10pt]{$t^{\nrep}_1$};

    \node at (0,-3.5) [anchor=west] {$MPI\_A$ with \msize\,\Byte message};

    \xdef\lastx{0}
    \xdef\y{-5}
    \foreach \x in {4,8,12,16,20}
    {
      \node at (\x, \y) {\textbullet};
      \draw (\lastx,\y) -- (\x,\y);
      \xdef\lastx{\x}
    }

    \node at (1.5,-4.5) [anchor=west] {$1$};
    \node at (5.5,-4.5) [anchor=west] {$2$};
    \node at (10.5,-4.5) [anchor=west] {$\cdots$};
    \node at (14.0,-4.9) [anchor=west,rotate=45] {$\nrep_{\msize}-1$};
    \node at (17.5,-4.9) [anchor=west,rotate=45] {$\nrep_{\msize}$};

    \draw[<->,yshift=0pt] (0, -5.5) -- (4, -5.5) node [midway,yshift=-10pt]{$t_{\msize}$};

  \end{tikzpicture}

  \caption{\label{fig:nrep_alg}Estimating the number of repetitions
    (NREP) based on the time obtained for \SI{1}{\Byte} messages.}
\end{figure}
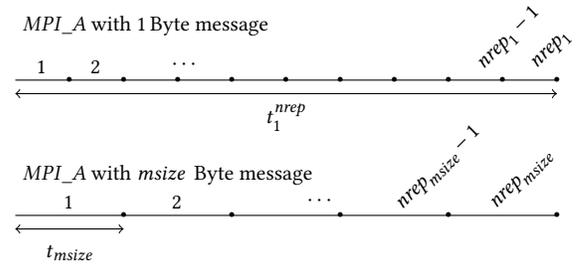

We use the following method to address the \textbf{NREP problem}: for
each MPI function \mpia, the \emph{general idea} is to determine the
time ($t^{\nrep}_1$) until the latency measurements with a
\SI{1}{\Byte} message have stabilized (\eg, a small variance). This
time is further used as the reference time for other message sizes
(\cf \fig~\ref{fig:nrep_alg}).  Our assumption is that relative system
noise decreases when the message size increases, as the \runtime of
each collective grows with the message size. Thus, we measure the
latency of \mpia with a different message size $\msize$
($\msize > \SI{1}{\Byte}$) for at least time $t^{\nrep}_1$. 

More precisely, in step~(1), we repeat measuring the latency of
function \mpia with \SI{1}{\Byte} messages until the current relative
standard error (\repRSE) over the measured latencies is below some
user-defined threshold. We repeat this process over several calls to
\mpirun and take the longest time that was required to make the
\repRSE drop below the threshold, and we denote this time as
$t^{\nrep}_1$. As this process can take hundreds of repetitions, we
only want to do that for a message size of \SI{1}{\Byte}. For all the
other messages sizes, we measure for at least time $t^{\nrep}_1$.  As
many benchmarking tools require the number of repetitions as an input,
we convert the time $t^{\nrep}_1$ into a number of repetitions
$\nrep_{\msize}$ of \mpia for any other message size \msize. To that
end, in step~(2), we run two batches of measurements called batches
$b_1$ and $b_2$, and this user-defined number of repetitions for both
batches should be relatively small ($<10$) or even zero in case of
$b_2$. We compute the \repRSE value of these $b_1$ measurements. If
that value is smaller than some predefined threshold (note this is a
different threshold than used for \SI{1}{\Byte} messages), measuring
is stopped. Otherwise, another batch with $b_2$ measurements is
started. For larger message sizes, taking one batch with $b_1$
elements leads to a very small variance, and thus, we can return
quickly. However, for smaller message sizes, we need a few more
measurements to get a reasonable value of the latency. In step~(3), we
compute the minimum latency of these $b_1+b_2$ measurements ($b_2$ may
be \num{0}), $t_{\msize}=\min_{1\le i \le b_1+b_2} l_i$, and use this
value as the ``expected'' latency when measuring.  Then, we compute
the estimated number of repetitions needed for \mpia and message size
\msize as
$\nrep_{\msize} = \max\left\{ \left\lceil
    \frac{t^{\nrep}_1}{t_{\msize}} \right\rceil, K \right\}$.
The value $K \ge 1$ ensures that at least $K$ latency measurements for
every collective are performed, especially when $t_{\msize}$ becomes
very large.

In our experiments, we use the following values to estimate the
$\nrep$ value for each collective or \mockup: we repeat measuring the
latency with $\msize = \SI{1}{\Byte}$ until the \repRSE value is
smaller than \num{0.01} (1\%).  We perform $b_1=\num{5}$ and possibly
$b_2=\num{5}$ more measurements for each collective (and \mockup) and
with larger message sizes and then compute $\nrep_{\msize}$. We
measure the latency of each collective and its \mockups for
$\nrep_{\msize}$ iterations and repeat that for $\nmpiruns=5$
different calls to \mpirun. The selection of which \mockup function to
use is done \emph{statically} on the command line (\pglibcli). We
check for guideline violations of each collective and write a
performance \pgprofile to disk, if violations have occurred. In our
particular case, we only replace a collective with its \mockup if the
\mockup is at least 10\% faster than the default implementation. We
can then run another set of experiments with the tuned version of the
MPI library. The selection of the best implementation (default or
\mockup) is done \emph{dynamically} at \runtime by \pglibtuned.

\

\List~\ref{lst:bench_file} shows the output of \reprompi when being
run and linked against \pgtunelib. The output is directly readable as
CSV data into data processing frameworks like R. The header contains
information about the specific benchmarking run, \eg, how many
processes, which clock, or which barrier implementation have been
used. However, the footer is written by \pgtunelib and shows whether
certain calls to MPI collectives have been replaced.  In the example,
for a message size of \SI{100}{\Bytes}, the default implementation of
\mpiallgather has been replaced by the \mockup implementation
$\mpigather + \mpibcast$.  In some other cases, for example with
\SI{8}{\Bytes}, the \pgdefault implementation has been used.  The
footer also contains information about how much memory has been
reserved for the temporary buffers in \pgtunelib.  Here, \pgtunelib
could use additional \SI{100}{\mega\Bytes} for allocating message
buffers and \SI{10}{\kilo\Bytes} for displacement and count vectors.

\noindent
\begin{minipage}{1.0\linewidth}
\begin{lstlisting}[label={lst:bench_file},caption={\reprompi output when benchmarking a tuned MPI library; some lines were omitted for better readability.},linerange={4-17,14-18,331-345,369-370},basicstyle=\ttfamily\scriptsize,frame=single,breaklines=true,numbers=left,xleftmargin=3em]
#@operation=MPI_BOR
#@datatype=MPI_CHAR
#@root_proc=0
#@reproMPIcommitSHA1=unknown
#@nprocs=1024
#@clocktype=local
#@clock=MPI_Wtime
#@sync=BBarrier
#@nrep=0
test nrep msize runtime_sec
MPI_Allgather 0 1 0.0002306175
MPI_Allgather 1 1 0.0001389663
MPI_Allgather 2 1 0.0001024431
MPI_Allgather 3 1 0.0001023700
MPI_Allgather 0 1 0.0002306175
MPI_Allgather 1 1 0.0001389663
MPI_Allgather 2 1 0.0001024431
MPI_Allgather 3 1 0.0001023700
MPI_Allgather 4 1 0.0001029038
#@pgmpi alg MPI_Allgather 16000 default
#@pgmpi alg MPI_Allgather 8 default
#@pgmpi alg MPI_Allgather 100 allgather_as_gather_bcast
#@pgmpi alg MPI_Allgather 32 default
#@pgmpi alg MPI_Allgather 32768 default
#@pgmpi alg MPI_Allgather 4 default
#@pgmpi alg MPI_Allgather 50000 default
#@pgmpi alg MPI_Allgather 8192 default
#@pgmpi alg MPI_Allgather 100000 default
#@pgmpi alg MPI_Allgather 2 allgather_as_gather_bcast
#@pgmpi alg MPI_Allgather 1024 allgather_as_gather_bcast
#@pgmpi alg MPI_Allgather 64 allgather_as_gather_bcast
#@pgmpi alg MPI_Allgather 4096 allgather_as_gather_bcast
#@pgmpi alg MPI_Allgather 512 allgather_as_gather_bcast
#@pgmpi alg MPI_Allgather 1 allgather_as_gather_bcast
#@pgmpi config size_msg_buffer_bytes 100000000
#@pgmpi config size_int_buffer_bytes 10000
\end{lstlisting}
\end{minipage}

\subsection{Experimental Results}

\begin{figure*}[tb]
\centering
\includegraphics[width=.32\linewidth]{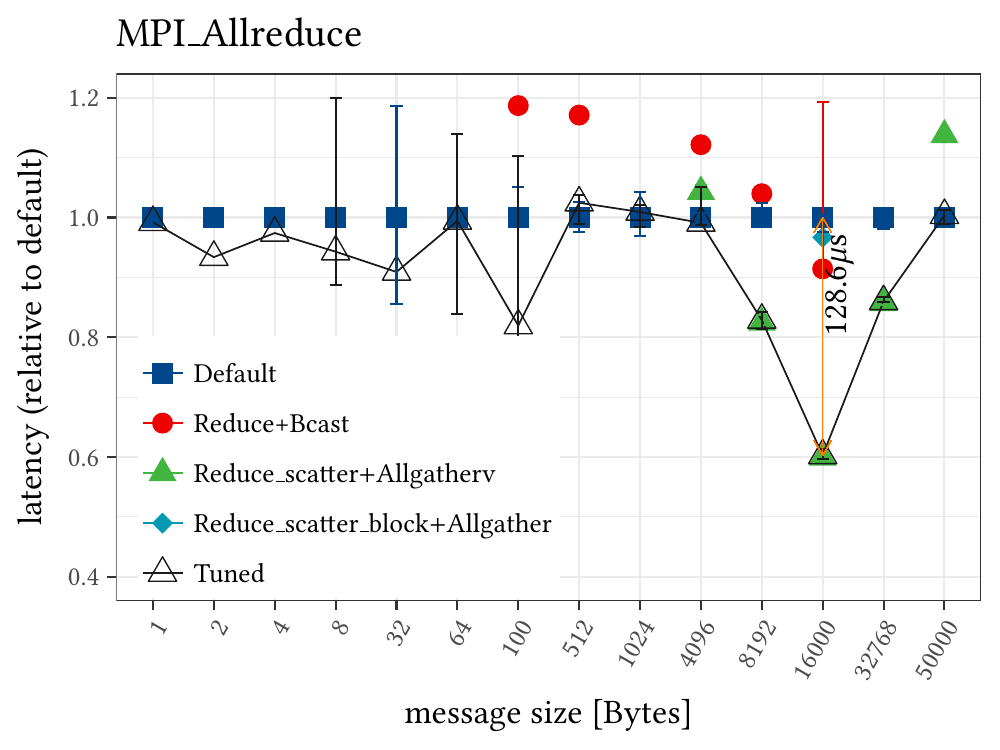}
\includegraphics[width=.32\linewidth]{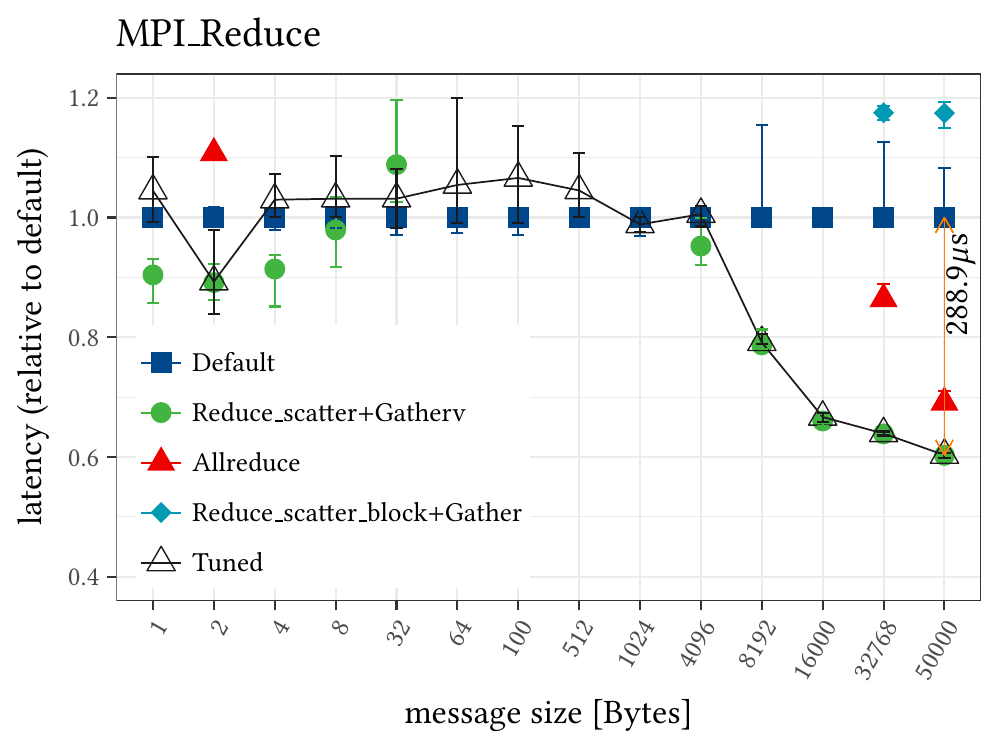}
\includegraphics[width=.32\linewidth]{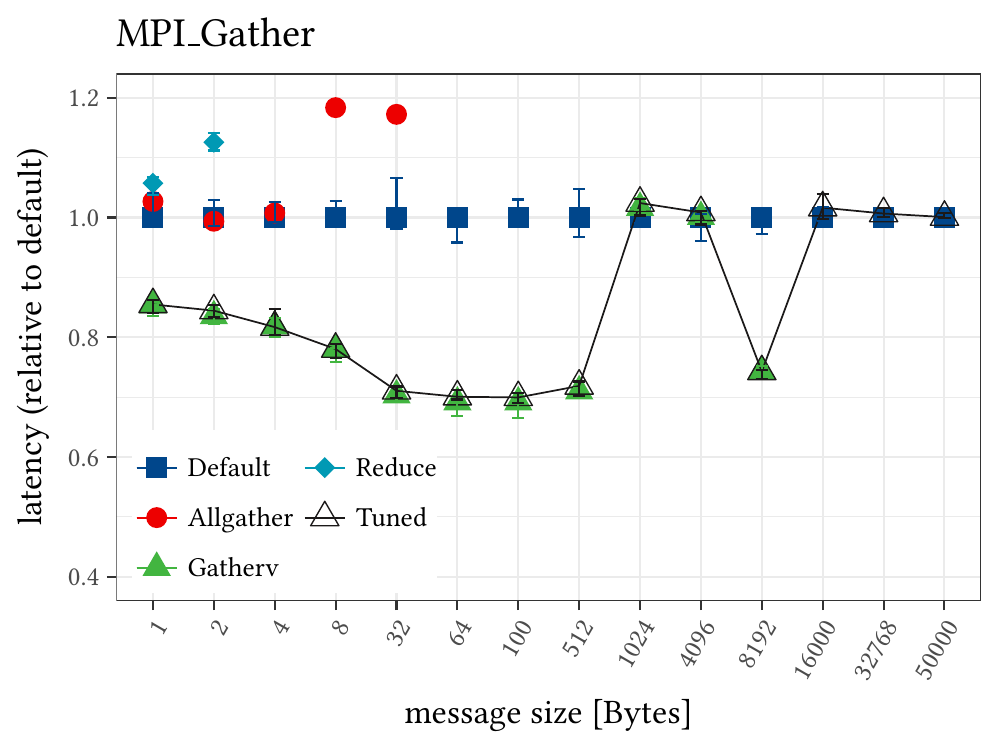}
\caption{\label{fig:perf_openmpi_32x1_jupiter}Performance comparison
  between \pgdefault and \pgtuned version of \jupiteropenmpilatest
  (\num{32x1}~processes, \machjupiter). 
}
\end{figure*}

\fig~\ref{fig:perf_openmpi_32x1_jupiter} summarizes the tuning results
that were obtained for \num{32x1}~processes and \jupiteropenmpilatest
on \machjupiter. Each plot contains the performance of the \pgdefault
algorithm, the \pgtuned version, and the individual \mockup
implementations.  As latencies for small and large messages differ by
orders of magnitude, we plot the relative performance of each
implementation, where the latency of the \pgdefault implementation is
used as reference. As we measure over multiple calls to \mpirun
(\nmpiruns), we use the median over the $\nmpiruns=5$ median latencies
measured. The error bars denote the minimum and the maximum of these
$\nmpiruns$ medians to reflect the variance of the data.
For a better comprehension, let us look at the plot on the right-hand
side of \fig~\ref{fig:perf_openmpi_32x1_jupiter}, which compares the
\pgtuned and the \pgdefault version of \mpigather.  The figure also
includes the performance data of three different \mockup
implementations of \mpigather (Allgather, Gatherv, Reduce). We
can observe that the \pgtuned version uses Gatherv as replacement up
to a message size of \SI{1024}{\Bytes}, for which \pgtunelib switches
back to the \pgdefault version. Except for \SI{8192}{\Bytes}, the
\pgdefault version has been found to perform best for larger message
sizes. As the scale of the y-axis is limited, not all individual
points are shown, \eg, the red points for the Allgather \mockup.

The data shown in \fig~\ref{fig:perf_openmpi_32x1_jupiter} suggest
that there is a large tuning potential to improve \mpireduce in
\openmpi, and this case will be considered in
\Sec~\ref{sec:fixing_violations}.

\begin{figure*}[tb]
\centering
\includegraphics[width=.32\linewidth]{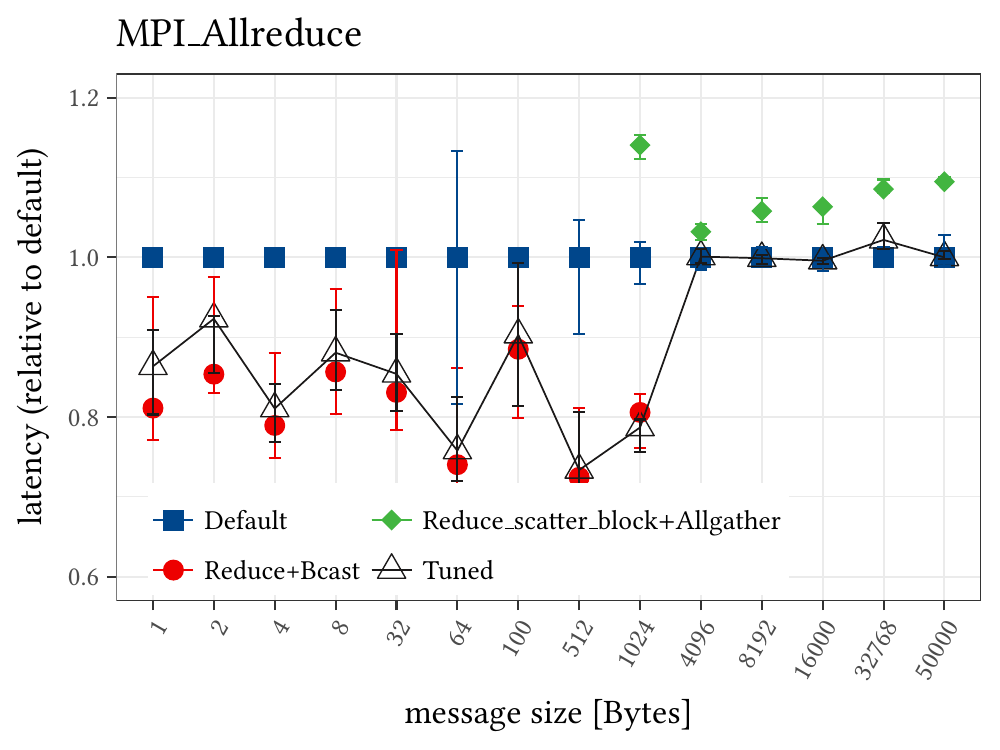}
\includegraphics[width=.32\linewidth]{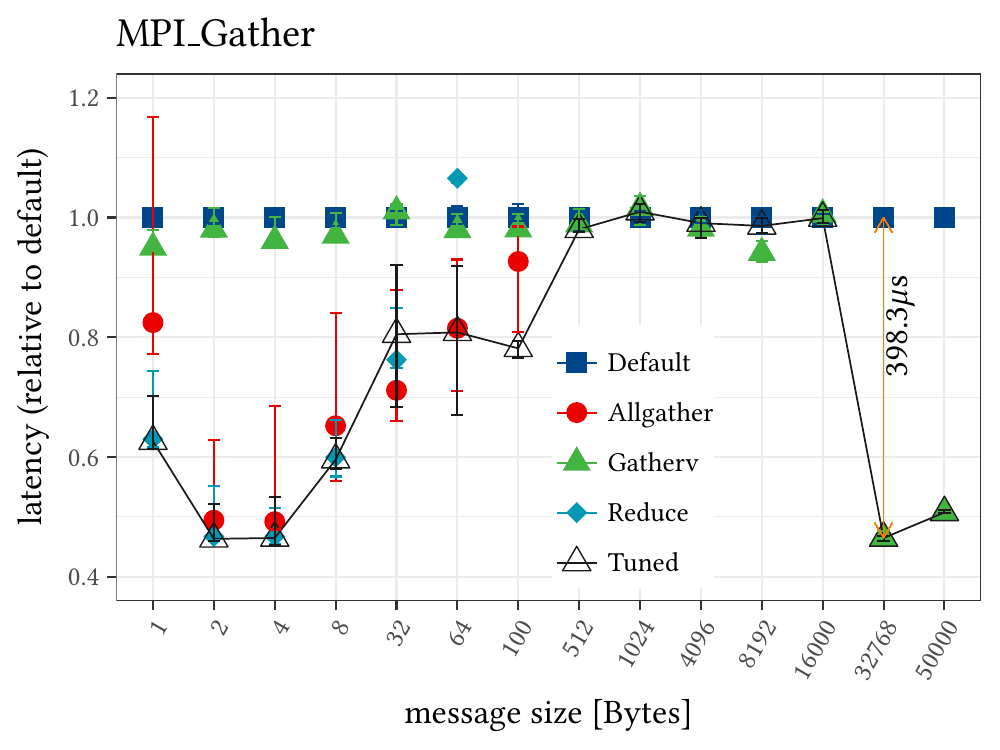}
\includegraphics[width=.32\linewidth]{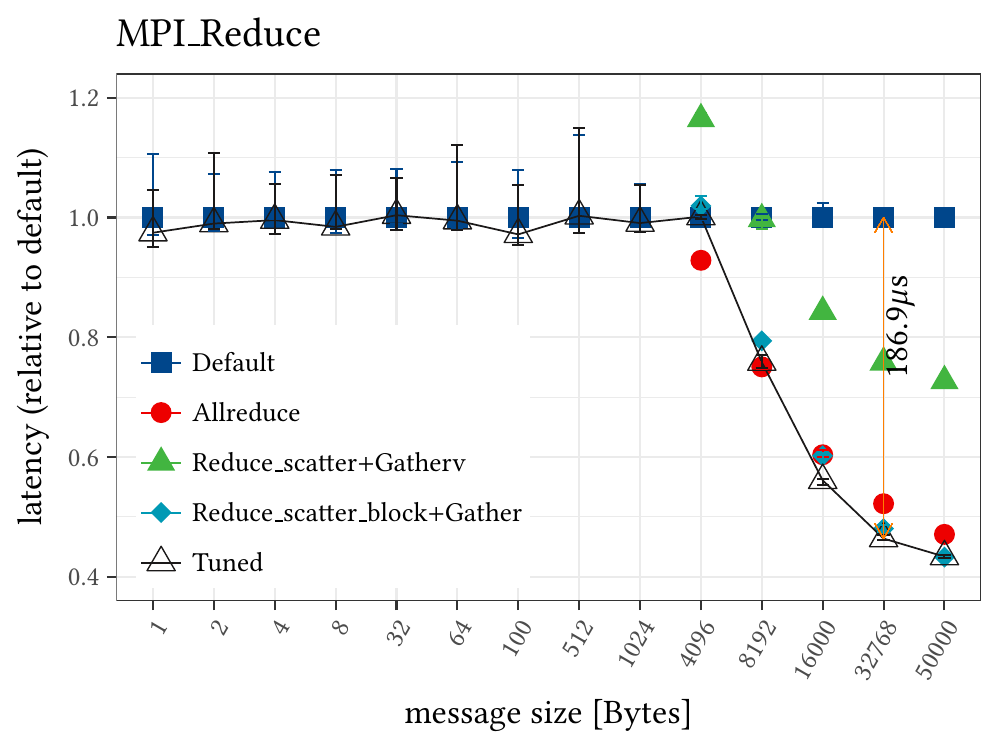}
\caption{\label{fig:perf_32x1_mvapich_jupiter} Performance comparison
  between \pgdefault and \pgtuned version of \jupitermvapich
  (\num{32x1}~processes, \machjupiter).}
\end{figure*}

With \jupitermvapich, different cases were detected for which
\pgtunelib can improve the performance (see
\fig~\ref{fig:perf_32x1_mvapich_jupiter}).
As the relative latency can sometimes be misleading, we also indicate
the absolute performance difference for a few cases. For example, the
latency of \mpireduce or \mpigather with \SI{32}{\kibi\Bytes} of data
can be reduced up to \SI{180}{\micro\second} or
\SI{400}{\micro\second}, respectively (an improvement of roughly
50\%).

\begin{figure*}[tb]
\centering
\includegraphics[width=.32\linewidth]{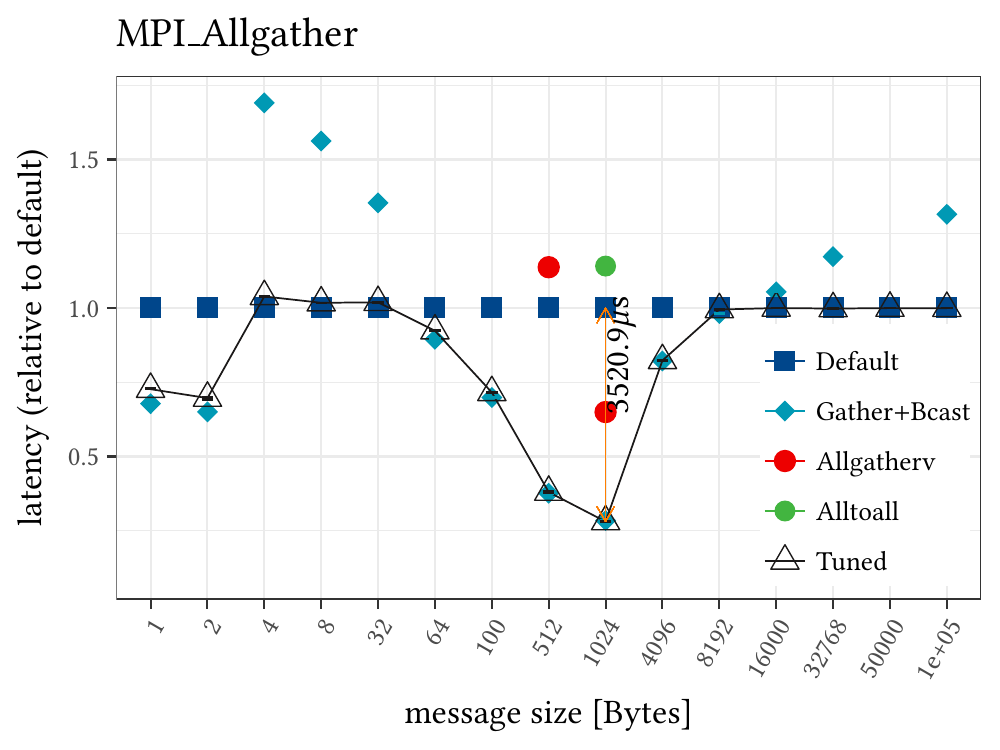}
\includegraphics[width=.32\linewidth]{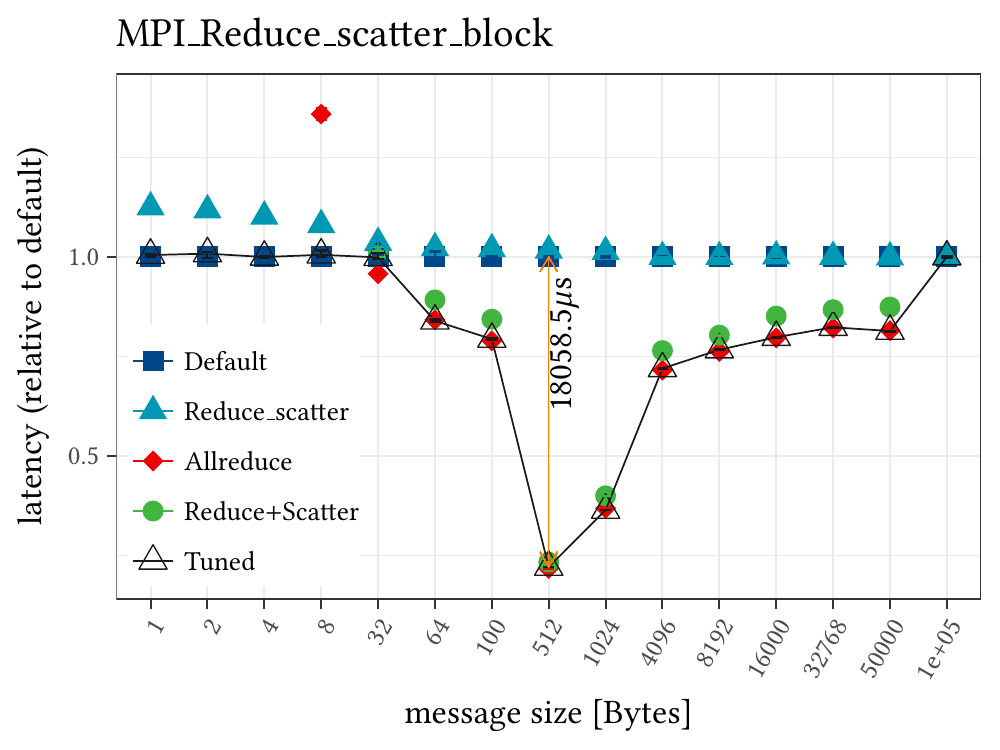}
\includegraphics[width=.32\linewidth]{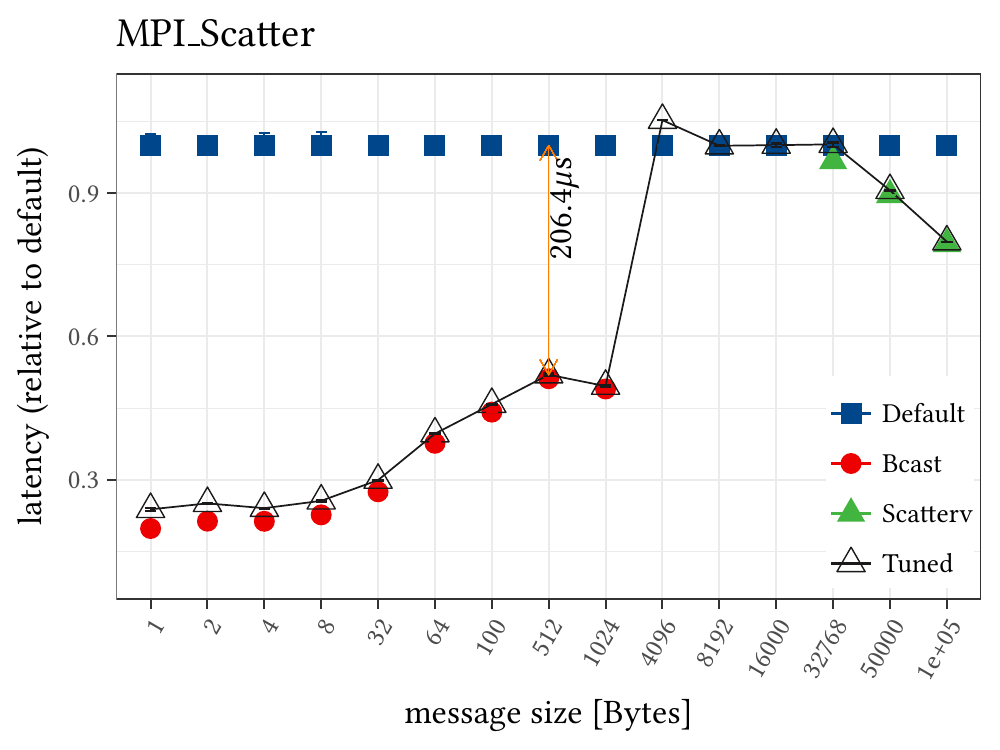}
\caption{\label{fig:perf_64x16_juqueen}Performance comparison between
  \pgdefault and \pgtuned version of \juqueenmpi
  (\num{64x16}~processes, \machjuqueen).}
\end{figure*}
\fig~\ref{fig:perf_64x16_juqueen} shows the performance improvement
achievable on \machjuqueen and \num{64x16}~processes.  It is
interesting to note, but not surprising, that many performance
violations have occurred when being tested against \mockup versions
that rely on \mpibcast, \eg, for
guideline~\eqref{gl:allgather_lt_gather_bcast}.  It seems often
beneficial to employ \mpibcast, for which the BlueGene/Q provides
hardware support. As a consequence of this, the performance of
\mpiallgather and \mpiscatter could significantly be improved.

\begin{taggedblock}{hpcasia}
Notice that we only present a selection of the performance plots,
showing the most significant results.
More performance graphs can be found in our technical
report~\cite{arxiv_pgtunelib}.
\end{taggedblock}
\begin{taggedblock}{arxiv}
Notice that we only present a selection of the performance plots,
showing the most significant results.
More performance graphs can be found in \append~\ref{sec:appendix}.
\end{taggedblock}

\subsection{Parameter vs. Guideline-based Tuning}
\label{sec:fixing_violations}

Now, we inspect two performance guideline violations, one of which
\tagged{arxiv}{we}%
had already %
\tagged{hpcasia}{been examined by \citet{pgmpi2016}.}%
\tagged{arxiv}{examined.}

\subsubsection{Case $\mpireduce \guidelt \mpiallreduce$}

\tagged{hpcasia}{\citet{pgmpi2016}}%
\tagged{arxiv}{We}
had shown that \mpireduce violates the Allreduce guideline with
\openmpi~1.10.1, for message sizes ranging from
\SIrange{128}{725}{\kilo\Bytes} and \num{32x16} processes on
\machjupiter. 
\tagged{hpcasia}{The authors}%
\tagged{arxiv}{We}
were able to overcome this
violation by implementing 
\tagged{hpcasia}{their}%
\tagged{arxiv}{our}
own \mpireduce function.
\begin{figure}[t]
  \centering
  \includegraphics[width=.75\linewidth]{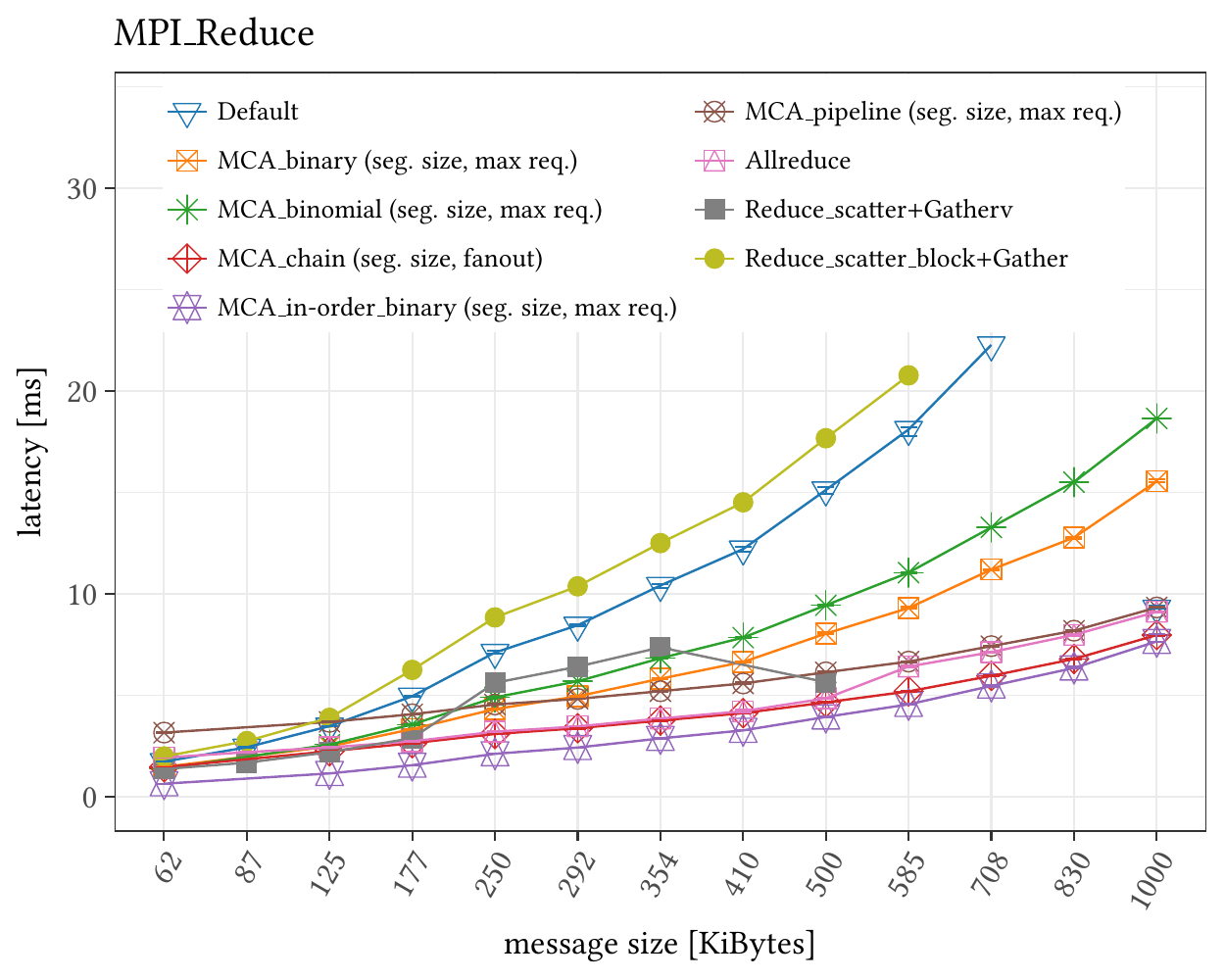}
  \caption{Violation of $\mpireduce \guidelt \mpiallreduce$. Comparing
    latencies of \mockups with algorithmic variants
    (\jupiteropenmpilatest, \num{32x16} processes, \machjupiter)}
  \label{fig:reduce_violations_openmpi}
\end{figure}
Now, we would like to go one step further and compare differently tuned versions of
\mpireduce: (1) the best \mockup algorithm found by \pgtunelib and (2)
the best algorithm found after an exhaustive search using the MCA
parameters of \openmpi. To that end, we have varied the relevant MCA
parameters (\eg, segment size, fan-out) for all Reduce algorithms
provided by \openmpi. The result of this brute-force tuning and the
results with \pgtunelib are compared in
\fig~\ref{fig:reduce_violations_openmpi}.  We can observe that the
\mpiallreduce \mockup is faster than the \pgdefault \mpireduce
implementation over the entire range of message sizes. However, the
latency can further be improved (although only moderately) by using
the \texttt{in-order\_binary}-algorithm of \openmpi.  This case
exemplifies that scanning for guideline violations and performing a
serious parameter tuning (\eg, MCA parameters in \openmpi) should
complement each other. In this case, a fully parameter-tuned version
of \openmpi would not have violated the \mpireduce performance
guidelines in the first place. The downside is that such an exhaustive
search is time-consuming.

\subsubsection{Tuning Potential}

\begin{figure}[t]
  \centering
  \includegraphics[width=.75\linewidth]{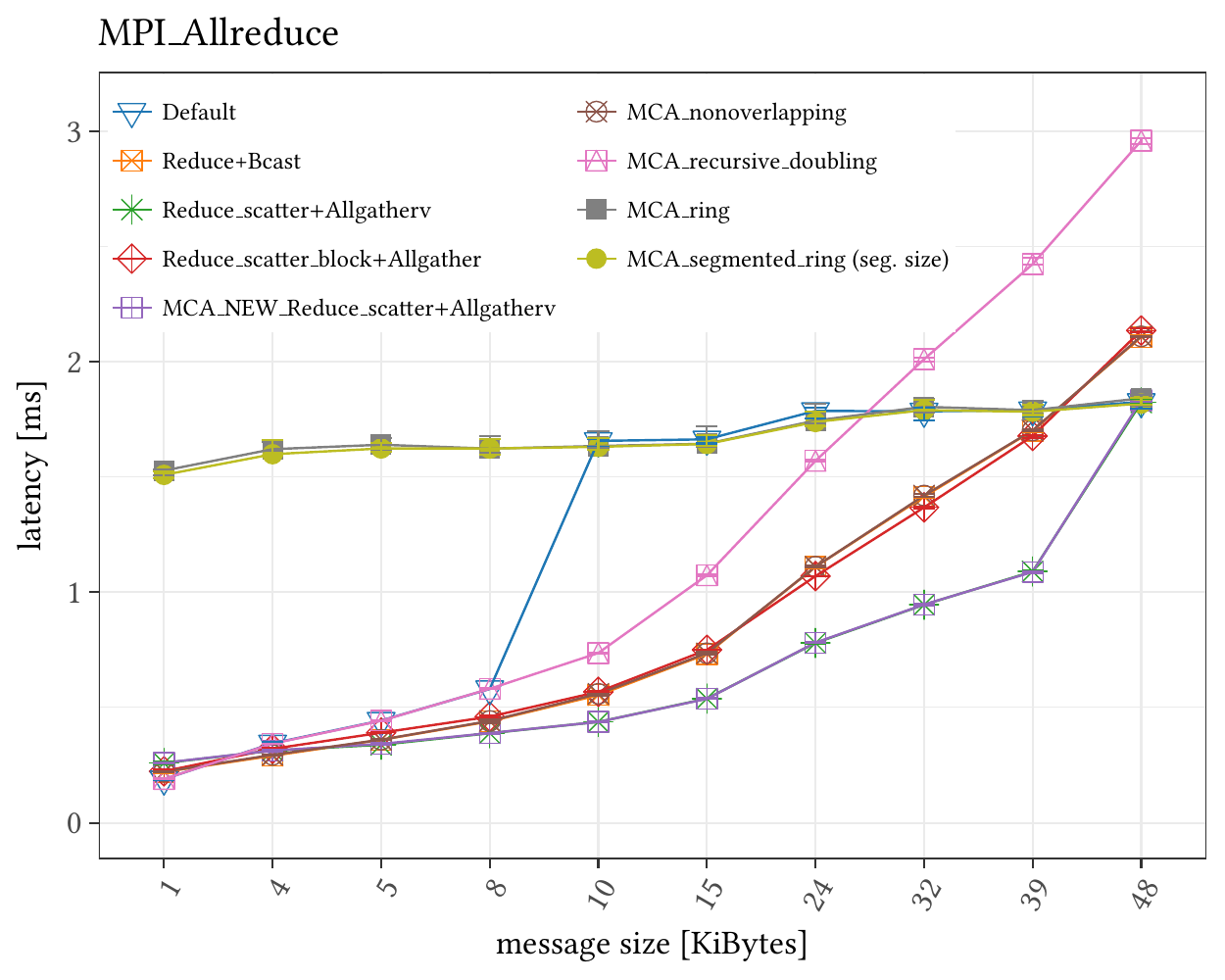}
  \caption{Performance comparison of \mpiallreduce, its \mockups, and
    MCA-tuned variants of \mpiallreduce (\jupiteropenmpilatest,
    \num{32x16} processes, \machjupiter).}
  \label{fig:allreduce_mca_tuned}
\end{figure}

Our extensive experimental analysis also revealed other interesting
cases.  One of them is shown in \fig~\ref{fig:allreduce_mca_tuned}.
The plot shows latencies measured for \mpiallreduce, its \mockup
variants implemented in \pgtunelib, and several algorithmic versions
found in \jupiteropenmpilatest (only the fastest ones). 
Here, the algorithmic version called \texttt{MCA\_nonoverlapping}
performs almost identical to our Reduce+Bcast \mockup variant. Indeed,
when inspecting the internals of \openmpi, this algorithmic variant
uses exactly these two collectives. Additionally, we discover that the
\mockup version combining \mpireducescatter and \mpiallgatherv
outperforms all other algorithms, even all versions provided by
\openmpi after the exhaustive search was done. Thus, \pgtunelib helps
developers to detect cases for which a better algorithmic variant
exists.%
\begin{taggedblock}{hpcasia}
We took the role of an \openmpi developer and implemented the
Allreduce variant based on Reduce\_scatter and Allgatherv within
\jupiteropenmpilatest~\cite{arxiv_pgtunelib}.
\end{taggedblock}%
\begin{taggedblock}{arxiv}
We took the role of an \openmpi developer and implemented the
Allreduce variant based on Reduce\_scatter and Allgatherv within
\jupiteropenmpilatest\footnote{The diff can be found at 
\url{http://hunoldscience.net/download/pgtunelib/allreduce_as_reducescatter_allgatherv.patch}}. 
\end{taggedblock}%
This version is denoted as
\texttt{MCA\_NEW\_Reduce\_scatter+Allgatherv} in
\fig~\ref{fig:allreduce_mca_tuned}. The plot shows that this new
algorithm in \openmpi exactly matches the expected latency achieved by
the \mockup combining Reduce\_scatter and Allgatherv and outperforms
all other variants.

\section{Conclusions}
\label{sec:conclusions}

Tuning MPI libraries can be extremely rewarding in terms of overall
efficiency of parallel machines, as MPI is the de-facto standard for
data communication on larger distributed memory machines.  Parameter
tuning is usually a valuable method for achieving the goal of an
improved MPI software layer. The downsides of parameter tuning are
twofold: (1) it is relatively expensive as libraries such as \openmpi
provide hundreds of possibly interacting parameters, and (2) a
performance baseline is often missing, \ie, how good is good enough
since global minima are usually unknown.

Tuning MPI libraries by using performance guidelines can complement
the traditional parameter-based approach. Self-consistent performance
guidelines define relations between the performance of a specialized
functionality and a less specialized functionality, both of which realize
semantically the same operation, \eg, the latency of \mpiallgather
should be smaller than using \mpigather and a subsequent call to
\mpibcast.

In the present paper, we have extended performance guidelines for
blocking, collective MPI operations. We have implemented each
semantically matching guideline as a \mockup function in a library
called \pgtunelib. With this library, it is possible to find
performance deficits of MPI libraries by scanning for guideline
violations. The library creates so-called performance profiles that
can be used to replace specific MPI functions by their \mockup version
at \runtime. 

Our experimental results show that \pgtunelib can indeed overcome
performance problems of MPI libraries on all systems that we have
tested on. In addition, our results also show that \pgtunelib also
reveals cases in MPI libraries (\eg, \openmpi) for which even better
algorithms exist. The biggest advantage of \pgtunelib, however, is the
fact that it can be used with any MPI library, whether or not it
exposes parameters for tuning purposes.

\begin{taggedblock}{arxiv}
\section*{Acknowledgments}

We thank Bernd Mohr (Julich Supercomputing Centre) for providing us
access to \machjuqueen.
\end{taggedblock}

\begin{taggedblock}{hpcasia}
\bibliographystyle{ACM-Reference-Format}
\bibliography{main}\end{taggedblock}
\begin{taggedblock}{arxiv}
\newpage
\bibliographystyle{ACM-Reference-Format}
\bibliography{main}\end{taggedblock}

\begin{taggedblock}{hpcasia}
\newpage
\appendix
\section{Appendix}
\label{sec:appendix}
\vspace*{.2ex}

\end{taggedblock}
\begin{taggedblock}{arxiv}
\appendix
\onecolumn 
\newgeometry{top=1in,bottom=1in,left=1in,right=1in}
\section{Appendix}
\label{sec:appendix}

We show a collection of performance graphs for \machjuqueen, \machvsc,
and \machjupiter. When no violations of performance guidelines are
detected, we mark these cases with a gray background.

\subsection{\machjuqueen}
\begin{figure*}[h!]
\centering
\begin{minipage}{ .38\linewidth }
\includegraphics[width=\linewidth]{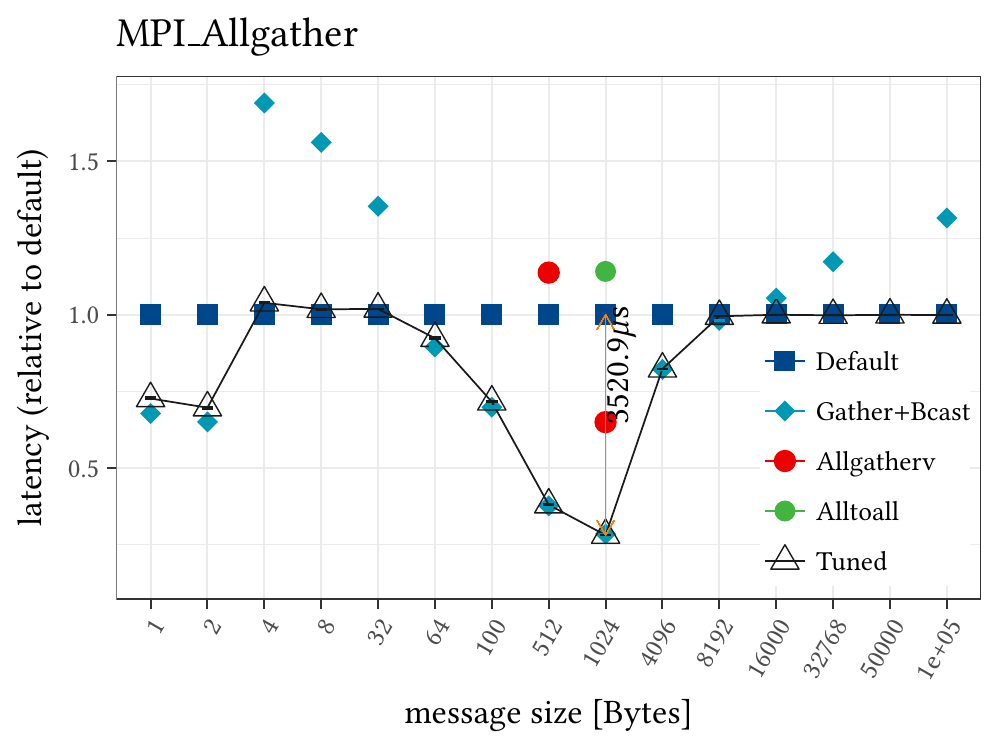}
\end{minipage}%
\hspace*{0.1\linewidth}%
\begin{minipage}{ .38\linewidth }
\includegraphics[width=\linewidth]{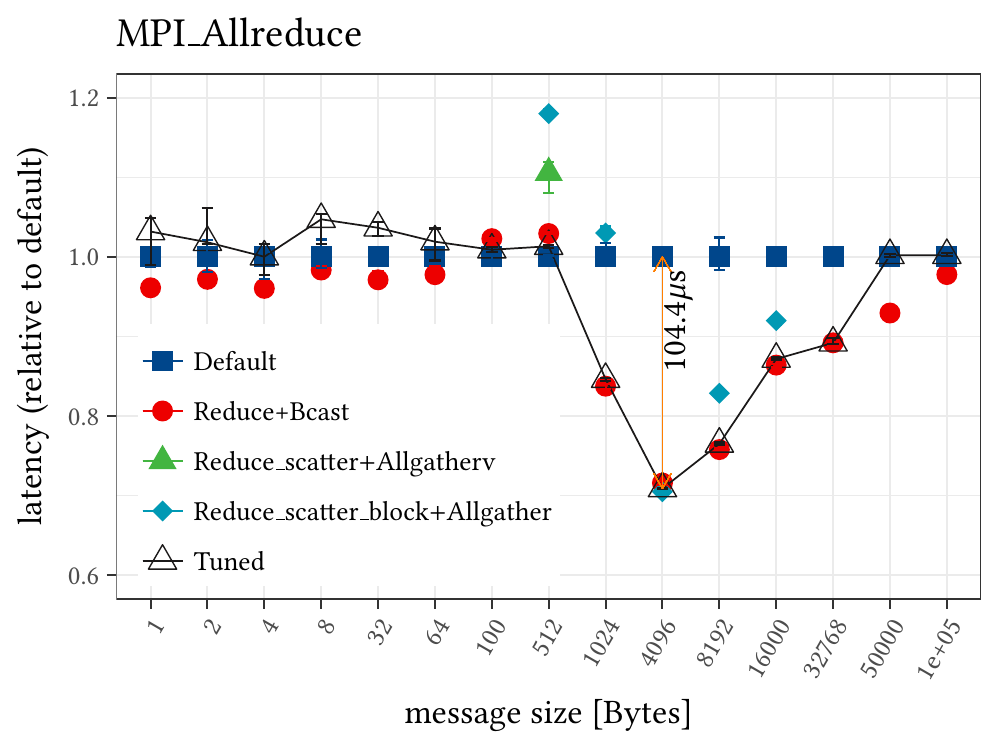}
\end{minipage}%
\\%
\begin{minipage}{ .38\linewidth }
\includegraphics[width=\linewidth]{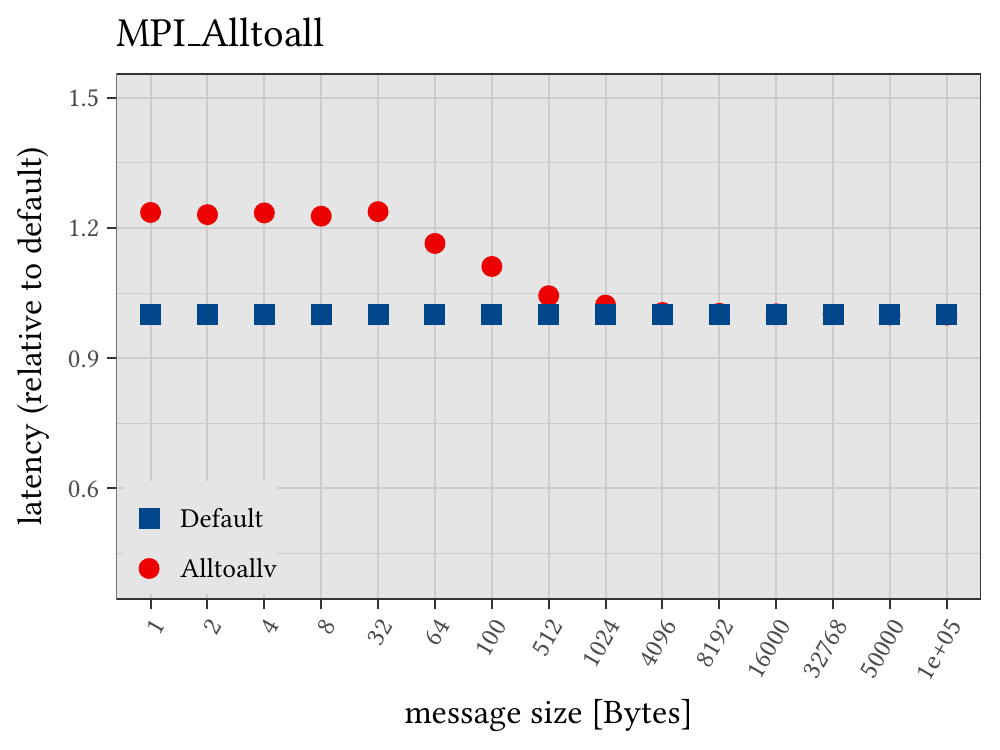}
\end{minipage}%
\hspace*{0.1\linewidth}%
\begin{minipage}{ .38\linewidth }
\includegraphics[width=\linewidth]{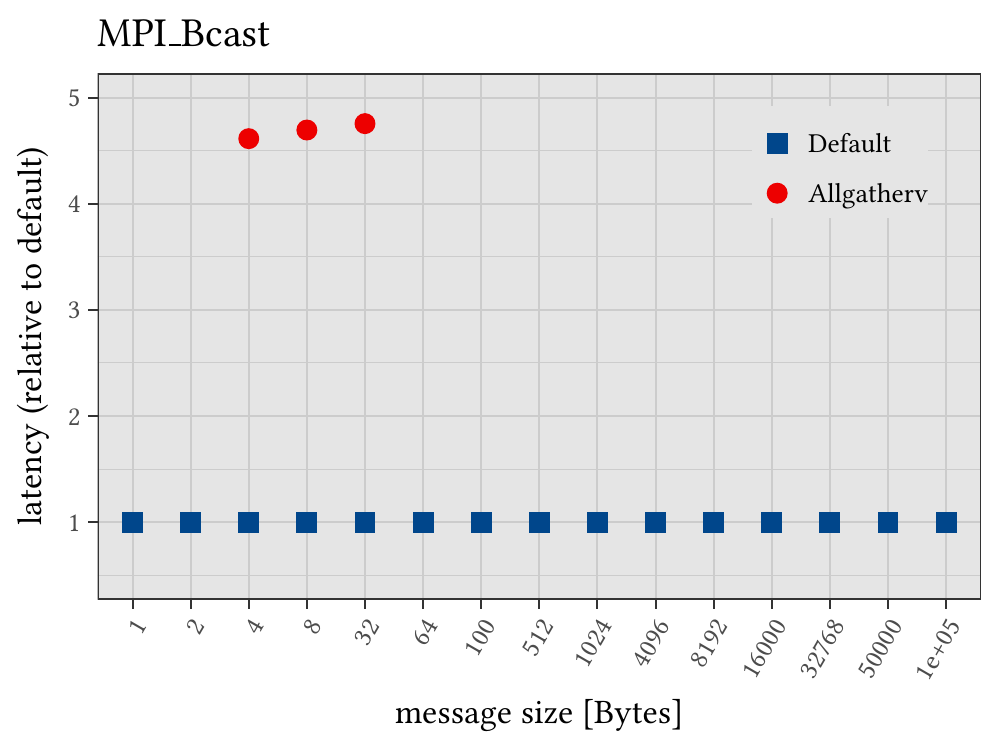}
\end{minipage}%
\\%
\begin{minipage}{ .38\linewidth }
\includegraphics[width=\linewidth]{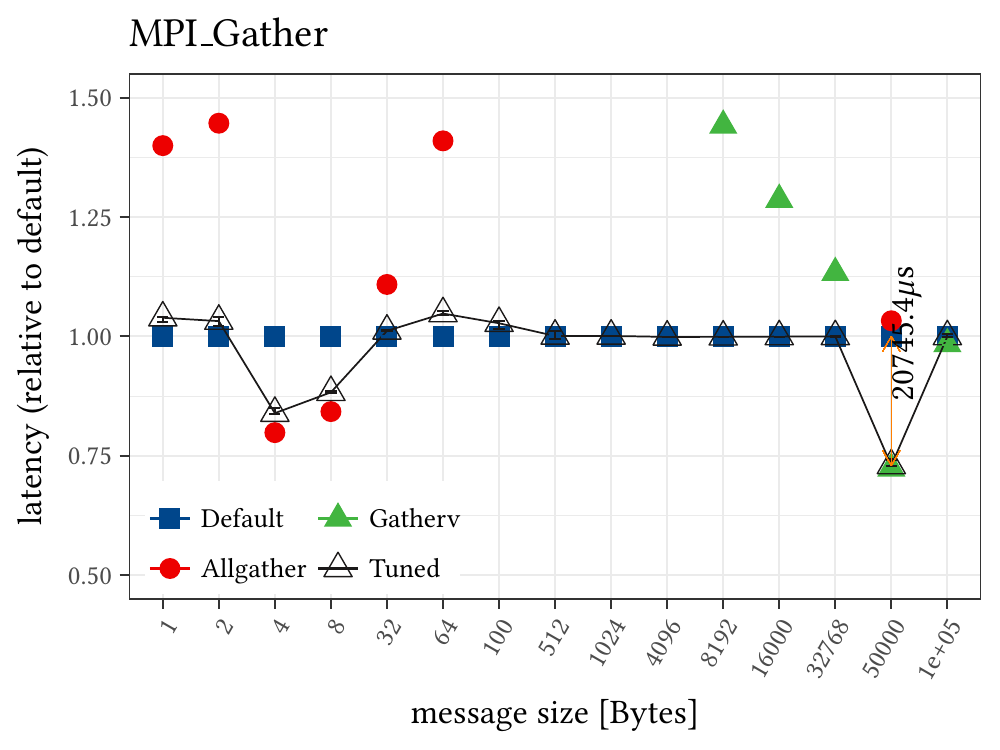}
\end{minipage}%
\hspace*{0.1\linewidth}%
\begin{minipage}{ .38\linewidth }
\includegraphics[width=\linewidth]{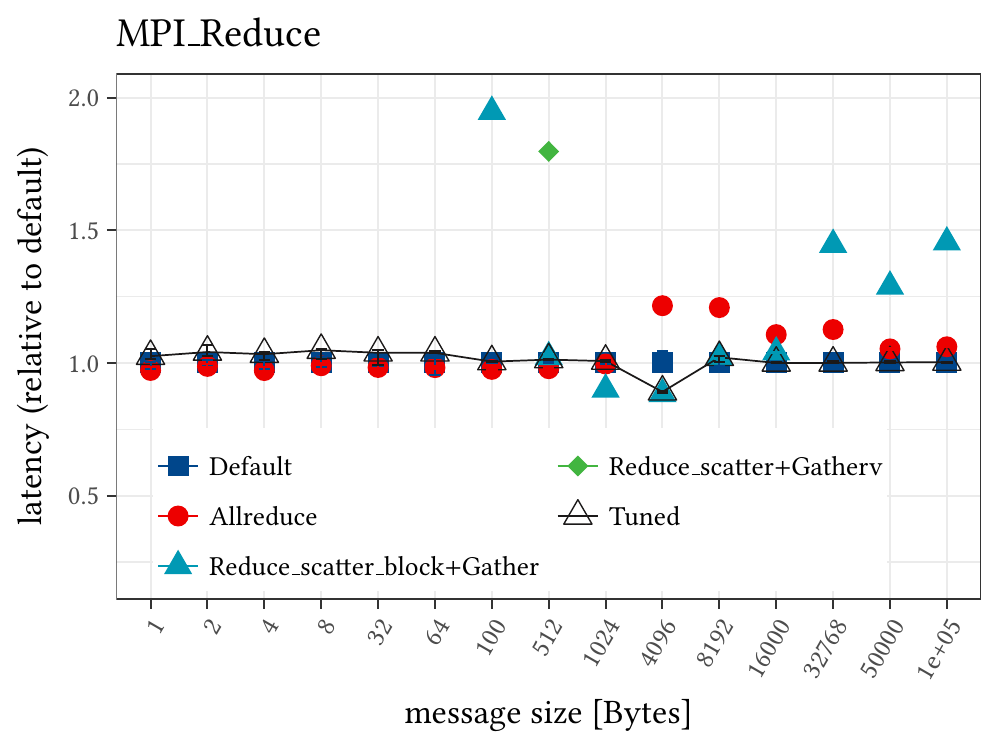}
\end{minipage}%
\\%
\begin{minipage}{ .38\linewidth }
\includegraphics[width=\linewidth]{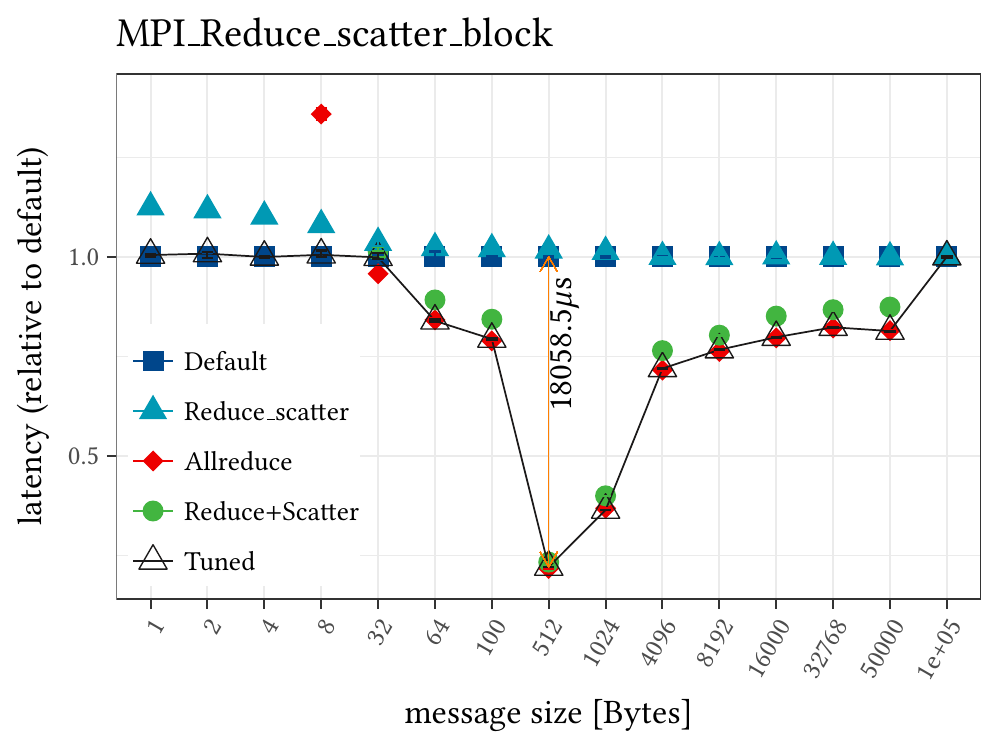}
\end{minipage}%
\hspace*{0.1\linewidth}%
\begin{minipage}{ .38\linewidth }
\includegraphics[width=\linewidth]{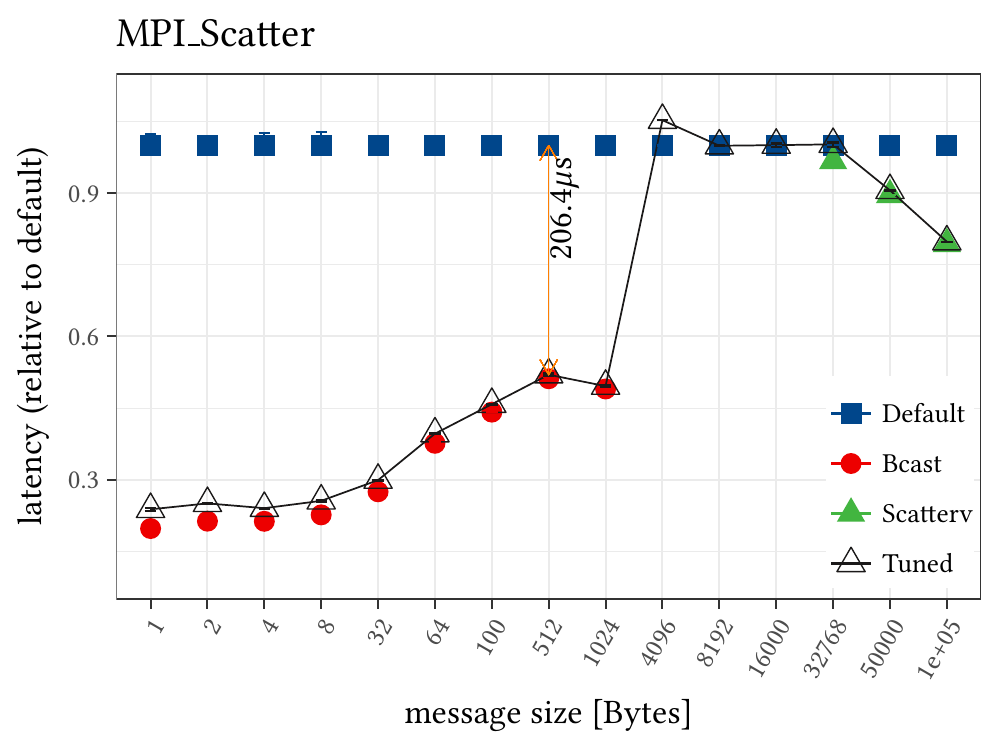}
\end{minipage}%
\caption{\label{fig:perf_64x16_juqueen_all}%
Performance comparison between \pgdefault and \pgtuned version of \juqueenmpi (\num{64x16} processes, \machjuqueen)%
}%
\end{figure*}

\FloatBarrier
\clearpage

\subsection{\machvsc}%
\begin{figure*}[htpb]
\centering
\begin{minipage}{ .38\linewidth }
\includegraphics[width=\linewidth]{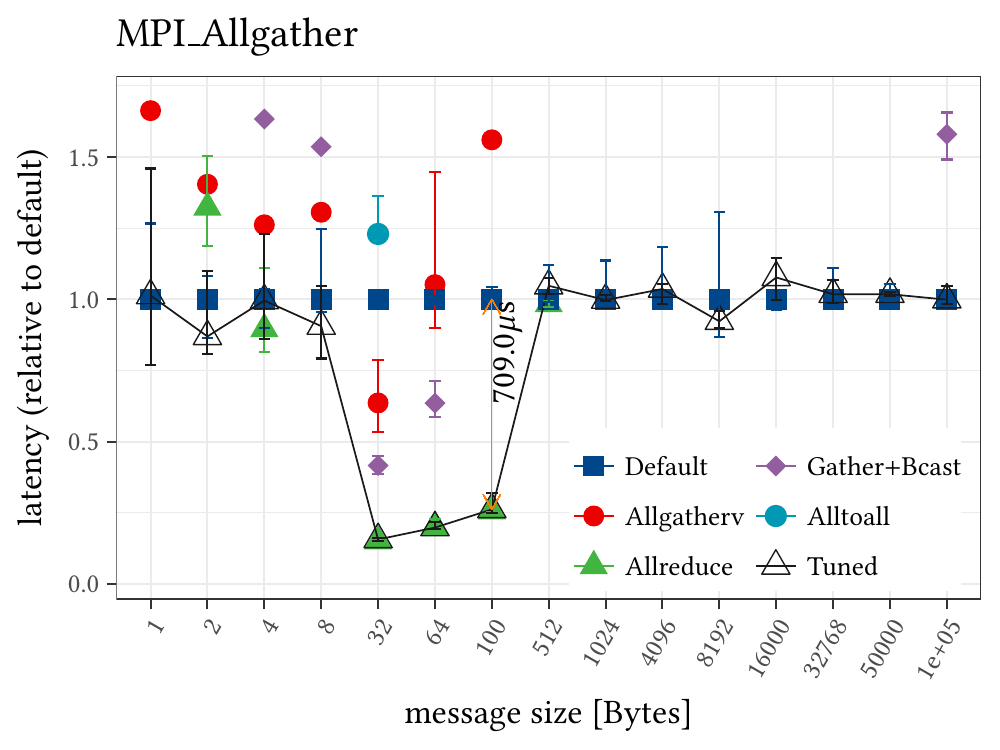}
\end{minipage}%
\hspace*{0.1\linewidth}%
\begin{minipage}{ .38\linewidth }
\includegraphics[width=\linewidth]{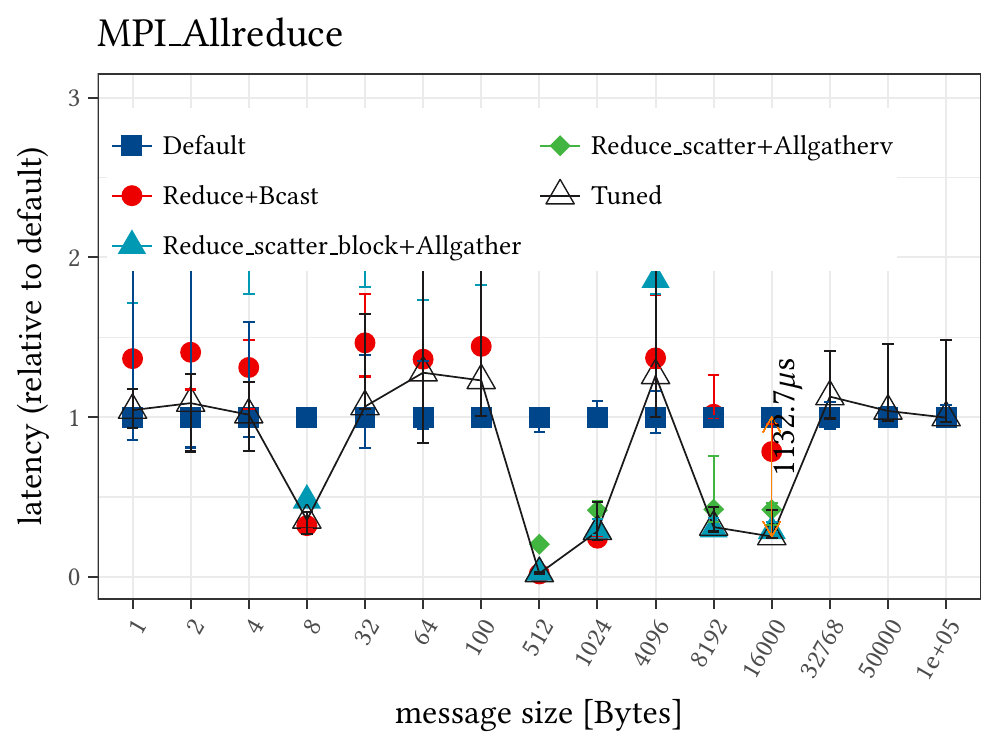}
\end{minipage}%
\\%
\begin{minipage}{ .38\linewidth }
\includegraphics[width=\linewidth]{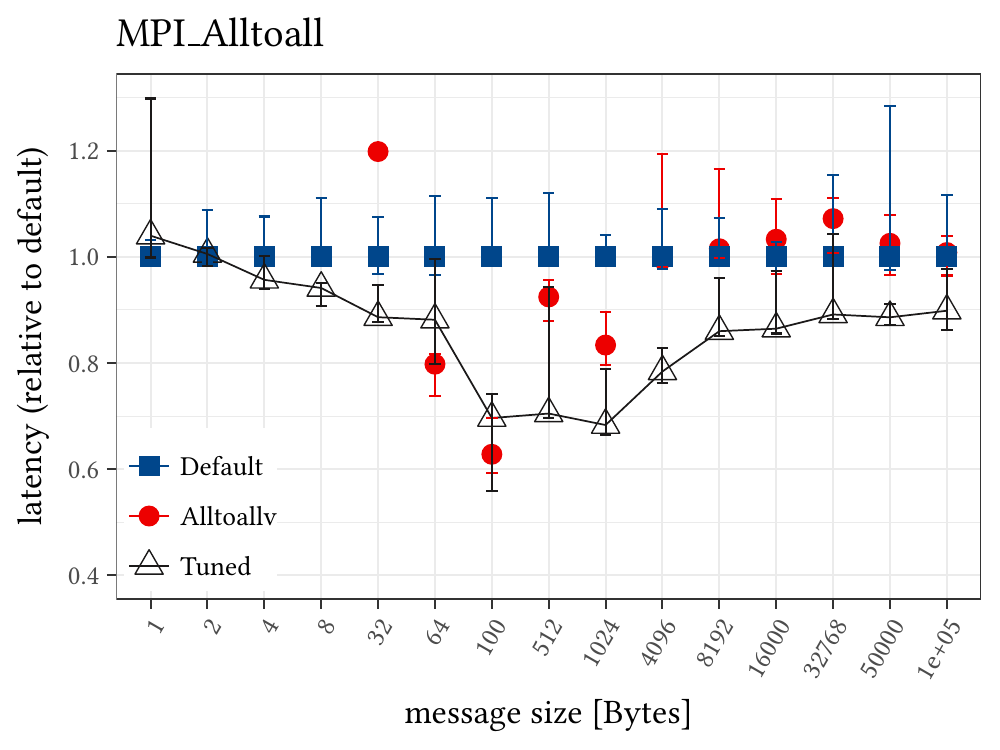}
\end{minipage}%
\hspace*{0.1\linewidth}%
\begin{minipage}{ .38\linewidth }
\includegraphics[width=\linewidth]{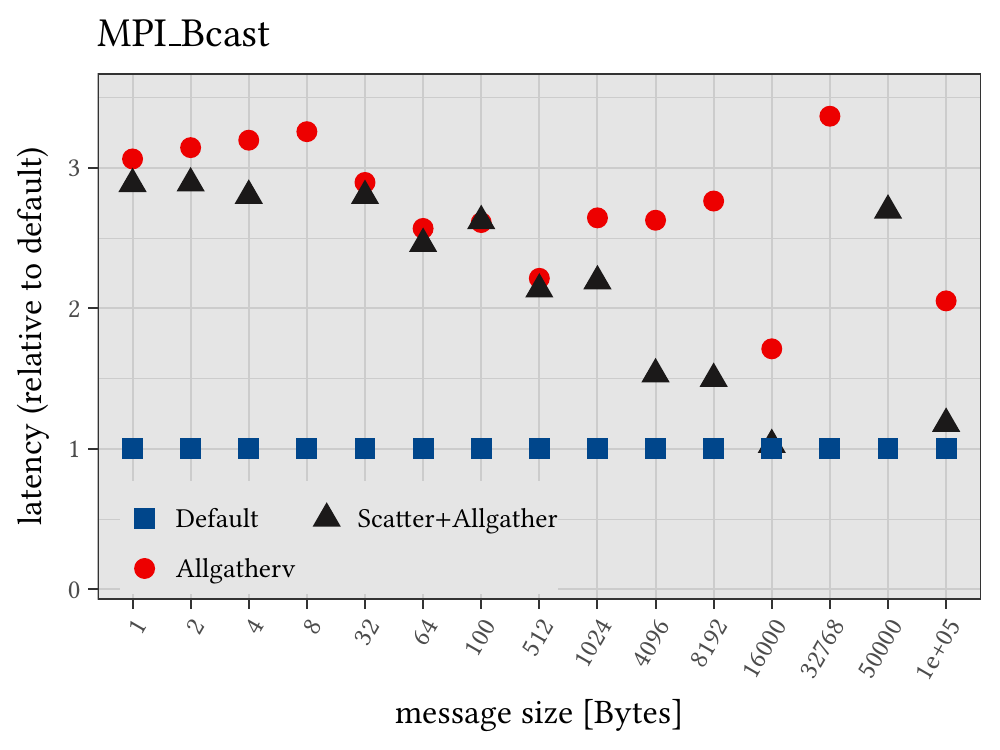}
\end{minipage}%
\\%
\begin{minipage}{ .38\linewidth }
\includegraphics[width=\linewidth]{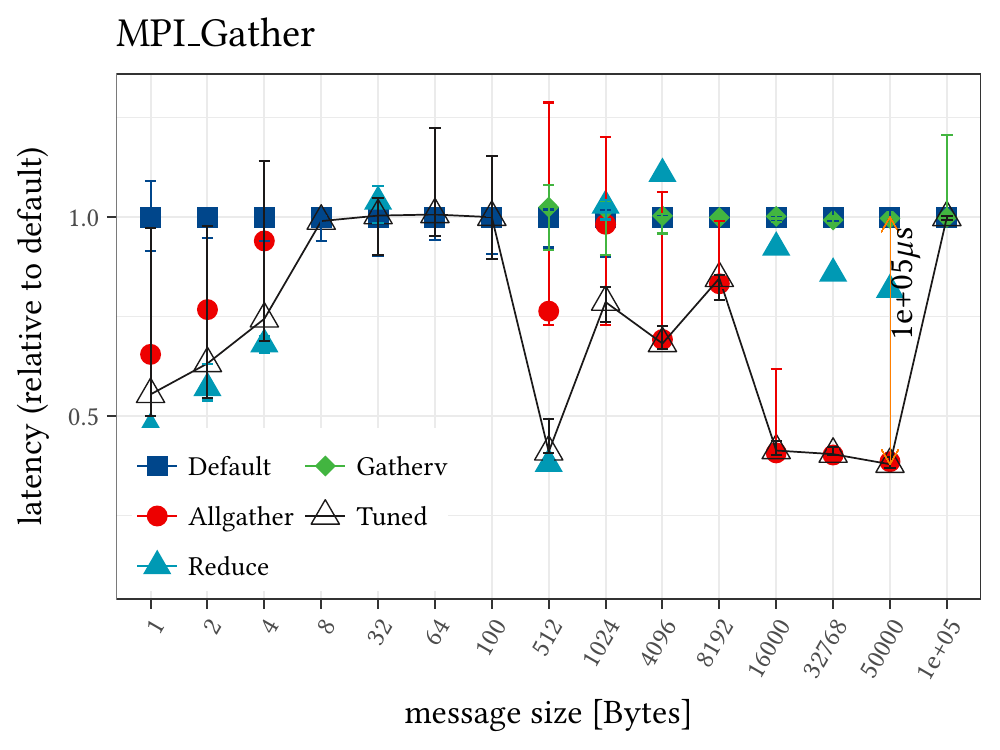}
\end{minipage}%
\hspace*{0.1\linewidth}%
\begin{minipage}{ .38\linewidth }
\includegraphics[width=\linewidth]{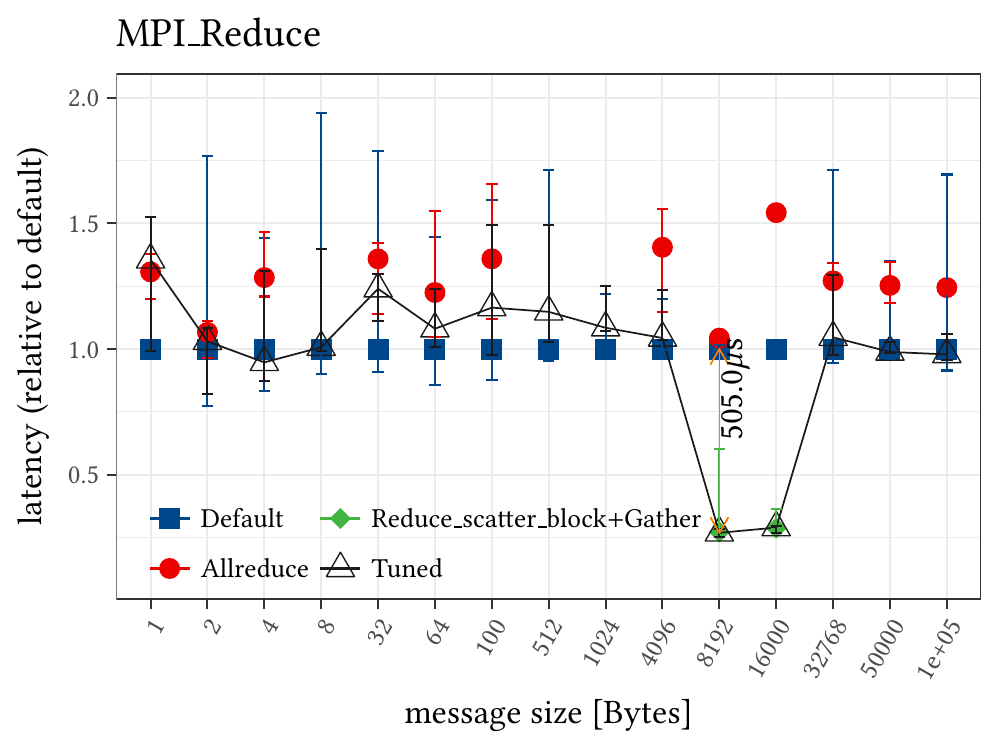}
\end{minipage}%
\\%
\begin{minipage}{ .38\linewidth }
\includegraphics[width=\linewidth]{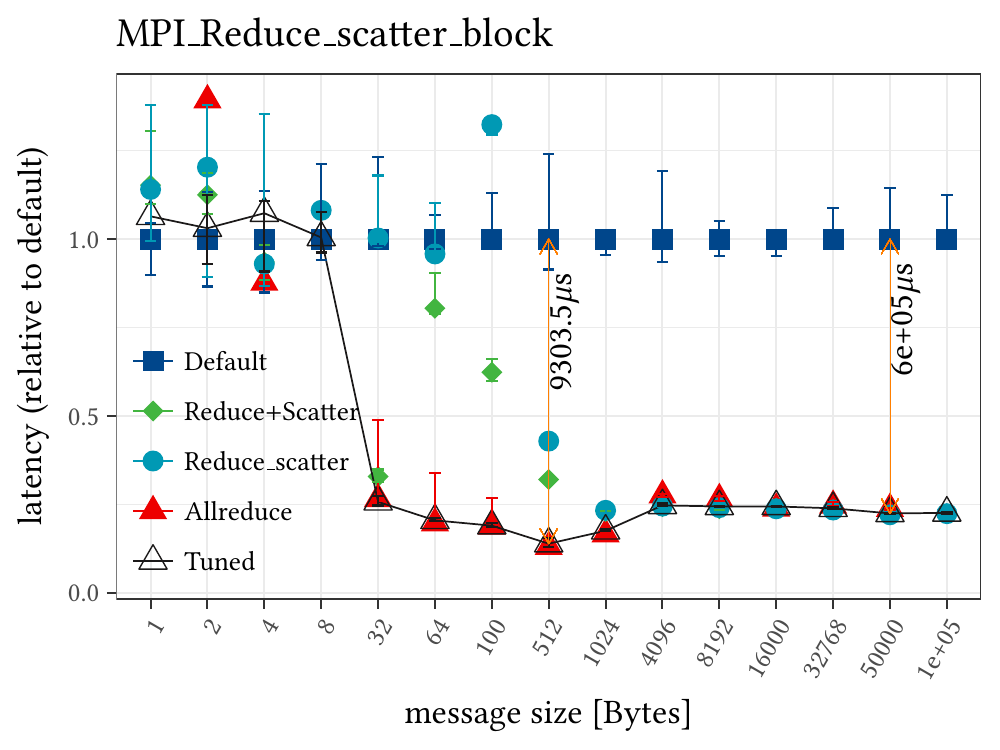}
\end{minipage}%
\hspace*{0.1\linewidth}%
\begin{minipage}{ .38\linewidth }
\includegraphics[width=\linewidth]{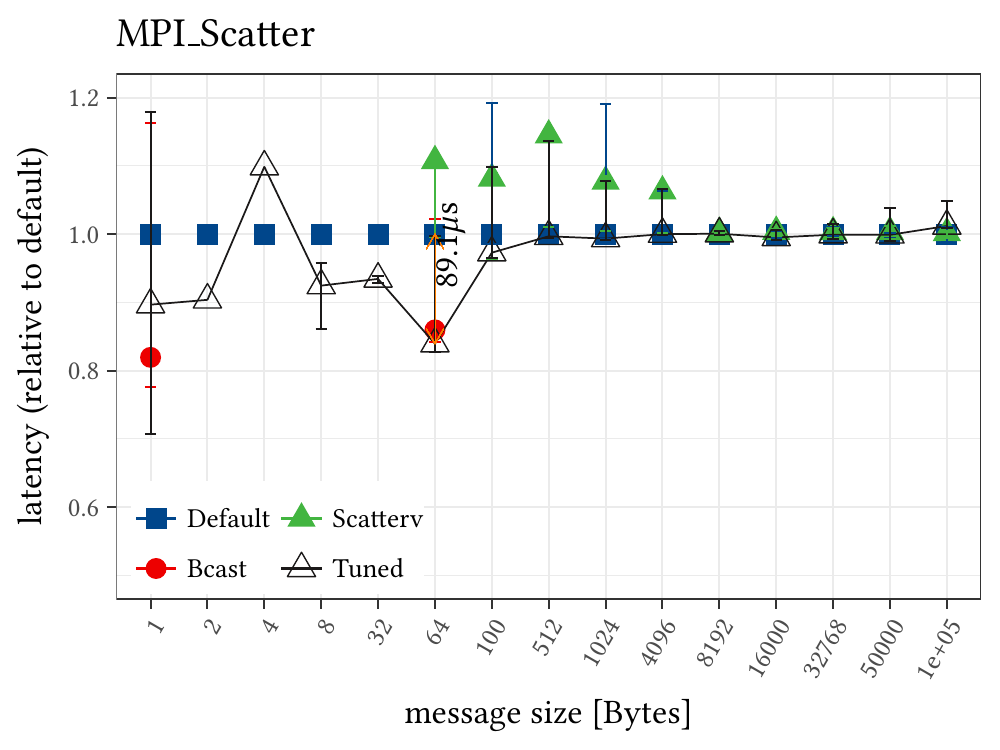}
\end{minipage}%
\caption{\label{fig:perf_64x16_vsc3_all}%
Performance comparison between \pgdefault and \pgtuned version of \intelmpi (\num{64x16} processes, \machvsc)%
}%
\end{figure*}

\FloatBarrier
\clearpage

\subsection{\machjupiter}%
\begin{figure*}[htpb]
\centering
\begin{minipage}{ .38\linewidth }
\includegraphics[width=\linewidth]{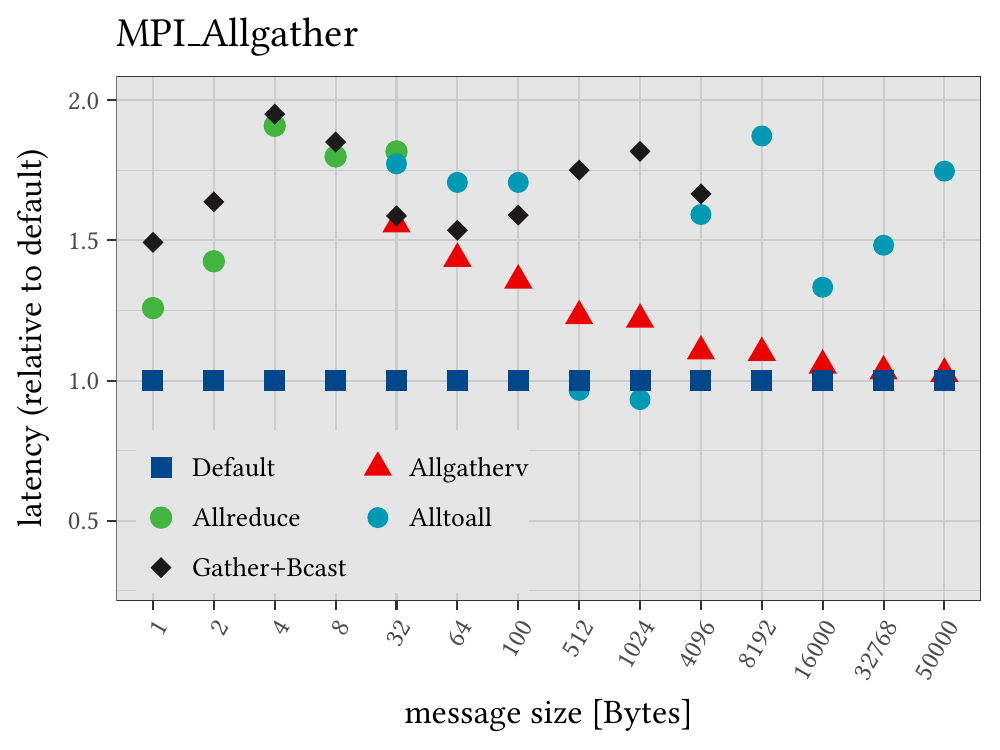}
\end{minipage}%
\hspace*{0.1\linewidth}%
\begin{minipage}{ .38\linewidth }
\includegraphics[width=\linewidth]{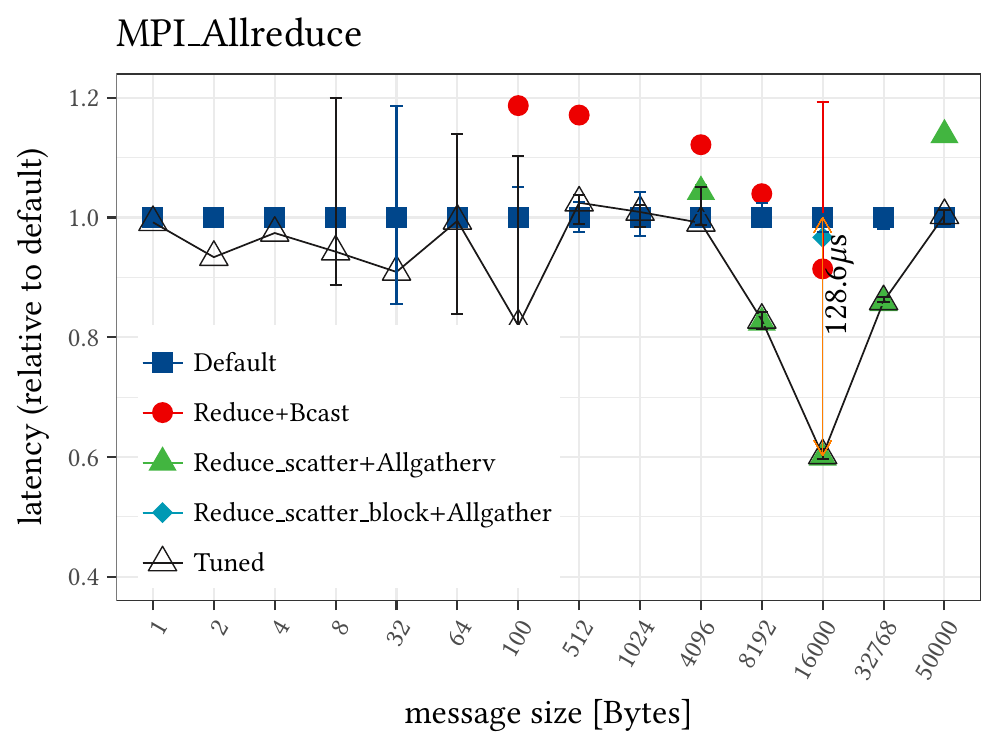}
\end{minipage}%
\\%
\begin{minipage}{ .38\linewidth }
\includegraphics[width=\linewidth]{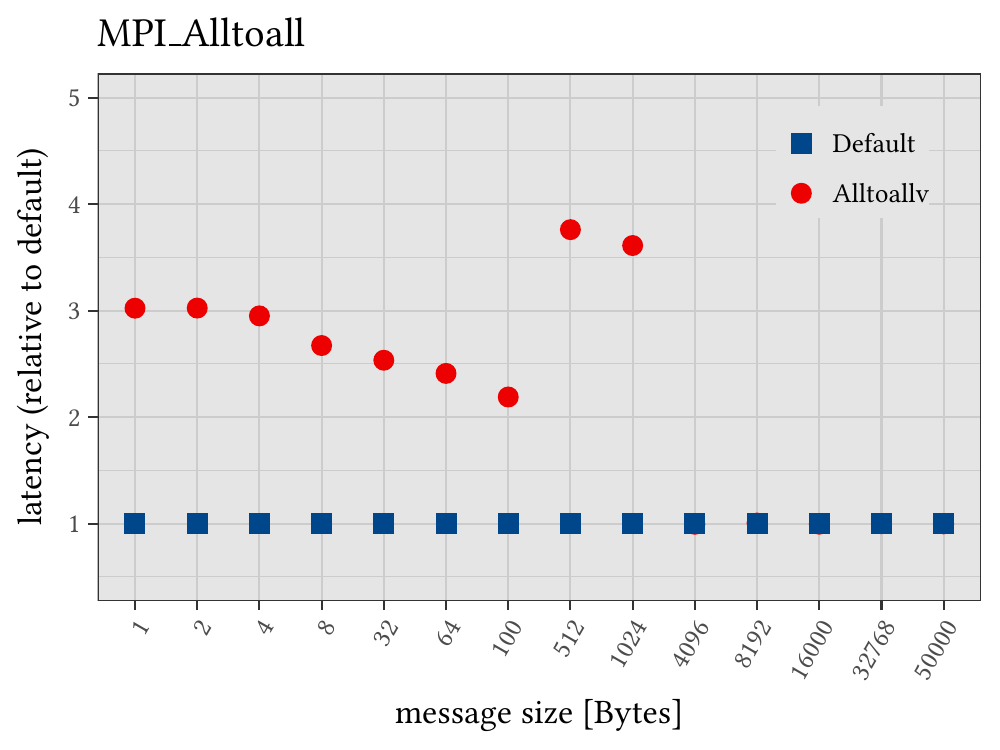}
\end{minipage}%
\hspace*{0.1\linewidth}%
\begin{minipage}{ .38\linewidth }
\includegraphics[width=\linewidth]{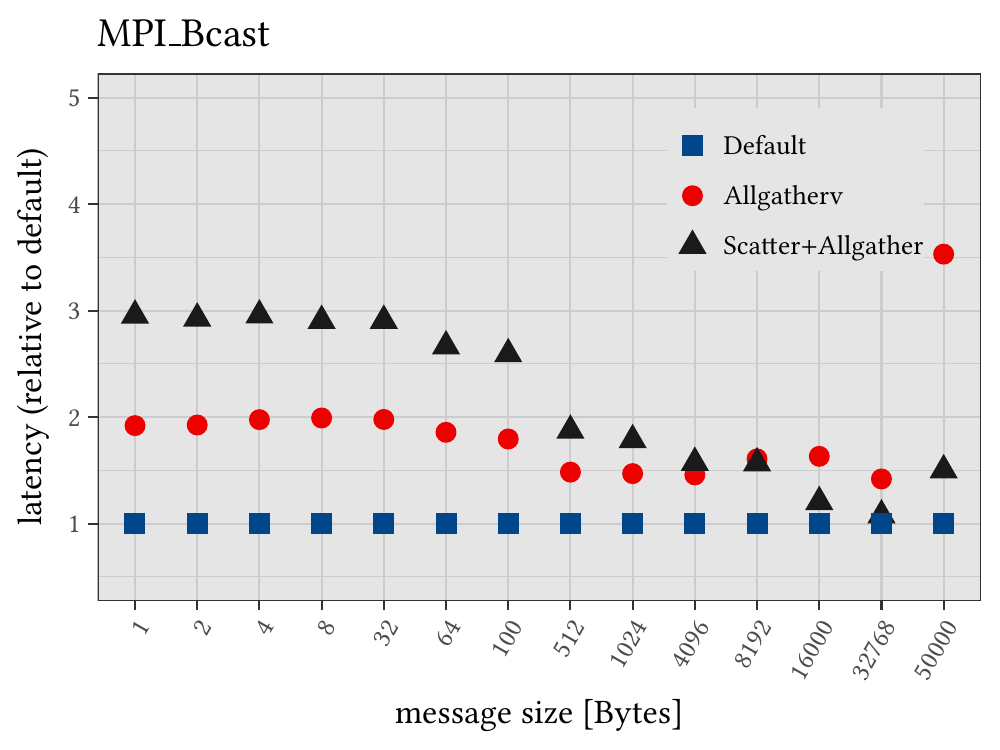}
\end{minipage}%
\\%
\begin{minipage}{ .38\linewidth }
\includegraphics[width=\linewidth]{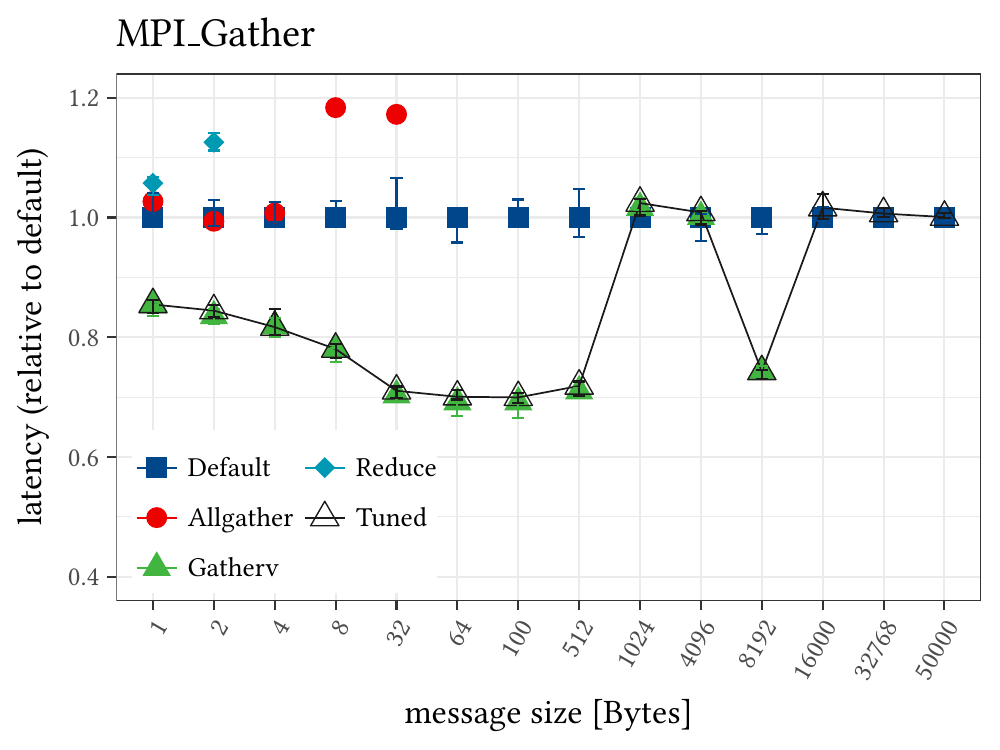}
\end{minipage}%
\hspace*{0.1\linewidth}%
\begin{minipage}{ .38\linewidth }
\includegraphics[width=\linewidth]{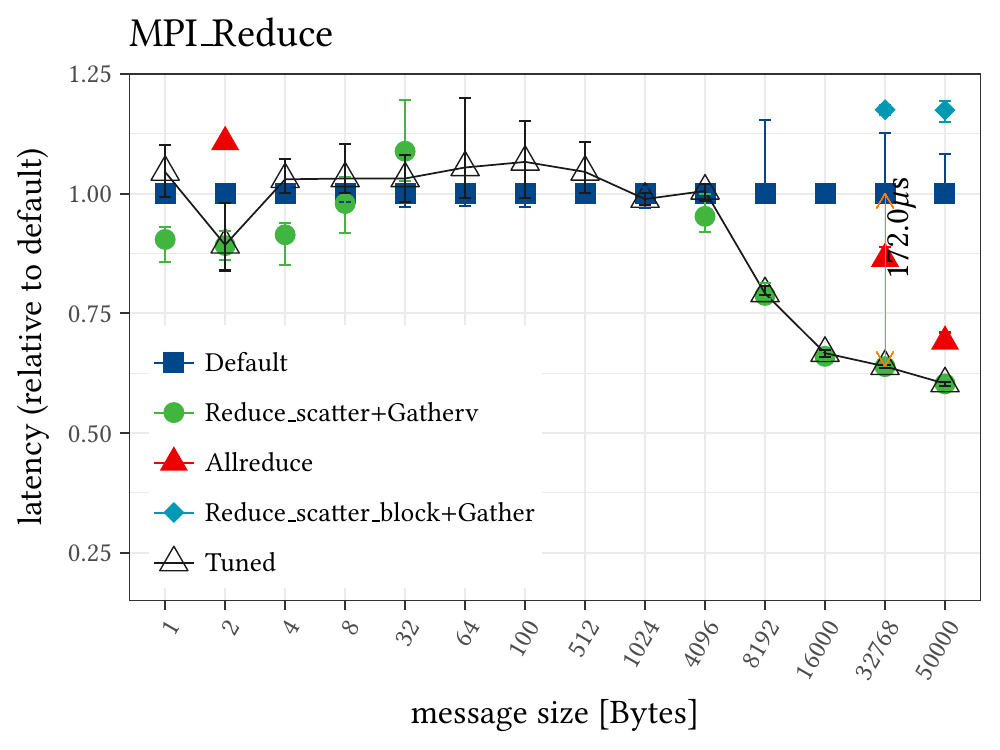}
\end{minipage}%
\\%
\begin{minipage}{ .38\linewidth }
\includegraphics[width=\linewidth]{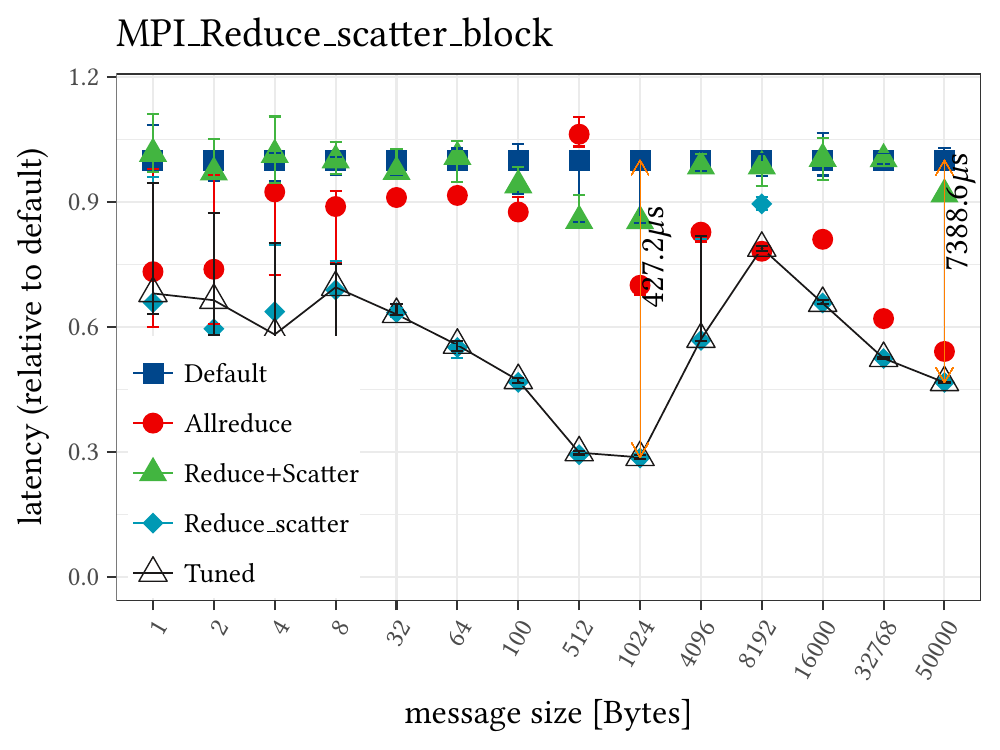}
\end{minipage}%
\hspace*{0.1\linewidth}%
\begin{minipage}{ .38\linewidth }
\includegraphics[width=\linewidth]{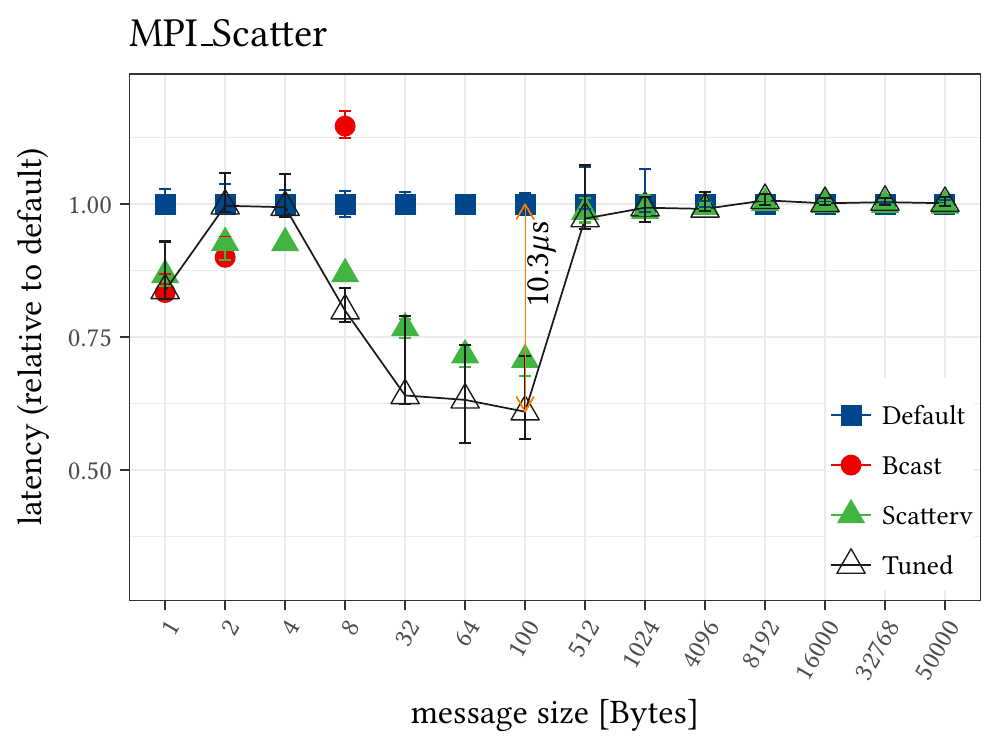}
\end{minipage}%
\caption{\label{fig:perf_32x1_jupiter_openmpi_all}%
Performance comparison between \pgdefault and \pgtuned version of \jupiteropenmpilatest (\num{32x1} processes, \machjupiter)%
}%
\end{figure*}

\begin{figure*}[htpb]
\centering
\begin{minipage}{ .38\linewidth }
\includegraphics[width=\linewidth]{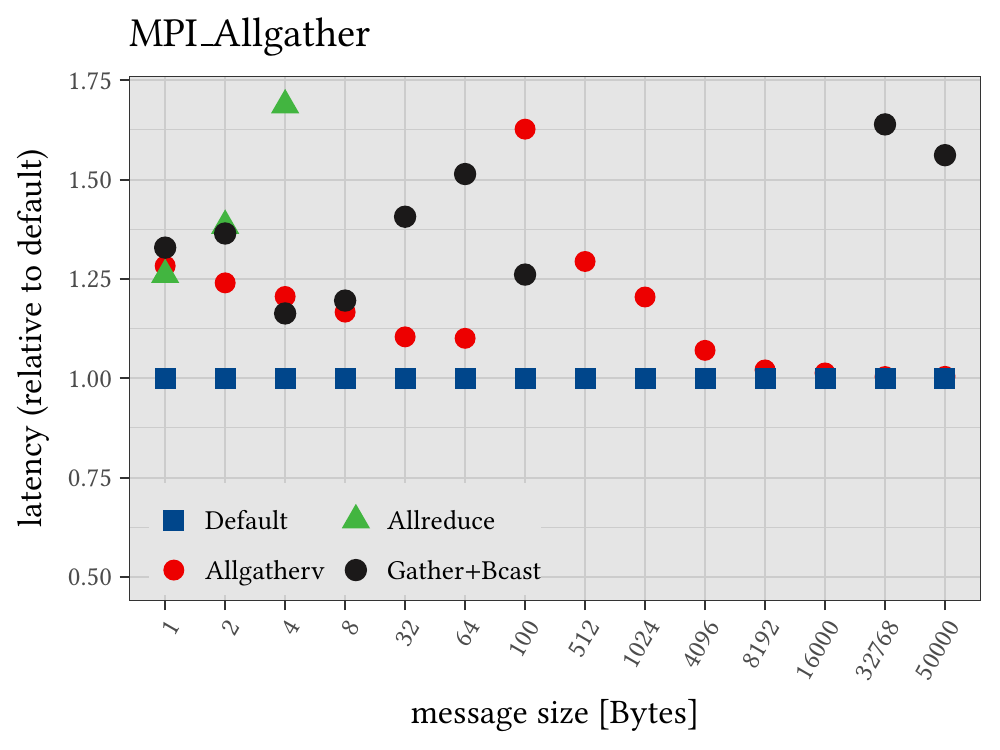}
\end{minipage}%
\hspace*{0.1\linewidth}%
\begin{minipage}{ .38\linewidth }
\includegraphics[width=\linewidth]{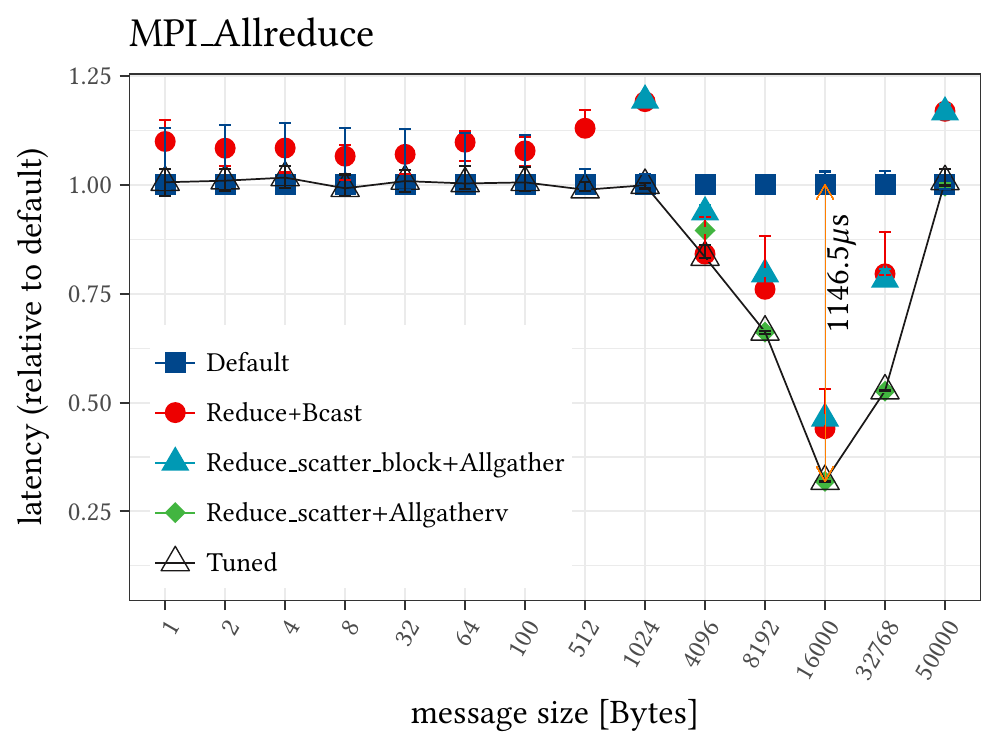}
\end{minipage}%
\\%
\begin{minipage}{ .38\linewidth }
\includegraphics[width=\linewidth]{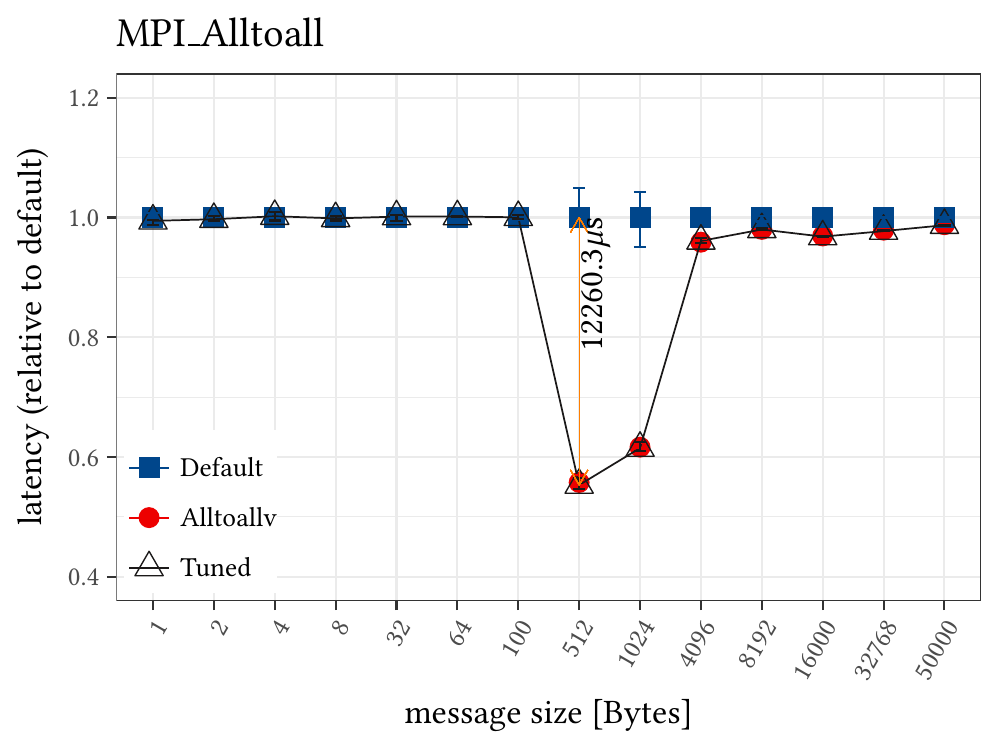}
\end{minipage}%
\hspace*{0.1\linewidth}%
\begin{minipage}{ .38\linewidth }
\includegraphics[width=\linewidth]{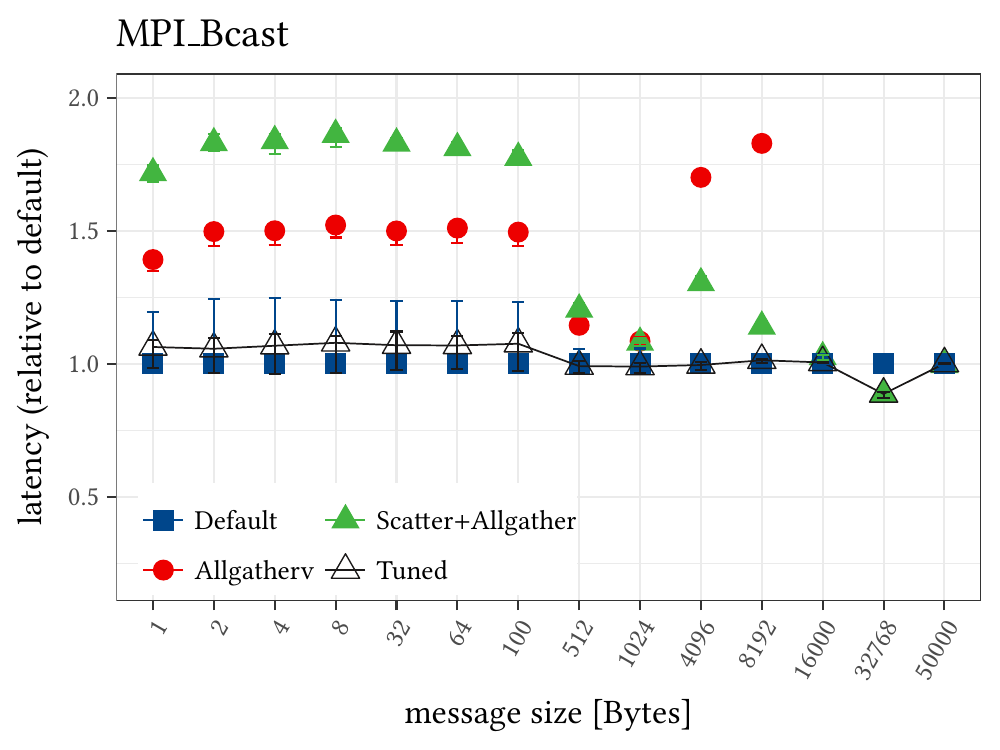}
\end{minipage}%
\\%
\begin{minipage}{ .38\linewidth }
\includegraphics[width=\linewidth]{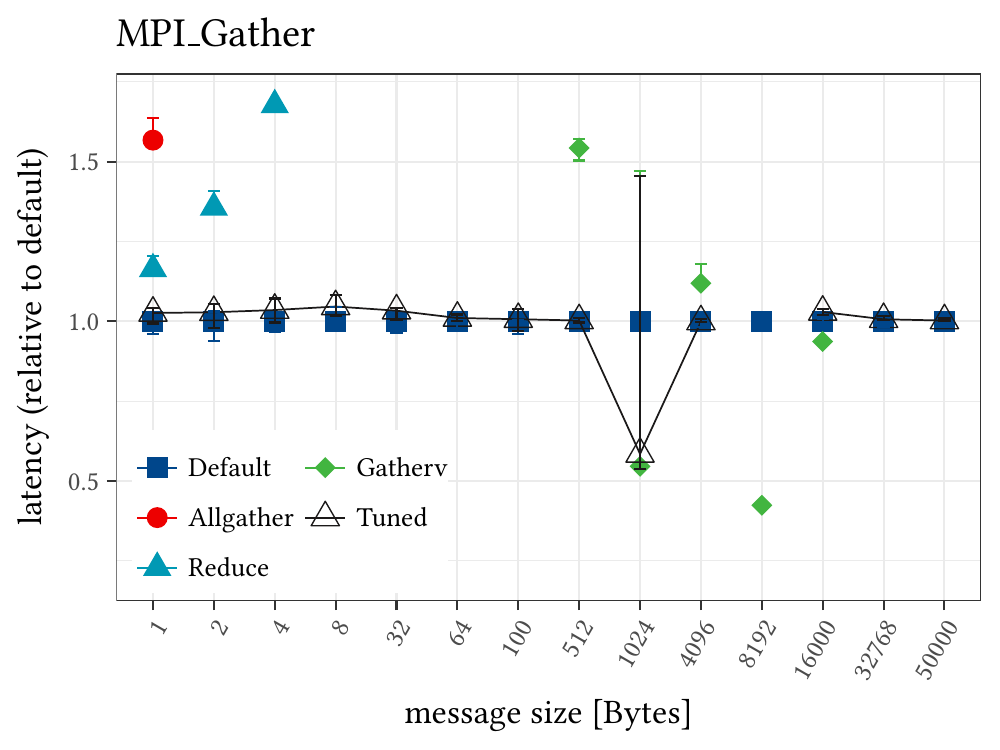}
\end{minipage}%
\hspace*{0.1\linewidth}%
\begin{minipage}{ .38\linewidth }
\includegraphics[width=\linewidth]{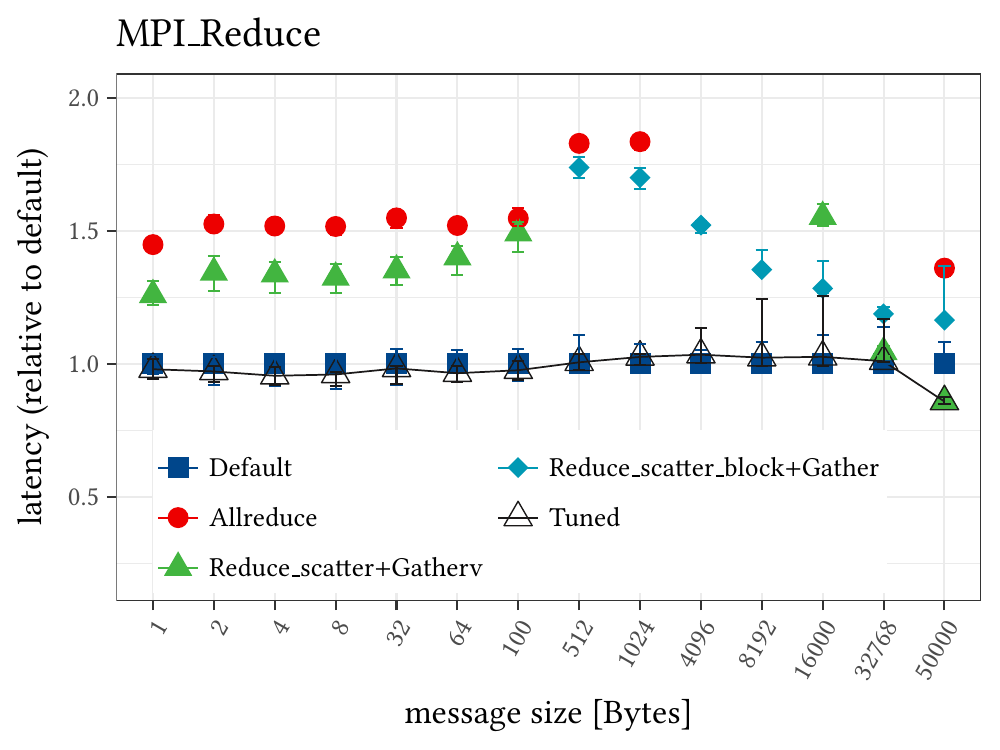}
\end{minipage}%
\\%
\begin{minipage}{ .38\linewidth }
\includegraphics[width=\linewidth]{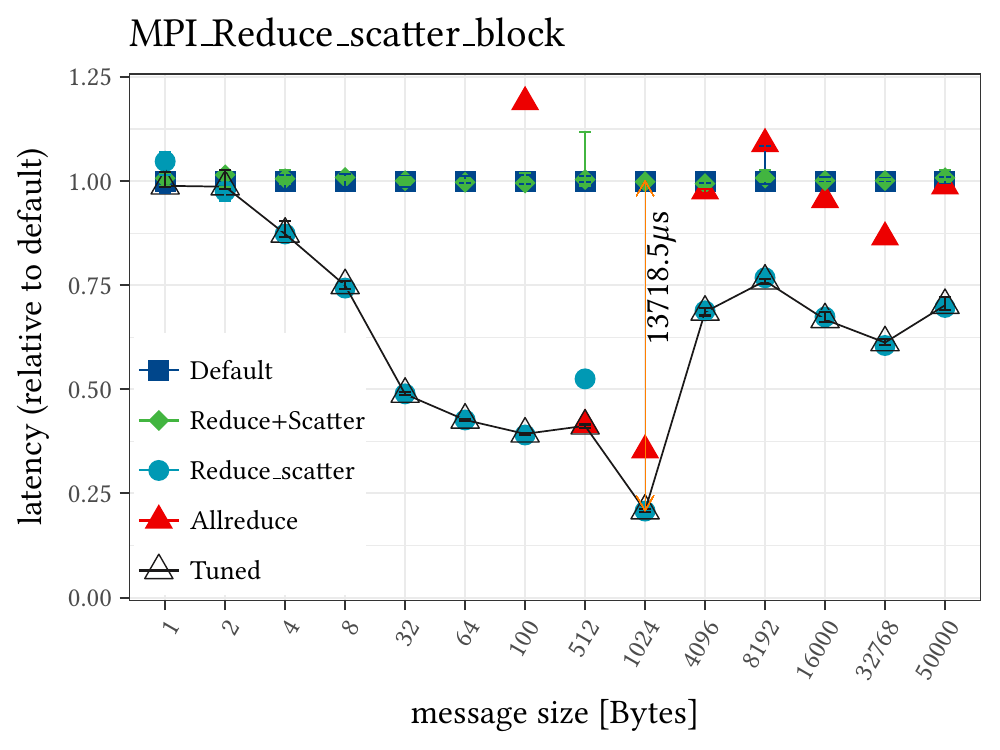}
\end{minipage}%
\hspace*{0.1\linewidth}%
\begin{minipage}{ .38\linewidth }
\includegraphics[width=\linewidth]{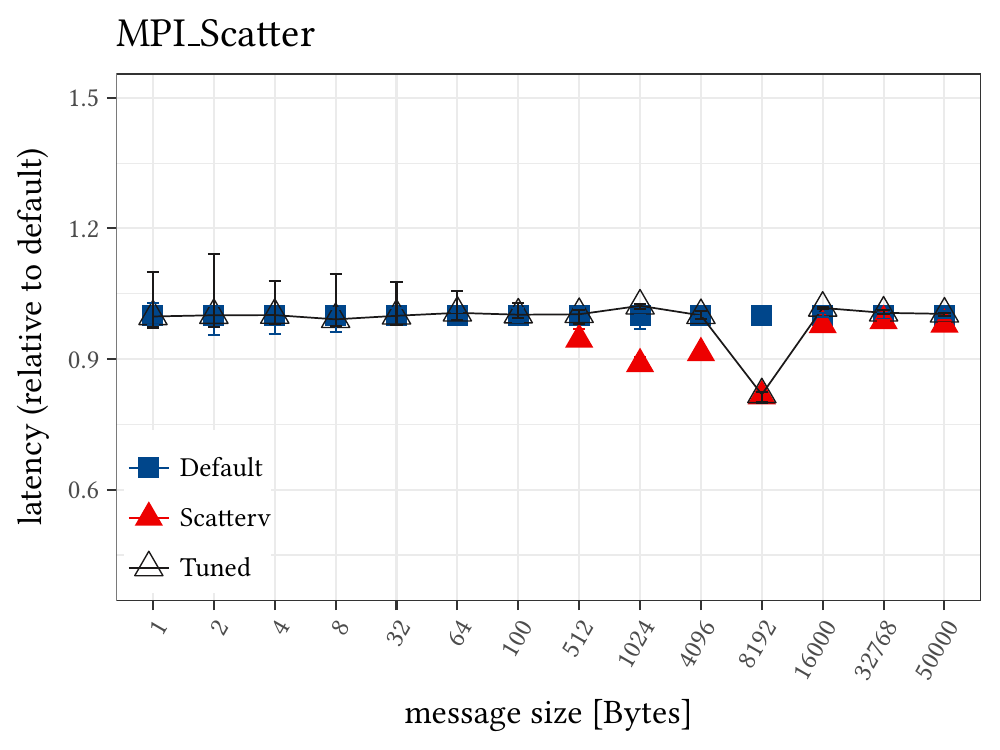}
\end{minipage}%
\caption{\label{fig:perf_32x16_jupiter_openmpi_all}%
Performance comparison between \pgdefault and \pgtuned version of \jupiteropenmpilatest (\num{32x16} processes, \machjupiter)%
}%
\end{figure*}

\begin{figure*}[htpb]
\centering
\begin{minipage}{ .38\linewidth }
\includegraphics[width=\linewidth]{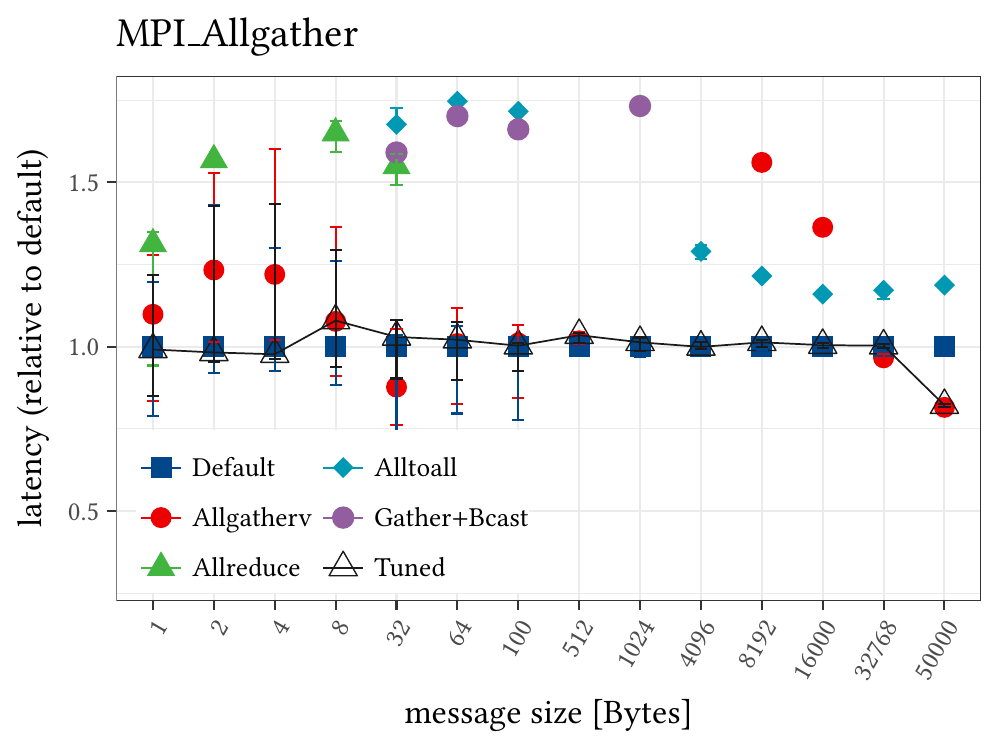}
\end{minipage}%
\hspace*{0.1\linewidth}%
\begin{minipage}{ .38\linewidth }
\includegraphics[width=\linewidth]{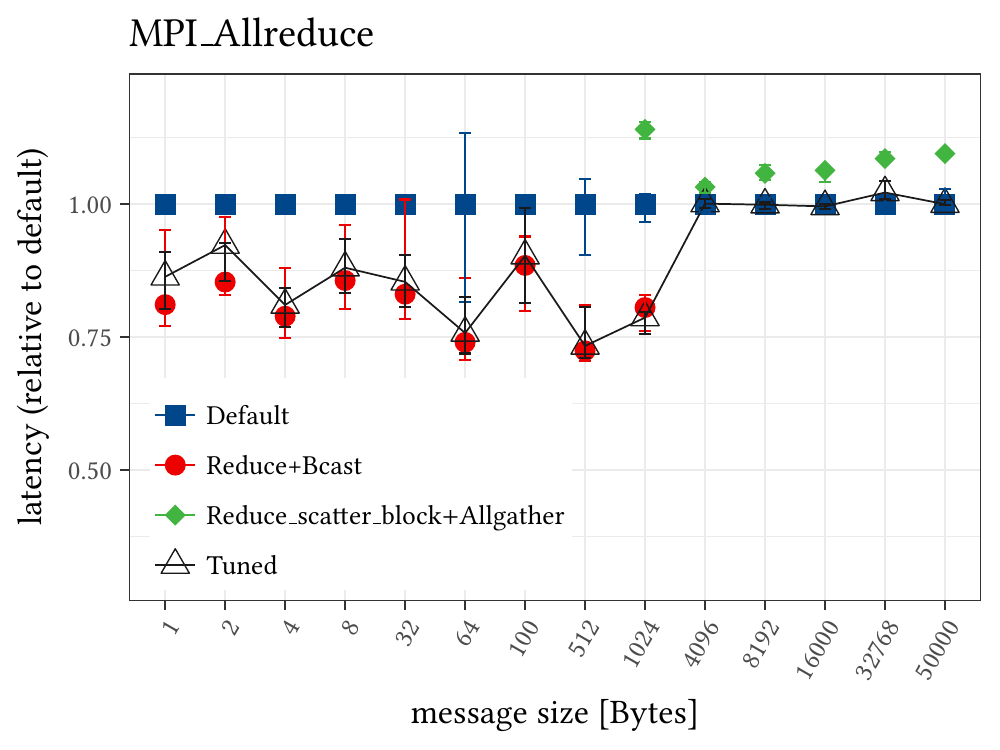}
\end{minipage}%
\\%
\begin{minipage}{ .38\linewidth }
\includegraphics[width=\linewidth]{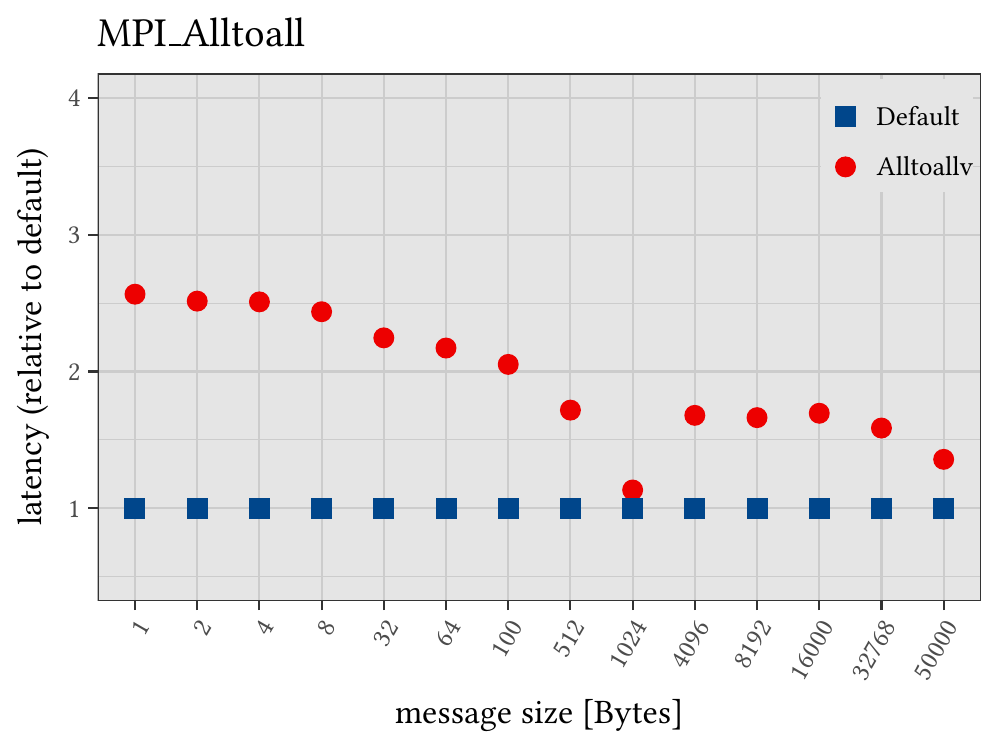}
\end{minipage}%
\hspace*{0.1\linewidth}%
\begin{minipage}{ .38\linewidth }
\includegraphics[width=\linewidth]{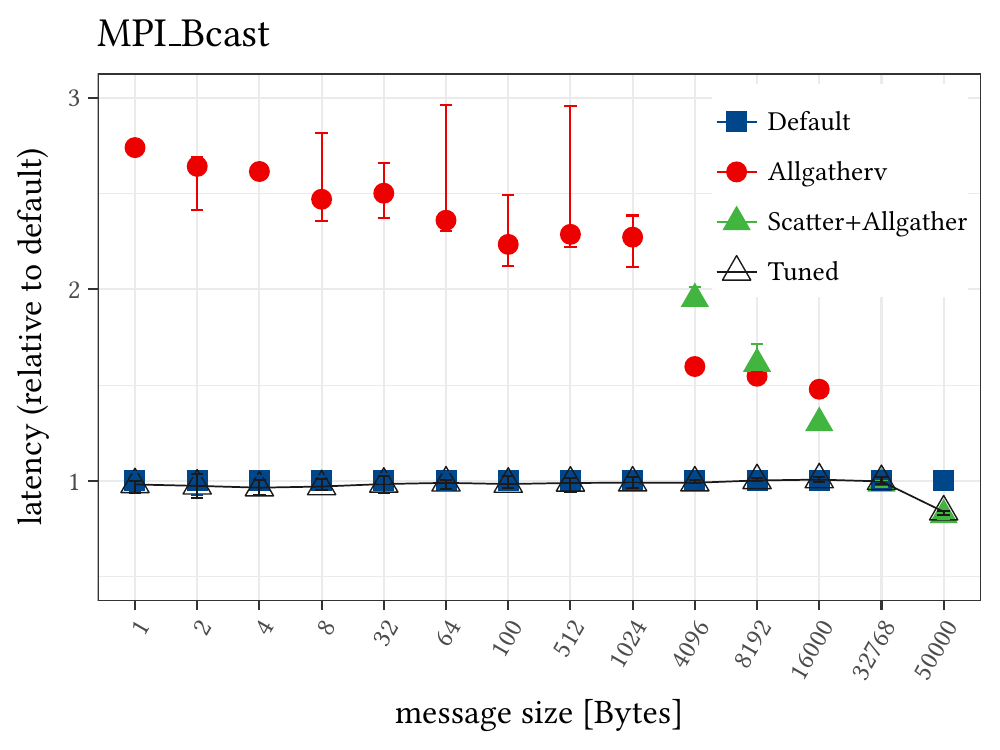}
\end{minipage}%
\\%
\begin{minipage}{ .38\linewidth }
\includegraphics[width=\linewidth]{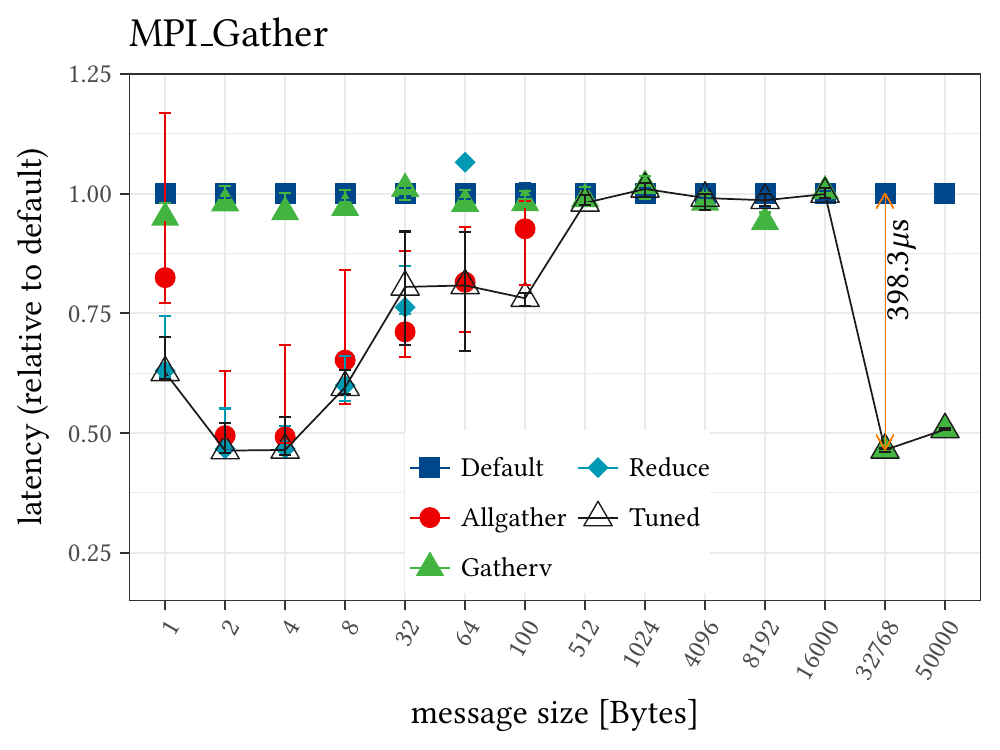}
\end{minipage}%
\hspace*{0.1\linewidth}%
\begin{minipage}{ .38\linewidth }
\includegraphics[width=\linewidth]{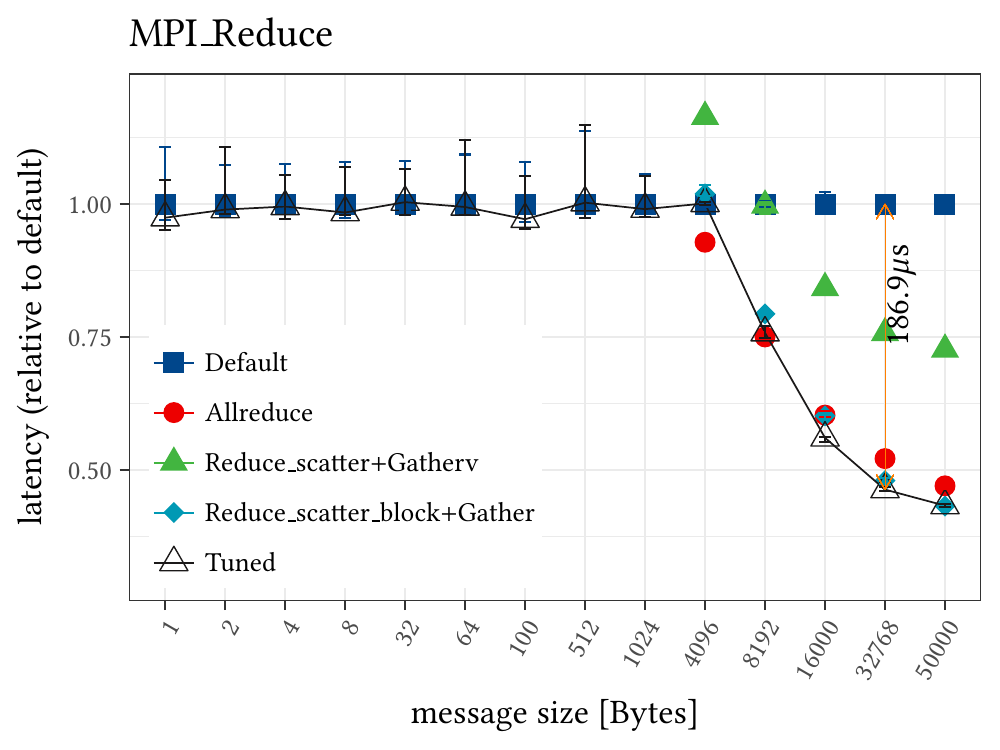}
\end{minipage}%
\\%
\begin{minipage}{ .38\linewidth }
\includegraphics[width=\linewidth]{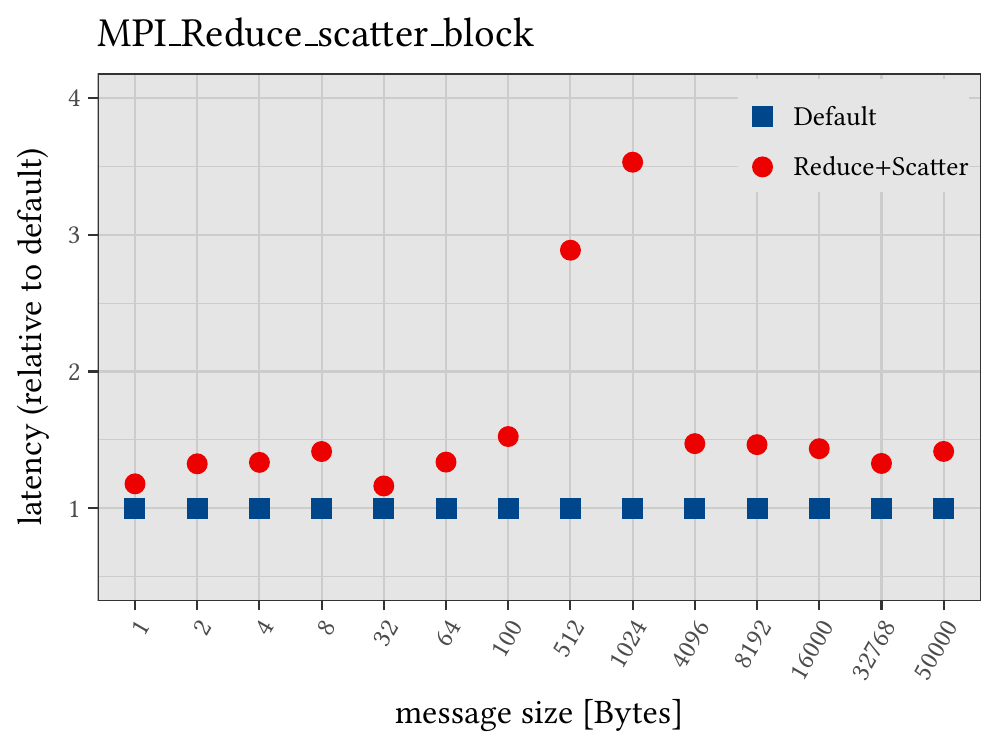}
\end{minipage}%
\hspace*{0.1\linewidth}%
\begin{minipage}{ .38\linewidth }
\includegraphics[width=\linewidth]{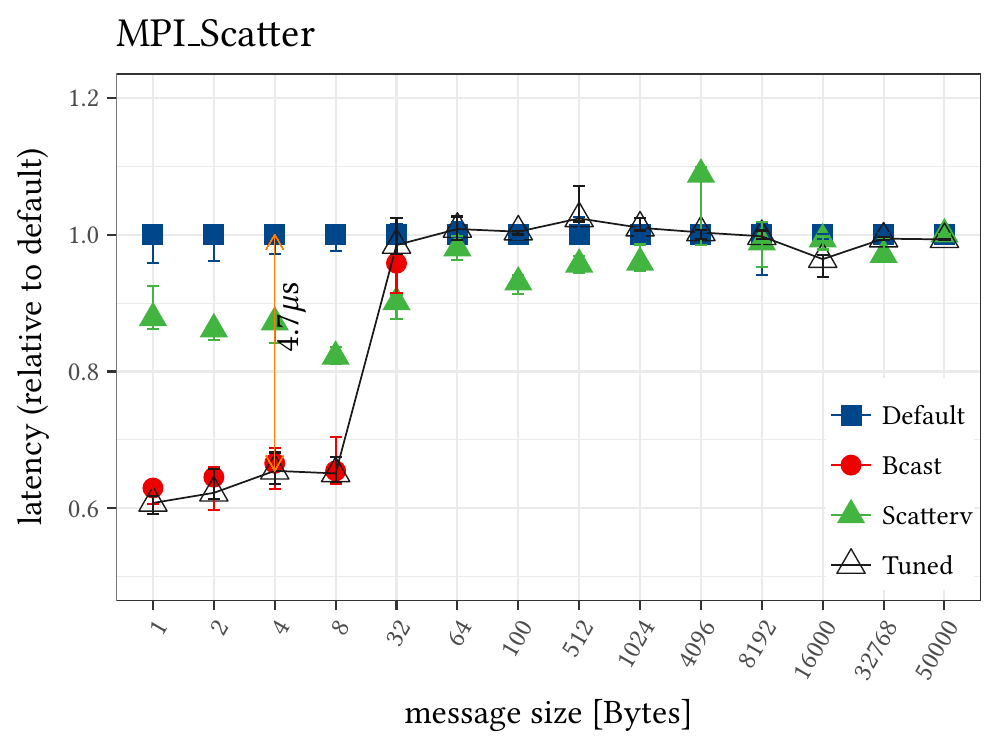}
\end{minipage}%
\caption{\label{fig:perf_32x1_jupiter_mvapich_all}%
Performance comparison between \pgdefault and \pgtuned version of \jupitermvapich (\num{32x1} processes, \machjupiter)%
}%
\end{figure*}

\begin{figure*}[htpb]
\centering
\begin{minipage}{ .38\linewidth }
\includegraphics[width=\linewidth]{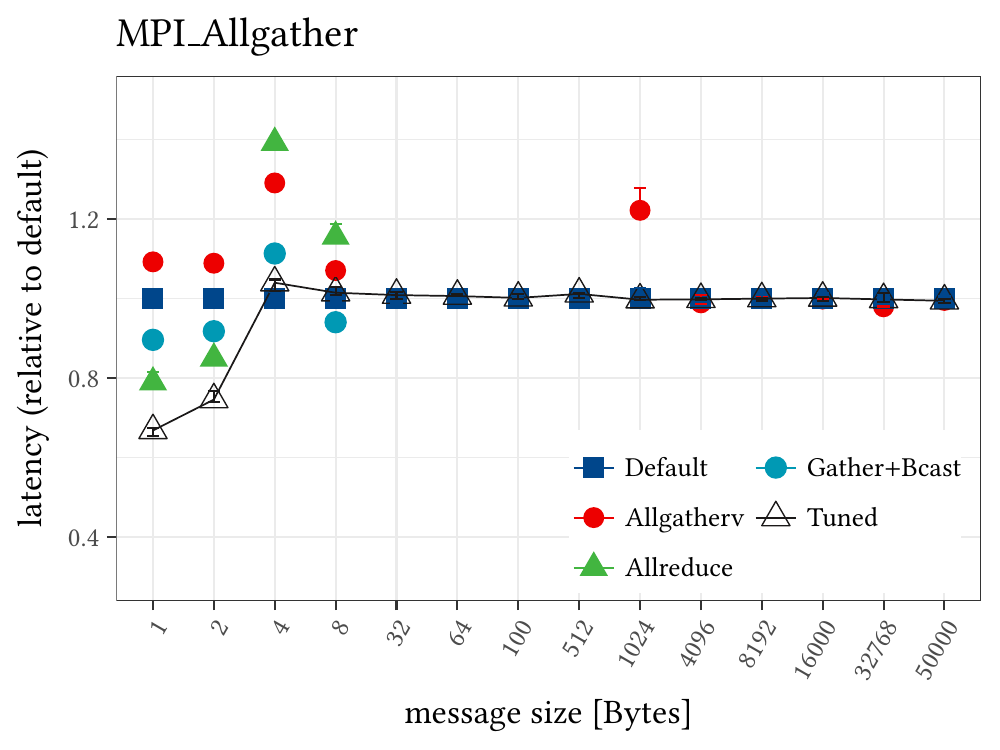}
\end{minipage}%
\hspace*{0.1\linewidth}%
\begin{minipage}{ .38\linewidth }
\includegraphics[width=\linewidth]{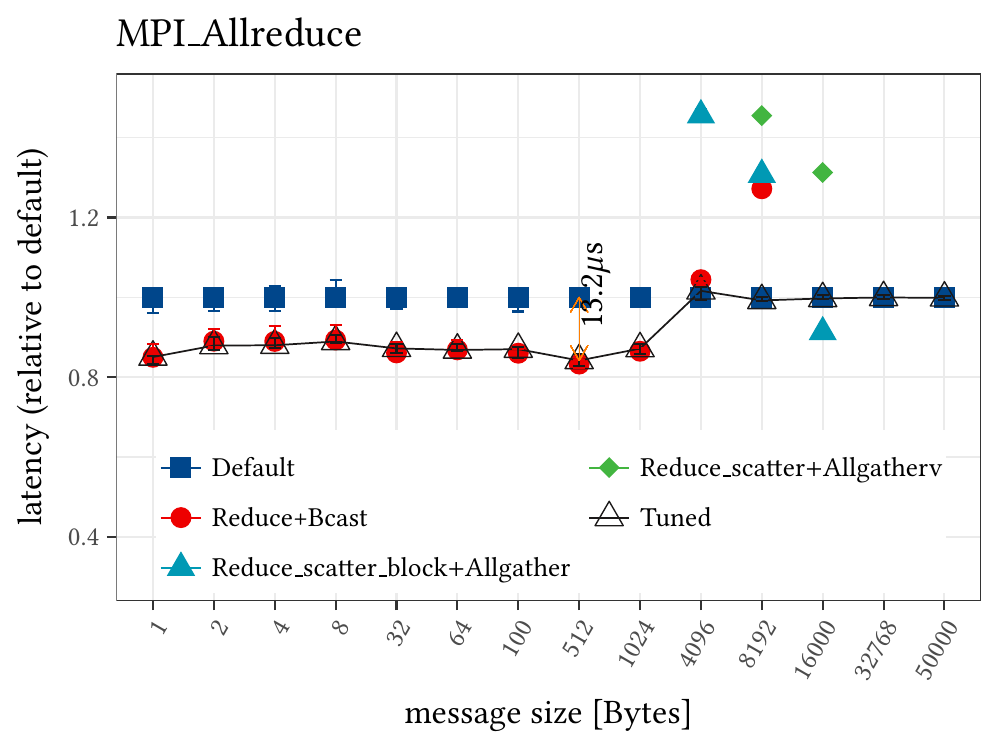}
\end{minipage}%
\\%
\begin{minipage}{ .38\linewidth }
\includegraphics[width=\linewidth]{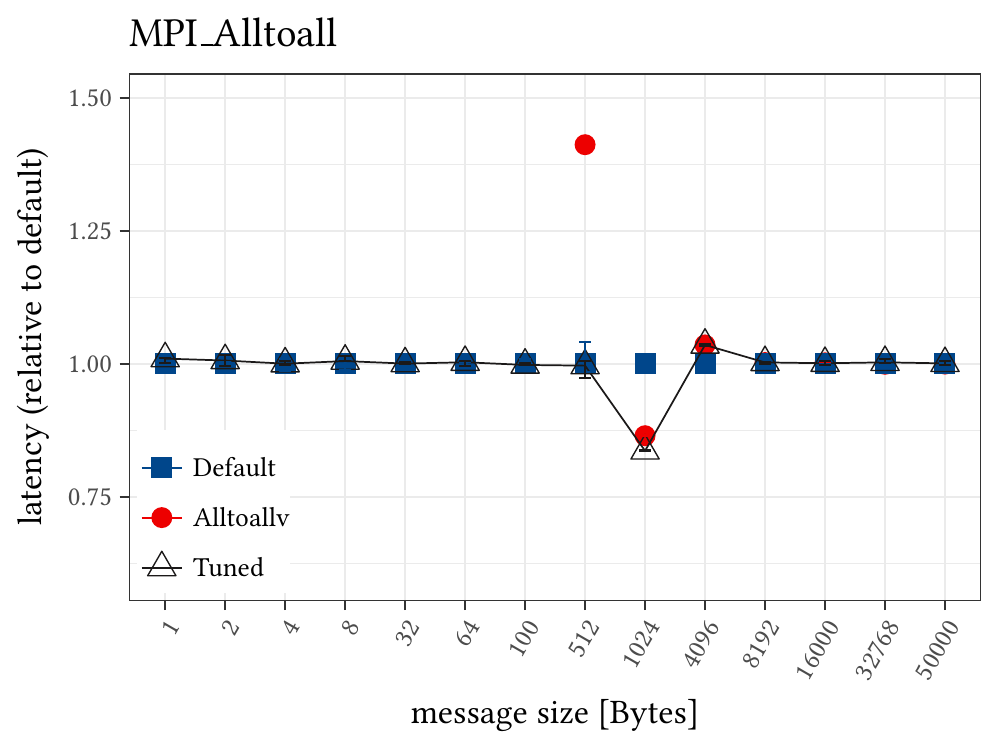}
\end{minipage}%
\hspace*{0.1\linewidth}%
\begin{minipage}{ .38\linewidth }
\includegraphics[width=\linewidth]{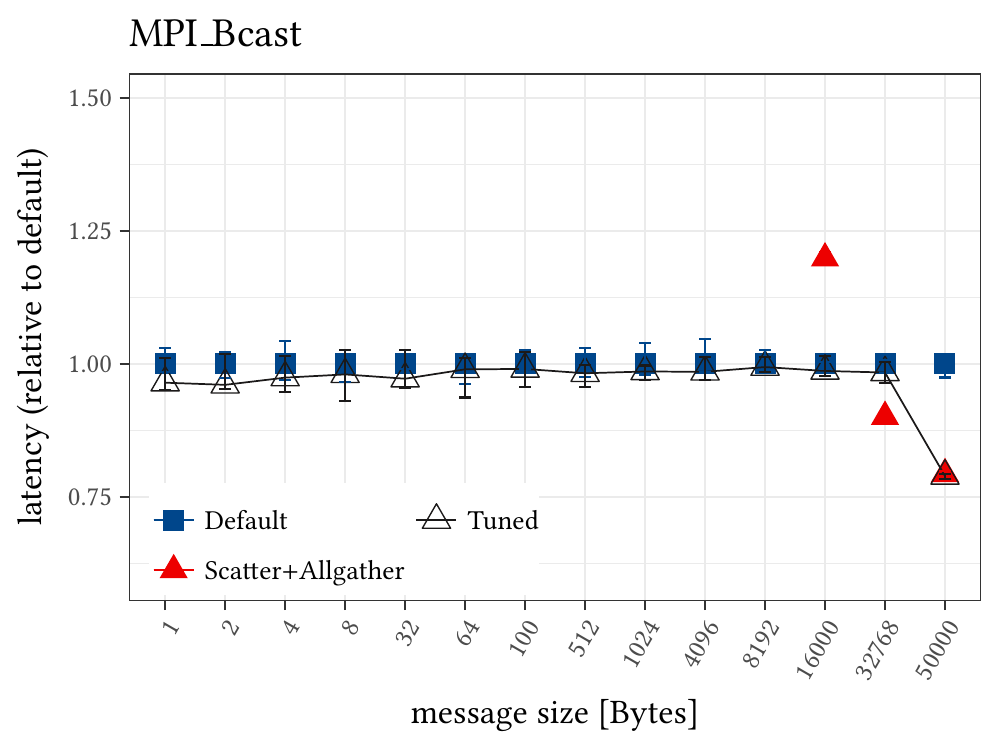}
\end{minipage}%
\\%
\begin{minipage}{ .38\linewidth }
\includegraphics[width=\linewidth]{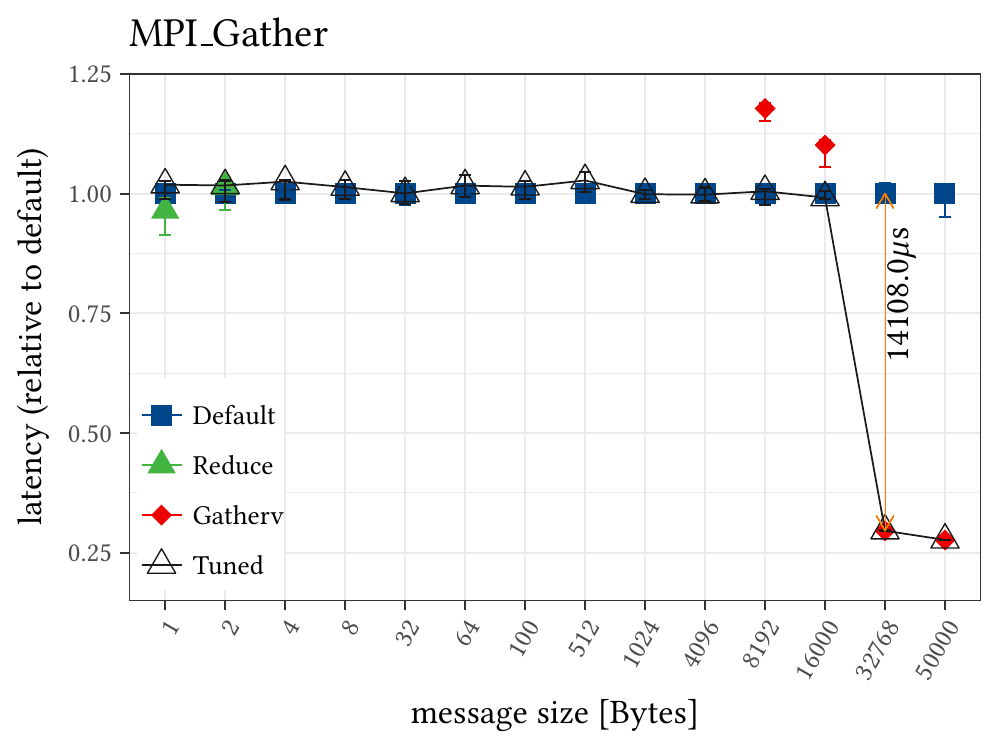}
\end{minipage}%
\hspace*{0.1\linewidth}%
\begin{minipage}{ .38\linewidth }
\includegraphics[width=\linewidth]{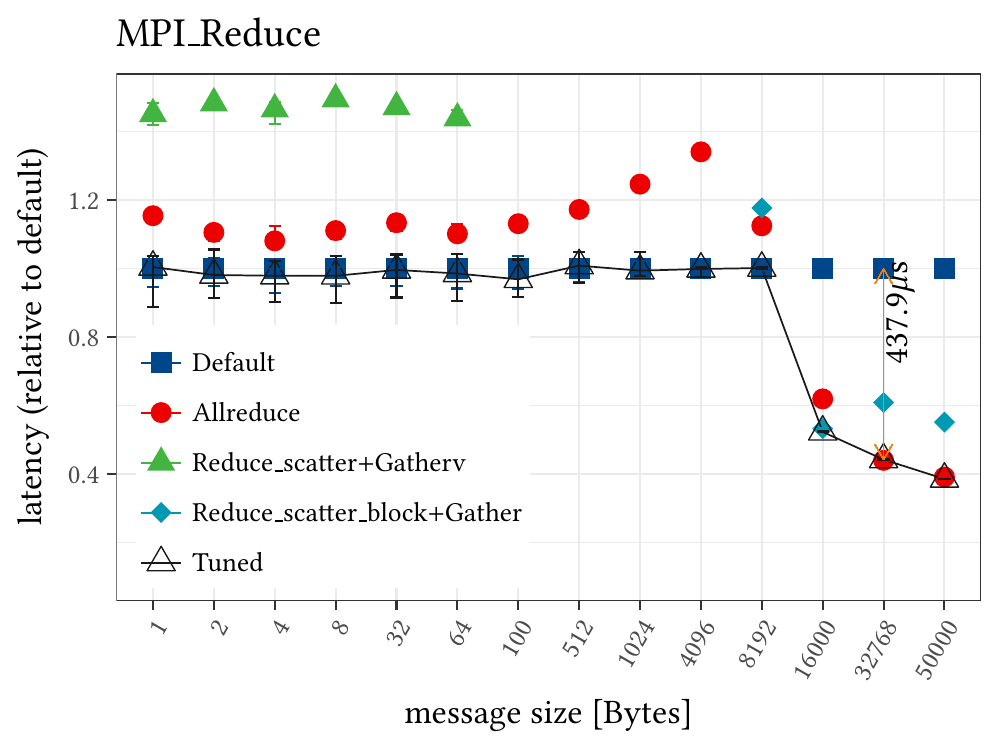}
\end{minipage}%
\\%
\begin{minipage}{ .38\linewidth }
\includegraphics[width=\linewidth]{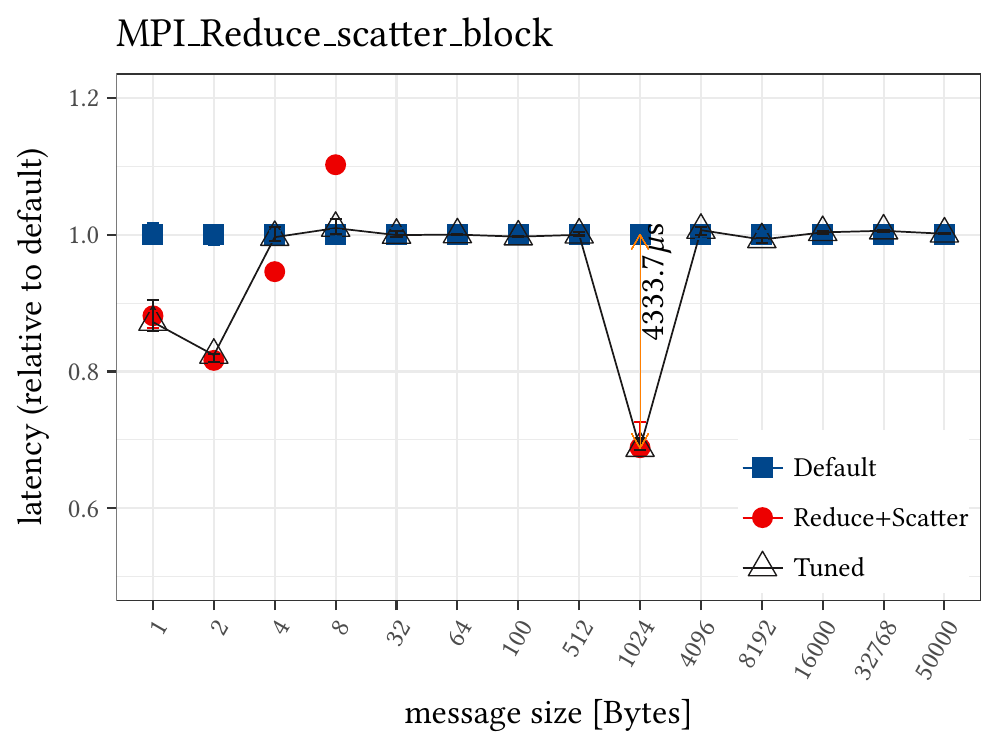}
\end{minipage}%
\hspace*{0.1\linewidth}%
\begin{minipage}{ .38\linewidth }
\includegraphics[width=\linewidth]{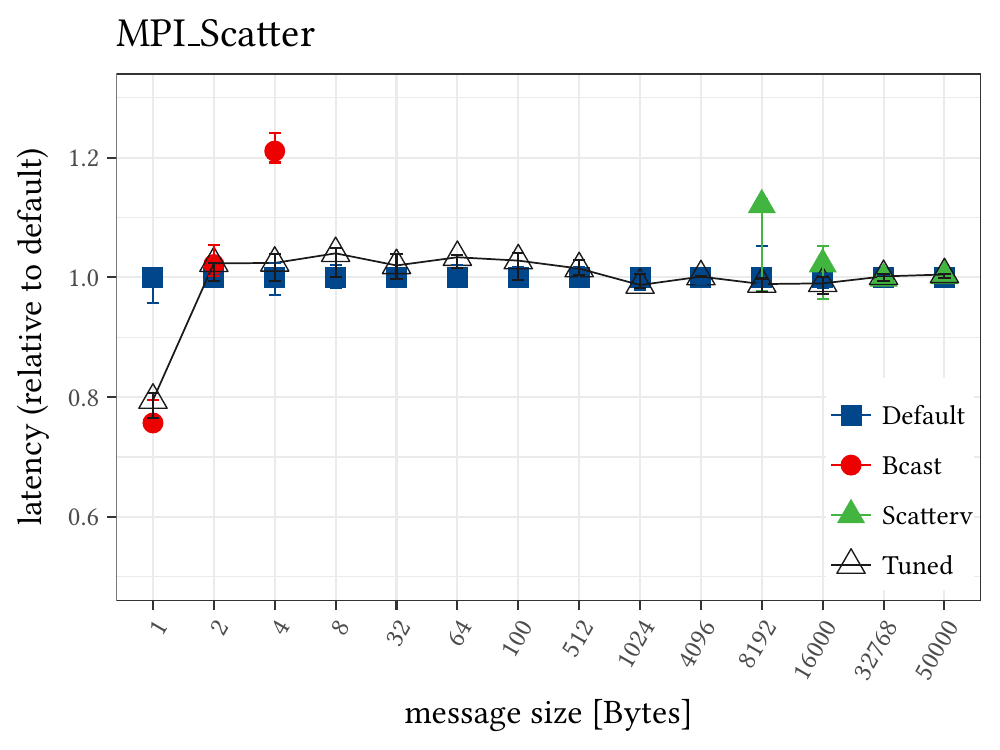}
\end{minipage}%
\caption{\label{fig:perf_32x16_jupiter_mvapich_all}%
Performance comparison between \pgdefault and \pgtuned version of \jupitermvapich (\num{32x16} processes, \machjupiter)%
}%
\end{figure*}

\end{taggedblock}

\end{document}